\documentclass[11pt,a4]{article}
\usepackage{epsfig,graphics}
\usepackage{caption}
\newcommand{\dbar}{\mathchar'26\mkern-12mu d}     
\begin{document}
\title{Probabilistic Foundations of Statistical Mechanics: a Bayesian Approach}
\author{B. Buck and A.C. Merchant \\ 
Department of Physics, University of Oxford, Theoretical Physics, \\
1 Keble Road, Oxford OX1 3NP, UK. 
}%
\bigskip
\baselineskip 25.5pt
\maketitle
\bigskip
\begin{abstract}
We examine the fundamental aspects of statistical mechanics, dividing the problem
into a discussion purely about probability, which we analyse from a Bayesian standpoint,
and a consideration of how the ensuing statistical results can be related to experimental
questions. Although mainly concerned with a quantum description of the mechanics, we
link to classical equivalent aspects where appropriate. We present in some detail a simplified 
and tractable version of the calculus of probabilities originally introduced by Cox
and use it to set up a convenient and consistent methodology for combining and manipulating 
probability symbols and subsequently to associate definite and unambiguous numerical
magnitudes with appropriate symbols of that calculus where possible. We argue that
the existence of a unique maximising probability distribution $\{p(j\vert K)\}$ for 
states labelled by $j$ given data $K$ implies that the corresponding maximal value of the 
information entropy $\sigma(\{(p_j\vert K)\}) = -\sum_j (p_j \vert K)\ln{(p_j\vert K)}$
depends explicitly on the data at equilibrium and on the Hamiltonian of the system. 
As such, it is a direct
measure of our uncertainty about the exact state of the body and can be identified 
with the traditional thermodynamic entropy. We consider the well known microcanonical,
canonical and grand canonical methods and ensure that the fluctuations about mean values
are generally minuscule for macroscopic systems before identifying these mean values 
with experimental observables and thereby connecting to many standard results from thermodynamics. 
Unexpectedly, we find that it is not generally possible for a quantum process to be both isentropic 
and reversibly adiabatic. This is in sharp contrast to traditional thermodynamics where
it is assumed that isentropic, reversible adiabatic processes can be summoned up on demand
and easily realised. By contrast, we find that linear relations between pressures 
$P_j$ and energies $E_j$ are necessary and sufficient conditions for a
quasi-static and adiabatic change to be isentropic, but, of course, this relationship 
only holds for a few especially simple systems, such as the perfect gas,
and is not generally true for more complicated systems. By considering the associated
entropy increases up to second order in small volume changes we argue that the consequences 
are in practice negligible.

\end{abstract}
\centerline{PACS numbers}
\vfil
\newpage
\section{Introduction}

\noindent Since the foundational work of Maxwell \cite{[Maxwell]}, Boltzmann \cite{[Boltzmann]}
and Gibbs \cite{[Gibbs]} the field of equilibrium statistical mechanics has evolved into
a mature discipline, yet certain aspects of its development and interpretation remain controversial. 
The strongest version of the ergodic hypothesis (in essence that all phase space states of a system 
are sampled during a process) held dear by Maxwell and Boltzmann is today generally considered untenable. 
In his classic textbook, Ma \cite{[Ma]} ignores it completely so that it does not even appear in the Index. 
Boltzmann's H-theorem, in attempting to explain how a favoured direction of time can emerge from 
microscopic equations that are themselves time reversible, has probably caused more confusion than 
enlightenment and raised more questions than it has answered. The arrow of time remains as
mysterious today as ever, with explanations in terms of initial conditions only begging the
question of how those conditions came into being. Even the idea of ensembles, introduced and
used to such exquisite effect by Gibbs, can be viewed as an inessential sop to those who will
only accept an interpretation of probability based on frequency of occurrence.

The field is well served by a series of high profile classic, even legendary, textbooks by
authors such as Pauli \cite{[Pauli]}, Schr"odinger \cite{[Schroedinger]}, Tolman, \cite{[Tolman]},
Landau and Lifschitz \cite{[LandauLifschitz]}, Feynman \cite{[Feynman]}, Kubo \cite{[Kubo]},
Huang \cite{[Huang]}, Reichl \cite{[Reichl]} and many, many more. Despite this ocean of high
quality material nobody would claim that the last definitive word has been said in this area.
We believe that the emergence and growing acceptance of the Bayesian interpretation of
probability warrants an attempt to integrate it into a coherent description of equilibrium
statistical mechanics, and we aim to take a step in this direction in this article.

We begin by pointing out that the probabilistic and mechanical aspects of the subject are
really quite distinct from one another, and that despite the undoubted technical difficulties of
the latter it does not present many unsolved questions of principle. Indeed, most of thermodynamics 
has evolved by avoiding any detailed discussion of the internal mechanics of its systems at all. 
We therefore concentrate our initial attention on the statistical problem. Although at first
sight the amount of information to be encoded in a probability distribution for a macroscopic system
appears overwhelming,
in practice we are usually only concerned with reproducible experimental results that depend on just a
few parameters. The restricted objective is then to find general relations between some small number of
macrosopic parameters and to go on to calculate values of observable quantities using a minimum
of additional microscopic information. Most of these results pertain to equilibrium situations, so 
we discuss this concept, and the related one of irreversibility, in some detail (although touching
only briefly on the mechanisms for relaxation to equilibrium). We conclude that the pivotal point
is to estimate the probabilities with which the energy eigenstates of the system occur, since
this leads to a means of calculating the final expectation values of physical quantities, which
may then be compared with their experimental counterparts.
We illustrate this conclusion with an early application of microcanonical probabilities,
which is to say basically that we take all states within a narrow energy range to be equally likely
to occur and so assign them equal probabilities of being the actually occurring state. This is
really just Bernoulli's principle of indifference - a means of reasoning when having incomplete
knowledge. In view of the enormous number of accessible states, the success of this postulate
must indicate that practically all such states correspond to nearly the same macroscopic properties,
and that the statistical averaging is merely preventing the accidental selection of an untypical state.
Now in a position to evaluate moments of the distribution, we demonstrate the important point that it 
only makes sense to identify the calculated mean value with the experimental observable if the standard 
deviation from the mean is sufficiently small. We do this by considering the pressure exerted 
by an isolated system, having a measured energy, on its confining walls. We find that the conditions
for statistical methods to give sharp numerical results are the same as those required for
experimental reproducibility and that the number of accessible states $\Omega$ is closely 
related to the experimental entropy.

Before going any further it is essential to give serious consideration to the question of
what we mean by probability. We take the Bayesian view that probability provides a numerical
measure of the degree of belief in a proposition based  on the available evidence. The major
early proponents of this view were Laplace \cite{[Laplace]} in the 19$^{\rm th}$ century 
and Jeffreys\cite{[Jeffreys]} in the first half of the 20$^{\rm th}$. 
E.T.Jaynes\cite{[ETJ_book]} has been
a stalwart champion of this approach in the second half of the 20$^{\rm th}$ century. We use 
probability calculus as a technique for plausible inference and a method for conjectural reasoning,
rather than viewing it as an intrinisic measurable property of a given physical event independent
of any prior knowledge. We follow Cox \cite{[Cox46]} in setting up a consistent calculus for 
combining and manipulating probability symbols. Cox's work is not widely known, but is entirely 
accessible to physicists, so we endeavour to explain it as comprehensibly as we can. Cox
argues that there is only one possible algebra of probable inference and deduces definite
quantitative rules for the combination of probabilities. We hope our explanatory efforts will 
lead the reader to agree with him in due course. A second major problem in probability theory
is to develop a tractable theory of probability assignments. For this we follow 
Shannon \cite{[Shannon]} and Jaynes \cite{[Jaynes1]} in associating a quantity called the
``uncertainty'' with a probability distribution over a set of propositions. With a few 
plausible assumptions about its behaviour and a condition of self-consistency Shannon
found a simple, unique function of the probabilities with the desired characterisitcs to
act as this ``uncertainty''. In Section 3 we shall sketch his proof before moving on to Jaynes' use
of ``uncertainty'' as a means of assigning probabilities by what he calls the principle
of maximum entropy.

The main idea is to make an impartial assignment of probabilities by maximising the information
entropy $\sigma(\{p_j\}) = -\sum_j p_j\ln{(p_j)}$,
subject to the constraints implied by the data $K$ and the normalisation condition. Mean
values of observables may then be calculated for comparison with experimental measurements.
This function $\sigma_{\rm max}(\{p(j\vert K)\})$ possesses all the properties of the
thermodynamic entropy $S$, and we aim to identify it as such, in suitable units.
In particular, this requires that we show it to be an extensive function of state, defined for 
equilibrium conditions, that does not decrease in adiabatic processes and is related
experimentally to heat and temperature. We address these questions in Section 4

The identification of the theoretical quantity $\sigma_{\rm max}$ with the phenomenological
entropy $S$ simplifies the development of statistical mechanics. The absolute temperature
$T$ can be then be introduced into the formalism via the traditional thermodynamical
arguments involving the second law. In fact, $T$ can be derived from theory using the relation
$\partial \sigma_{\rm max}/\partial\bar{E} = 1/(k_BT)$. This requires an explanation
of how to calculate the absolute maximum of the information entropy, and the
associated probability distribution in terms of the assumed data. This requires a consideration
of how to apply a constrained maximisation and how to ascertain that it is indeed the globally
largest maximum achievable. In section 5 we will examine this question within the contexts
of the traditional microcanonical, canonical and grand canonical methods where only probability 
is constrained to be normalised, then the mean energy is constrained to a certain value and 
finally the mean particle number is contrained to a given value. Lagrange multipliers
are introduced to handle these calculations and are themselves found to have important
physical significance, being related to the equilibrium temperature and chemical potential.
We will touch briefly on the third law and the interpretation of negative temperatures,
but do not go deeply into these areas since they are more specialised than our current
scope warrants.

With the identifications mentioned above in place, we are in a position to derive
equations relating the macroscopic and experimentally accessible functions of state
to quantities derived by probabilisitic methods from the microscopic structure and 
equilibrium data. In many cases, these are simply differently focused derivations of
already well known relations. However, there is one tantalising surprise requiring deeper
investigation, concerning the question of whether quasi-static and reversible adiabatic
processes are also necessarily isentropic. Such processes are easily defined in thermodynamics,
making up essential legs of a Carnot cycle, for example, and are assumed always to be
possible. Nevertheless, it is not clear that, for an arbitrary system, their existence
is compatible with quantum mechanics and our interpretation of entropy as maximised 
uncertainty. In section 6 we show that such processes, requiring the entropy to remain 
constant, are only possible if the energy eigenspectrum at the end is uniformly
expanded or compressed with respect to the initial eigenspectrum in the ratio of the 
final to initial temperatures. This can happen for some very simple, idealised systems 
such as the perfect gas, but cannot be achieved more generally. In fact, we show that
it is equivalent to requiring the correlation coefficient between energy and pressure 
deviations to vanish, which in turn requires $P_j = aE_j + b$ (where $a$ and $b$ are constants)
for all states $j$, which is a very special restriction unlikely to be obeyed by any real systems.
We discuss this surprise more thoroughly and resolve the apparent discrepancy between
thermodynamics and statistical mechanics at the end of section 6,  

\vfil
\newpage
\section{Physics and probability}
\subsection{Separation of problems}

\noindent There are two outstanding conceptual difficulties in the study of statistical mechanics.
The first is to understand why the ideas of probability are needed at all, and the second is to
see why the statistical methods which have been invented can ever give clear-cut
answers to experimental questions. It should be firmly grasped at the 
outset that the probabilistic and mechanical aspects of the subject are
quite distinct and that the statistical problem can be solved completely
once and for all. The application of this solution to real physical systems
usually involves great technical difficulties having to do with
the mechanics, but there are few questions of principle left to worry
about. Most often we have in the end to be content with rather drastic 
approximations or model assumptions when trying to predict and
understand actual experimental observations.

On the other hand it is frequently possible to understand how various
observables of a system are related to each other without using any
specific knowledge of the internal mechanics. This is the subject matter 
of thermodynamics, which relies heavily on the non-mechanical notions of
entropy and temperature. The implication is that these latter concepts
are fundamentally statistical in character and that the basic laws of
thermodynamics follow naturally from considerations of probability. A
demonstration of the truth of these statements is an important secondary
aim of statistical mechanics.

The typical problem in statistical mechanics is to elucidate the
measurable  and reproducible properties of a macroscopic system. In
principle, mechanics alone should suffice. To see what is involved,
consider in outline how a calculation of physical observables might
conceivably be attempted using either classical mechanics or quantum
mechanics. The methods employed in these two theories are very
different, but the logical structure of the two types of calculation is
the same. Corresponding steps in the classical and quantum approaches
can be set out for comparison as follows:
\begin{center}
\begin{table}[htb]
\caption{Classical and Quantum Correspondence}
\smallskip
\begin{tabular}{lll}
\hline
\smallskip
& ${\underline{\rm CLASSICAL\ MECHANICS}}$ & ${\underline{\rm QUANTUM\ MECHANICS}}$ \\
Specification of system & Hamiltonian\ function & Hamiltonian\ operator \\
& $H=H(\{x_i,p_i\},t)$ & $\hat{H} = \hat{H}(\{\hat{x_i},\hat{p_i}\},t)$ \\
& & \\
Initial state (at time $t=0$) & Canonical variables & State function \\
& $ \{x_i(0), p_i(0) \}$ & $\Psi(\{x_i\},t=0)$ \\
& & \\
Equations of motion & Hamilton's equations & Schr\"odinger equation \\
& $\dot{x_i}={\partial H \over \partial p_i }, \ \dot{p_i} = -{\partial H \over \partial x_i}$ &
$\hat{H}\Psi = i\hbar{\partial \Psi \over \partial t}$ \\
& & \\
Final state (at time t) & Evolved variables & Evolved state \\
& $ \{x_i(t), p_i(t) \}$ & $\Psi(\{x_i\},t)$ \\
& & \\
Observable property & Phase function & Expectation value \\
& $Q( \{ x_i(t), p_i(t) \})$ & $\langle \Psi(t) \vert \hat{Q} \vert \Psi(t) \rangle $ \\
\hline
\end{tabular}
\end{table}
\end{center}
These familiar expressions need only a little in the way of explanatory
comment. The specification of the nature of the system is considered
complete when we have enough information to construct the Hamiltonian
function or operator, which is in most examples an expression for the 
energy. This Hamiltonian may have an explicit time dependence arising
from variable external forces applied to the system. To start the
calculation it is also necessary to know the initial state. This is
specified classically as the initial position, in a many-dimensional
phase space, of the point representing the system. In quantum theory the
state-vector at time $t=0$ is needed.

The equations of motion in both theories are of first order in time and
hence can, in principle, be integrated to give the state of the system
at any later time $t > 0$. Finally, the evolved value of an observable may
be computed from the variables in classical mechanics by finding the
magnitude of the appropriate phase function. The corresponding quantity in 
quantum mechanics is estimated as the expectation value of a definite
operator with respect to the evolved final state-vector.

Since these operations all seem perfectly well defined the question
arises as to why statistical theory is needed. One occasionally
suggested reason is that the equations of motion are so impossibly
complicated that the introduction of probability considerations will
somehow enable us to avoid solving them. The dynamical equations are
indeed almost always intractably difficult; however, they can not be
simplified by probability arguments. In many applications of statistical
mechanics it is in fact possible to dispense with the equations of
motion entirely, but the reasons for this have nothing to do with
probability theory {\it per se\/}; and in any serious calculation there will
still usually be a non-trivial mathematical problem of pure mechanics to
be faced squarely.

The true situation is that any approach using mechanics only is bound to
be an empty exercise since the information necessary to define the
initial state is never available. Setting up a classical initial state
would require knowledge of something of the order of $10^{23}$ pairs of
canonically conjugate variables $ \{x_i(0), p_i(0) \}$; a similar amount of data
in the form of quantum numbers would be needed to specify an initial
quantum state $\Psi(t=0)$. But the actual knowledge available usually 
consists of the values of just a few, say two or three, macroscopic
observables of the system, and we are forced to guess what state occurs
on the basis of this very slender data. It is at this point that 
probability and statistics enter our subject since these are the tools
for informed guesswork. Still, it certainly appears as if the problem is
absurdly underdetermined. Any guesses we could possibly make must have
an extremely small probability of being correct and it is at first sight
miraculous that firm predictions of other observables can be attempted.

Before proceeding to technical details it is perhaps worthwhile to give
a rough preliminary answer to the question of how it is possible for our
crude guesses to be useful, based as they are on such laughably small
amounts of information. The clue is that we choose to record in
experimental work only the observationally {\it reproducible \/} results and
relations. Hence if it is true experimentally that measurement of just a 
few parameters of a system suffices for accurate and reproducible
estimates of other measurable quantities, then all the fine details of the 
exact state of the system must be {\it irrelevant\/}.

What is needed theoretically is some highly efficient machinery for
bypassing all the redundant details and it is precisely such a beautiful
technique that is provided by the general theory of probabilistic
inference. We shall apply this method almost exclusively in the context
of the quantum mechanics of discrete states. A corresponding conceptual
framework can be developed in connection with classical mechanics, but
it is much harder to justify and understand; it is also basically
unnecessary, since the characteristic classical expressions can usually
be deduced straightforwardly from the quantum results.

The sorts of reproducible information about a system that are 
practically obtainable can be classified into the three main types
mentioned below, though we do not attempt to give an exhaustive list
of examples in each category.

\begin{itemize}
\item{(a)} Macroscopic Internal Parameters: including such quantities as
the total energy (measured relative to the energy of an arbitrary reference 
state), the total mass of each constituent, the pressure 
exerted on the retaining walls, possible electromagnetic moments and
various kinds of mechanical strains in the material. Something may
also be known about the previous internal history.
\item{(b)} Macroscopic External Parameters: which generally define the 
environment of the system; whether, for example, it is confined in 
some definite volume, is acted upon by force fields and mechanical
stresses, or is placed in contact with another body with which it can
exchange energy or matter. A record of previous manipulations may also 
be available.
\item{(c)} Microscopic Specification: comprising a knowledge of the kinds
and numbers of atoms in a system, with their masses, spins, magnetic
moments, spectra, mutual interactions and spatial arrangements,
together with the forces acting on them.
\end{itemize}

Our object is first to establish general relationships between the
macroscopic parameters listed under (a) and (b) above (that is, to
derive thermodynamics); and second to calculate definite values for the
observables using the microscopic information (this of course goes
beyond thermodynamics).

\subsection{Relaxation to equilibrium}

\noindent The main accepted results in thermodynamics and statistical mechanics
refer to a special condition of systems called equilibrium, and the next
order of business is to discuss this important and slippery concept. A 
related and controversial topic is irreversibility, the indisputable
tendency for systems to end up in the condition of equilibrium even
though other behaviour is not forbidden by the laws of mechanics.

The basic experimental fact is that if a macroscopic body is isolated
from external influences, except for the presence of static fields and
stresses, and it is left to itself for long enough, then its global
observable properties will eventually cease to change. The time taken to
reach this apparently stationary condition may be fractions of a second
or many years, but some period of aging is generally necessary.
Furthermore, for the same kind and size of system under the same
external conditions, the observed macroscopic quantities have the same
final values when equilibrium is reached, which is to say that the
results are reproducible. Restricting attention to such a static
situation will clearly simplify matters enormously since the equations 
of motion can be discarded. The crucial point is that when our
information no longer changes with time any guess as to what microscopic
quantum state actually obtains should also be time independent. Once we
have assigned probabilities to the possible states on the basis of some
measurements, we can proceed immediately to the final step of estimating
observables by means of expectation values. Hence the only conceptual 
problem left is how to guess the likely states.

These simplifications occur only for the particular global parameters
that we have used in judging whether systematic changes in the body have
stopped. If we make measurements on too fine a scale in space and time,
as for instance rapidly repeated observations of the number of particles
in a small part of the system, the results may well fluctuate in time in
an unpredictable way. It can happen that the patterns of fluctuation 
themselves have static and reproducible features, but a detailed
treatment of noise spectra and similar topics is outside our scope.

The further questions of why systems relax towards equilibrium and how
to calculate the rates at which they do so after some disturbance are
also difficult problems. We can, if we like, just take the existence of
effective equilibrium situations as a brute fact of nature, and try to
describe such conditions, while at the same time hoping to discover how
the different equilibria are related. But it is useful and illuminating
to discuss briefly the origin of the time dependent relaxation and
fluctuation phenomena.

Consider then an isolated system prepared at time $t=0$. It will have a
definite Hamiltonian operator $\hat{H}$ and a definite state vector $ \vert \Psi(t) \rangle$
obeying the Schr\"odinger equation
\begin{equation}
\hat{H}\vert \Psi(t) \rangle = i\hbar {\partial \vert \Psi(t) \rangle \over \partial t}.
\end{equation}

We take the initial state to be normalized by $\langle \Psi(0) \vert \Psi(0) \rangle = 1$ and it
then follows easily from the Schr\"odinger equation that the state remains 
normalized at later times, i.e. $\langle \Psi(t) \vert \Psi(t) \rangle = 1$ for $t>0$. Typically,
the system will have no particular rotational or translational
symmetries, but there always exists at least one other constant of the
motion apart from the normalization, namely the expectation value of the
total energy:
\begin{equation}
\langle E \rangle_t = \langle \Psi(t) \vert\hat{H} \vert \Psi(t) \rangle = {\rm constant}.
\end{equation}
This also follows directly from the Schr\"odinger equation but is easier
to understand if we make use of the special solutions of that equation
represented by:
\begin{equation}
\vert \Psi_j(t) \rangle = \vert \psi_j \rangle\exp{(-iE_jt/\hbar)},
\end{equation}
where the $\vert \psi_j \rangle$ satisfy the time-independent equations
\begin{equation}
\hat{H} \vert \psi_j \rangle = E_j\vert \psi_j \rangle,
\end{equation}
and 
\begin{equation}
\langle \psi_j \vert \psi_k \rangle = \delta_{jk},
\end{equation}
A general normalized state function may be written as a linear
combination of these energy eigenstates in the form:
\begin{equation}
\vert \Psi(t) \rangle =  \sum_j a_j \vert \psi_j \rangle\exp{(-iE_jt/\hbar)}
\end{equation}
where $\sum \vert a_j \vert^2 = 1$. Hence we have easily that
\begin{eqnarray}
\langle \Psi(t) \vert \hat{H} \vert \Psi(t) \rangle&=&\sum_{jk} a_j^{\ast}a_k
\langle \psi_j \vert \hat{H} \vert \psi_k \rangle \exp{(i(E_j-E_k)t/\hbar)} \nonumber \\
&=&\sum_{jk} a_j^{\ast}a_k E_k
\langle \psi_j \vert \psi_k \rangle \exp{(i(E_j-E_k)t/\hbar)} \nonumber \\
&=&\sum_j \vert a_j \vert^2 E_j = {\rm constant} 
\end{eqnarray}
If some other observable corresponds to an operator $\hat{C}$ which commutes
with $\hat{H}$ then the energy eigenstates may be chosen to be simultaneous
eigenstates of $\hat{C}$, and a similar argument shows that $\langle C \rangle_t = 
\langle \Psi(t) \vert \hat{C} \vert \Psi(t) \rangle$
is also a constant of the motion. Note that these results depend only on 
the system being isolated and not on any assumed condition of equilibrium.

When an observable has an operator $\hat{Q}$ which does not commute with $\hat{H}$ it is
a little harder to see how it could eventually come to have an almost
constant expectation value. For in this case the energy eigenstates can 
not be simultaneous eigenstates of $\hat{Q}$  and we are left with the general 
expression for $\langle \hat{Q} \rangle_t$ given by
\begin{eqnarray}
\langle \Psi(t) \vert \hat{Q} \vert \Psi(t) \rangle &=&\sum_{jk} a_j^{\ast}a_k
\langle \psi_j \vert \hat{Q} \vert \psi_k \rangle \exp{(i(E_j-E_k)t/\hbar)} \nonumber \\
&=&\sum_j \vert a_j \vert^2 \langle \psi_j \vert \hat{Q} \vert \psi_j \rangle +
\sum_{j\not=k} a_j^{\ast}a_k
\langle \psi_j \vert \hat{Q} \vert \psi_k \rangle \exp{(i(E_j-E_k)t/\hbar)}, 
\end{eqnarray}
which contains a constant part and a term depending on time.

Now suppose that the system is prepared at $t=0$ in some special non-equilibrium
condition such that $\langle Q \rangle_{t=0}$ deviates appreciably from the 
constant part of $\langle Q \rangle_t$. This non-negligible deviation is represented by
$\sum_{j\not=k} a_j^{\ast}a_k \langle \psi_j \vert \hat{Q} \vert \psi_k \rangle$,
the initial value of the time dependent part of $\langle Q \rangle_t$.
Provided that the various energy differences are largely incommensurate, 
the individual terms of the time dependent sum will thereafter get out
of phase with each other and stay out of phase indefinitely. If in
addition the series is not dominated by just a few large terms it is
intuitively plausible that as time goes on the initial coherence will be
destroyed. That is, it becomes less and less likely that the numerous
small terms, with effectively random phases, will add up to give an
appreciable contribution. Thus if we wait long enough only the constant
part $\sum_j \vert a_j \vert^2 \langle \psi_j \vert \hat{Q} \vert \psi_j \rangle$
of the general expression for $\langle Q \rangle_t$ survives. It is
certainly possible that some partial restorations of coherence will
occasionally happen, but for global properties of a system they are so
small as to be detectable only with difficulty, and they are in any case
fundamentally irreproducible and unpredictable in unique detail, such as experiments on
spin-echoes \cite{[spin_echoes]}, which demonstrate spectacular atomic memory phenomena; but
in general it remains true that our best estimate of the value of an
observable in the long term is given by the constant part of $\langle Q \rangle_t$.

This dephasing argument, though extremely qualitative, contains the
essence of the matter and we see roughly the mechanism for relaxation to
equilibrium. At the same time we understand that a quantity which is not
a constant of the motion will always, in principle, show fine time-
dependent fluctuations about its equilibrium value. Actually, for
discrete quantum states, the expectation $\langle \hat{Q} \rangle_t$ is quasi-periodic, which
means that it will eventually return arbitrarily closely to its initial
value; but for most macroscopic observables the relevant recurrence time
is enormously long compared with all feasible observation times. Thus
the apparently irreversible nature of relaxation processes in an
isolated body is to some extent an illusion. However, for gross
collective motions in such a system, which are always observed to die
out ultimately, it is a most convincing one.

Our simplified discussion has a bonus in that it strongly suggests that
the central concern of equilibrium statistical mechanics is to estimate
the probabilities with which the various energy eigenstates occur, for
then the final expectation values of physical quantities are computable.
The remaining mechanical part of the calculation reduces to the
construction and enumeration of the allowed energy states and the
evaluation of matrix elements of suitable operators. What is not so
clear from the above is why the same system may reach  the same
reproducible equilibrium condition from a wide range of different
initial states. The quantities $\sum_j \vert a_j \vert^2$ appearing in the formulae are, of
course, interpreted as the probabilities for finding the energy states
$\vert\psi_j\rangle$, given that the actual state is $\vert\Psi(t)\rangle$, and they obviously depend
on exactly what the state vector is. An additional difficulty is that we
never really know what $\vert\Psi(t)\rangle$ occurs, so this must also be guessed
before the quantum probabilities $\sum_j \vert a_j \vert^2$ are even calculable. Thus there
is a second level of probability to be compounded with the first.

The way out is to recall that these compounded probabilities will in the
end have to be assigned directly from a knowledge of the observed
properties of the final equilibrium state. In particular a measurement
of the total energy turns out to be of great utility for this purpose,
and for an isolated system the final energy expectation value is 
necessarily the same as that of the initial state, whatever that may be.
We therefore conclude tentatively that all initial conditions which lead
to the same given final equilibrium must imply state vectors $\vert\Psi(t)\rangle$ with
similar probability distributions for the energy eigenstates appearing
in them. Discrimination between the state vectors corresponding to the
great variety of possible non-equilibrium initial states depends on the 
phase relations among the terms of their eigenfunction expansions. As we
have seen, it is likely that these phase relations become effectively
washed out when equilibrium is reached. The last link in the argument is
that the energy eigenstates most likely to occur, compatibly with a
given total energy, must almost all imply essentially the same set of
expectation values for other macroscopic quantities. If it were not so,
the equilibrium properties would not be reproducibly related or have
sharply reproducible magnitudes.

Conversely, this suggests that knowledge of the equilibrium values of
properties other than the energy may be sufficient for the assignment of
probability distributions, and hence make possible the prediction of
the remaining macroscopic observables, including the energy expectation
value itself. Thus the validity of the above remarks does not depend in
any essential way on the supposition that we have measured the energy directly.

\subsection{Microcanonical probabilities}

\noindent We delay a full exposition of how to assign probabilities until some
necessary theory has been developed in the next section. However, we can
with profit consider an important special case in order to form some idea
of the kind of problem to be solved. The method has the drawback of
being rather artificial, but is valuable because the required
probabilities are obvious to common sense.

Here and in later treatments of an isolated body we assume that enough
is known about the environment and constitution of the system that its
Hamiltonian can be written down, at least approximately. This is
sufficient data, in principle, to derive its possible energy eigenstates
and eigenvalues by solution of the time-independent Schr\"odinger equation
\begin{equation}
\hat{H} \vert \psi_j \rangle = E_j\vert \psi_j \rangle,
\end{equation}
where the symbol $j$ stands for all the labels necessary to specify the
state completely.

Now suppose that an isolated system has reached a condition of
equilibrium and that its internal energy has been measured. In practice
this energy is determined relative to that of some equilibrium reference
state by doing macroscopically measurable amounts of work on the body.
The work is performed by altering the applied mechanical stresses or
fields in a controllable way and care is taken that the body is not
otherwise affected by the environment. Specifically, no part of the
resultant energy change can be attributed to the uncontrolled random 
perturbations called heat transfer. Under these conditions we can assign
an internal energy to a given equilibrium state by direct experiment. We
have, of course, assumed the law of energy conservation so that the
measured external work corresponds to an equal amount of energy gained
by the body.

According to the rules of quantum mechanics, a measurement of the energy
results in the state of the body becoming an eigenstate of energy, an
example of the well-known ``collapse of the wavepacket''. The state vector 
subsequently has the form
\begin{equation}
\vert \Psi_j(t) \rangle = \vert \psi_j \rangle\exp{(-iE_jt/\hbar)},
\end{equation}
for some $j$, and the constant energy of the system is given by
\begin{equation} 
E_j = \langle \Psi(t) \vert \hat{H} \vert \Psi(t) \rangle =  
\langle \psi_j \vert \hat{H} \vert \psi_j \rangle,
\end{equation}
since the harmonic time factors cancel out. Any other observable $Q$ for
an isolated system will correspond to an operator $\hat{Q}$ independent of time
and our estimate for its magnitude can be taken as the similarly
constant expectation value
\begin{equation} 
\langle Q \rangle_j = \langle \Psi(t) \vert \hat{Q} \vert \Psi(t) \rangle =  
\langle \psi_j \vert \hat{Q} \vert \psi_j \rangle.
\end{equation}

The quantum probabilities $\vert a_j \vert^2$ of the last subsection are no longer
needed, but for two reasons this is not the whole story. One reason is
that the eigenstates may be degenerate, which means that there may be
many states $\vert \psi_j \rangle$ belonging to the energy $E_j$; the other is that there
will inevitably be some experimental error in the measurement of the
energy so that the exact value of $E_j$ is not known. Since our knowledge
is not sufficient to pin down the exact state, there still appears to be
considerable uncertainty left in our estimates for other observables $Q$,
even apart from the spread of values implied by quantum theory when the 
operators $\hat{Q}$ do not commute with $\hat{H}$. Clearly we need to assign
probabilities to the eigenstates $\vert \psi_j \rangle$ which {\it could occur\/} consistently
with the knowledge we do in practice obtain.

Let us therefore assume as our basic information that the energy has 
been definitely observed to lie somewhere between a fixed energy $E$ and
another fixed energy $E + \Delta E$, where $\Delta E$ is small ($\Delta E \ll E$). Equivalently
\begin{equation}
E < E _j < E + \Delta E.
\end{equation}
Any state $\vert \psi_j \rangle$ with an energy $E_j$ obeying these inequalities will be
called {\it accessible\/}, and we define a quantity: \\
\makebox[\textwidth]{$\Omega = $ The Number of Accessible States.}
It is also convenient to let the labels $j$ 
of these states range over the
values $j = 1 \to \Omega$. The word accessible is used for historical reasons and
there is no implication that the system will eventually visit every such
state. It means only that our knowledge is consistent with any one of 
them being the actual state, but that we do not know which one occurs.

Probabilities for the accessible states must be assigned in the light of 
all available information. This includes the above energy condition and
it is also relevant that the states form an exhaustive and mutually
exclusive set of possibilities. Exhaustive means that at {\it least\/} one of
the $\Omega$ states must occur, which is true by hypothesis. Mutual exclusion
signifies that at {\it most\/} one of the $\Omega$ states is the actual state of the
system. This follows from the orthogonality of eigenstates of different
energy together with the hitherto tacit assumption that the independent
states of a degenerate energy level have also been chosen to be
orthogonal, as is always possible. For if the energy state $\vert \psi_j \rangle$ occurs
then the condition $\langle \psi_k \vert \psi_j \rangle = 0$ definitely excludes the possibility of
finding any other state $\vert \psi_k \rangle$. Thus we know that one, and only one, of
the $\Omega$ accessible eigenstates must be present as a result of the energy measurement,
but our information gives no further help  in choosing the
right one. As far as we can see, none of the possible states is picked
out as being more likely than any other. We must therefore at this point
resort to reasonable guesswork and make an unadorned postulate. A 
plausible statistical theory can be set up on the following \hfil

\begin{itemize}
\item{}
{\bf Basic Hypothesis:} Given that an isolated system has an energy between $E$ and
$E + \Delta E$, every one of its energy eigenstates $\vert \psi_j \rangle, \ j=1 \to \Omega$,
with energy eigenvalue $E_j$ lying in that interval, has the same 
probability of being the actually occurring state.
\end{itemize}

To apply this, let the probability for the state $\vert \psi_j \rangle$ be $p_j$. Then the
total probability for occurrence of an accessible state must be unity 
and all the $p_j$ are equal. The two conditions
\begin{equation}
\sum_j p_j = 1 \ \ {\rm and} \ \ p_j = p_k \ \ \ \forall \ j,k
\end{equation} 
lead easily to the probability assignments:
\begin{eqnarray} 
p_j &=& {1 \over \Omega} \ \ {\rm if} \ E < E_j < E + \Delta E \nonumber \\
&=& 0 \ \ {\rm otherwise},
\end{eqnarray}
and we have derived what is called the Microcanonical Distribution.

Averages are now defined by expressions of the form
\begin{equation}
\bar{X} = \sum_j p_j \langle X \rangle_j, \ \ j = 1 \to \Omega,
\end{equation}
so that our final estimates for the energy and other observables are
\begin{eqnarray}
\bar{E} &=& \sum_j {\displaystyle E_j \over \Omega} \ {\rm and} \\
\bar{Q} &=& \sum_j {\displaystyle \langle \psi_j \vert \hat{Q} \vert \psi_j \rangle \over \Omega}.
\end{eqnarray}
It is obvious from these equations that $\bar{E}$ must lie somewhere between $E$
and $E + \Delta E$, as we already knew; but it is not at all obvious that $\bar{Q}$ will
also have well-defined limits. Indeed, as remarked earlier, such an
expression for the estimated value of an observable $Q$ will be useful 
only if almost all states $\vert \psi_j \rangle$ compatible with our information yield
very similar expectation values for $\hat{Q}$. But before discussing this vital
point further it is useful to make some comments on the Basic Hypothesis.

The bald postulate announced above has been called the ``Principle of Indifference'' 
or ``Insufficient Reason'' on the grounds that the information
available provides no reason to prefer any one of the allowed
possibilities over any other. It is usually attributed to Bernoulli and is discussed
at length by Keynes\cite{[Keynes]}. Some such principle is clearly required, 
in order to make progress, but the formulation just given has a negative
sound to it. To emphasize its positive r\^ole in the theory, it has been
suggested that it should rather be called a ``Principle of Consistency"\cite{[Jaynes2]},
since it would certainly seem inconsistent to assign greater probability
so some subset of the accessible states if the data we have indicate no
compelling reason to do so.

It may be objected that some of the states may really be more probable
than others; but this implies a view of the concept of probability
different from the one adopted here. Our view is basically that
probability is an encoding of our knowledge of possibilities, so that it
appears at first sight as somewhat subjective and capricious.
Objectivity is restored if we insist that anyone with the same data 
should arrive at the same probability assignments. We discuss these 
things more fully in the next section, in which the above ``Principle of
Consistency'' will emerge as just a special case of a more powerful
algorithm. This more general principle applies even when there {\it is\/} reason
to prefer some possibilities over others.

Returning now to the special assumptions of this subsection, it is
important to realise that our probability postulate (or Basic
Hypothesis) is not a {\it physical\/} principle, but only a way of reasoning in
the face of uncertainty, and hence requires no justification from
physics. The system is not in a probability distribution but in some 
definite state that we do not happen to know. Thus we have to form an
opinion about which state really occurs and this is most consistently
represented by the assumption of equal probabilities for the states
permitted by the data. If we want to do any better than this then more
information must be acquired.

Obvious though all this may appear, we emphasize it because many
attempts {\it have\/} been made to justify the principle by physical arguments. 
The idea is that the system, perhaps under the influence of tiny but
uncontrollable perturbations, continually makes transitions between its
various eigenstates, so that it spends an equal amount of time in each
possible state. There, it is implied, since an observation takes a
finite time, a time-averaged observed value can be calculated as an
average over the ensemble of accessible system states at a fixed time,
each occurring with equal probability. This sounds plausible but has
never been proved even for averages over indefinitely long times, let 
alone for the quite short times involved in typical experiments. Very
much more to the point, it is easy to see that only a vanishingly small
fraction of the accessible states could be sampled in any reasonable
observation time.

For if we assume, conservatively, that each of the $3\times 10^{22}$ degrees of 
freedom of even a quite moderately sized body could have only two likely 
quantum number values at equilibrium, then the number of possible states
would be $2^{(3\times10^{22})} \approx 10^{(10^{22})}$. Assuming further that the transition rate
for each degree of freedom reaches as high as $10^{20}$ per second (thus
approaching nuclear rates \cite{[Satchler]}), then the transition rate for the whole body 
would be of the order of $10^{42}$ per second. The time required to visit 
each state even once would then be $10^{(10^{22}-42)} \approx 10^{(10^{22})}$ seconds and
this is so long that it hardly matters whether we express it in seconds or in 
units of the age of the universe, which is a mere $10^{17}$ seconds.
The time-averaging suggestion is seen to be quite implausible.

The real problem to be understood is why our statistical averaging over
the almost incomprehensibly huge number of accessible states leads to
values of observables in agreement with experiment, even though the
system will never actually be in the overwhelming majority of such
states. The only conceivable reason for the practical success of our
postulate is that virtually every accessible state has nearly the same
macroscopic properties and the averaging is necessary merely to avoid
picking out an untypical state accidentally.

\subsection{Estimation of observables} 

\noindent We now investigate in a little more detail how the estimate
\begin{equation}
\bar{Q} = \sum_j { \langle \psi_j \vert \hat{Q} \vert \psi_j \rangle \over \Omega}
\end{equation}
for the value of an observable could be useful. What we require is that 
any value likely to be found as a result of measurement should be near
to the number $\bar{Q}$, i.e. that the magnitude $\vert \Delta Q \vert$ of the observed deviation
from $\bar{Q}$ should virtually always satisfy the relation
\begin{equation}
{ \vert \Delta Q \vert \over \bar{Q}} \ll 1.
\end{equation}
In principle, the various quantum expectation values $\langle \psi_j \vert \hat{Q} \vert \psi_j \rangle$
could be wildly different from each other and, although the average $\bar{Q}$ would then
still be a definite number, actual observations would show a large
spread of results about the mean. Sharply reproducible results can be
expected only if it is possible to demonstrate that
\begin{equation}
\langle \psi_j \vert \hat{Q} \vert \psi_j \rangle \approx \langle \psi_k \vert \hat{Q} \vert \psi_k \rangle
\end{equation}
for nearly all arbitrarily selected pairs of accessible states $\vert \psi_j \rangle $ and
$\vert \psi_k \rangle $, and in addition that these matrix elements themselves represent
sharp quantum predictions given the energy states involved. Obviously it is
impossible to give a general proof of such properties and each
suggested observable for a particular system must be treated separately.
Alternatively we can choose to calculate only those macroscopic
quantities which are in fact observed to be reproducible experimentally,
for then the conditions mentioned must be satisfied.

As a specific example consider the possibility of estimating the
pressure exerted by an isolated system, of measured energy, on its 
confining walls. We assume that the system always exactly fills the 
container even when the volume $V$ is varied over some range of values. An
important special case of this is a gas contained in a cylinder which is
fitted with a movable piston. The parameter $V$ will certainly appear in
the defining Hamiltonian operator $\hat{H}$ and it follows that the accessible 
energy eigenstates and eigenvalues will also in general depend on $V$,

To calculate the pressure $P$, imagine that the system is in a definite
state $\vert \psi_j(V) \rangle$ belonging to energy $E_j(V)$ and that the volume is changed,
adiabatically and slowly, by a small amount $\Delta V \ll V$. An adiabatic change
is defined as one brought about by adjustment of the force fields or
mechanical stresses acting on a body, which remains otherwise isolated
from its environment. It is implied that the Hamiltonian changes in a
definite way, e.g. $\hat{H}(V) \to \hat{H}(V + \Delta V)$. By a slow change we mean that the
change proceeds so gradually that the system effectively stays at all 
times in an eigenstate of its gently evolving Hamiltonian. This slowly
applied perturbation causes the accessible eigenstate $\vert \psi_j(V) \rangle$ of $\hat{H}(V)$ to
go smoothly and continuously to a corresponding accessible eigenstate 
$\vert \psi_j(V + \Delta V) \rangle$ of $\hat{H}(V+\Delta V)$; there are no transitions to other
eigenstates, and the total number of accessible states is unchanged.
After these explanations it should be sufficiently clear that in our
slow, adiabatic process the work done on the system is equal to the
change in its internal energy $E_j(V)$ and we may write
\begin{equation}
\Delta E_j \approx  \left ( { \partial E_j(V) \over \partial V} \right ).\Delta V = -P_j\Delta V
\end{equation}
The second equality records the accepted relation between pressure and
work done in a small change of volume, and we have at once that the 
pressure $P_j$ in state $\vert \psi_j \rangle$ is
\begin{equation}
P_j = -{ \partial E_j(V) \over \partial V}.
\end{equation}
The pressure observable may also be written, in conformity with previous
expressions, as the expectation value of an operator in the state $\vert \psi_j \rangle$.
A straightforward exercise in quantum mechanics soon shows that for
discrete, normalized, energy eigenstates of $\hat{H}$ we have
\begin{equation}
P_j = \left \langle \psi_j \left \vert  -{ \partial \hat{H}(V) \over \partial V} \right \vert \psi_j \right \rangle =
 -{ \partial E_j(V) \over \partial V}.
 \end{equation}
A similar theorem holds for any observable whose effective operator can
be represented as the derivative of the Hamiltonian $\hat{H}$ with respect to a
parameter. We mention in passing that such operators, and in particular
the pressure $\hat{P} = -{ \partial \hat{H} \over \partial V}$, usually do not commute with the Hamiltonian
and are therefore not constants of the motion. As explained before,
nearly steady expectation values would then be attained only at 
equilibrium and fluctuations are inevitable.

Our statistical estimate for the equilibrium pressure $\bar{P}$ is computed by
averaging the pressures $P_j$ over the microcanonical probability 
distribution appropriate to accessible states:
\begin{equation}
\bar{P} = { [\sum_j P_j] \over \Omega} = { \left [ \sum_j \left (-{ \partial E_j \over \partial V} \right )
\right ] \over \Omega}.
\end{equation}
Now the change envisaged in the definition of pressure in an isolated 
system was such that the number of accessible states remains constant.
Therefore on rearranging, we have
\begin{equation}
\bar{P} = -{ \partial \over \partial V} \left [ { (\sum_j E_j ) \over \Omega } \right ] = 
-\left ( {\partial \bar{E} \over \partial V } \right )_{\Omega = {\rm constant}}.
\end{equation}
Hence our estimate of the pressure will be good only if the observed
energy is a smooth, reproducible function of volume in a slow, adiabatic
process, which is very often true experimentally. On the theoretical 
side we would have to show that most accessible energy eigenvalues $E_j(V)$
have closely equal derivatives ${ \partial E_j(V) \over \partial V}$. 
Again, this is found to be true in many analytically tractable 
models. It is certainly true, for
example, for a gas of structureless, non-interacting particles over a
wide range of internal energy. It will be shown later, for this model,
that the pressures are given by $P_j = 2E_j/3V$, from which it follows that
\begin{equation}
{2E \over 3V} < \bar{P} < {2(E + \Delta E) \over 3V},
\end{equation}
implying that the estimated pressure $\bar{P} = 2\bar{E}/3V$ will lie between close
limits if $\Delta E \ll E$.

In this subsection we have pointed out the basic preconditions for
statistical methods to give sharp answers and shown that they are the 
same as those for experimental reproducibility. The main conclusion is
that when we estimate a physical quantity we should at the same time
attempt to assess the expected deviations of actual observations from
our result, for only then is it possible to judge whether the
calculation will provide useful predictions and systematizations of experience.

\subsection{Significance of $\Omega$-number}

\noindent It may seem from the development so far that the quantity $\Omega$, the number
of accessible states, is of somewhat secondary importance - a sort of
mathematical prop to be discarded eventually. For we have been at pains
to emphasize that averaging over the uniform probability distribution of
accessible states is reasonable only if most such states have similar 
macroscopic properties. A closer look soon shows that $\Omega$ is in fact a
highly significant function of the data used in its construction.
A hint of this appeared already in our general expression for the 
estimated pressure of an isolated system, which was
\begin{equation}
\bar{P} = -\left ( { \partial \bar{E} \over \partial V} \right )_{\Omega}.
\end{equation}
where the derivative is taken with $\Omega$ held constant. This should be
compared with the well-known thermodynamical formula for pressure in
terms of the volume derivative of the internal energy $\bar{E}$, i.e.
\begin{equation}
\bar{P} = -\left ( { \partial \bar{E} \over \partial V} \right )_S.
\end{equation}
in which the quantity held constant is the entropy $S$.

There is here a strong suggestion that $\Omega$ is closely related to the
experimental entropy. The idea is well founded and we shall eventually
present detailed arguments to show that for an isolated system of known
internal energy the exact connection has the form
\begin{equation}
S = k_B\ln{(\Omega)},
\end{equation}
where $k_B$ is Boltzmann's constant. Analogous relations between entropy 
and purely statistical constructs hold also under other experimental
conditions and they provide the fundamental links connecting microscopic
mechanics with phenomenological thermodynamics. Although statistical
mechanics can be developed without reference to thermodynamics, greater
understanding is achieved by extracting such connections as exist, both
for their intrinsic interest and in order to get some insight into that
other mysterious concept of thermal physics, namely, the temperature.

There is, however, a striking conceptual difficulty in any attempt to
identify the apparently distinct notions of experimental entropy and
statistical uncertainty. In many accounts entropy appears as a physical
property of bodies, as definite as mass or energy, to be measured with
laboratory instruments, and admitting no ambiguity apart, perhaps, from
an arbitrary additive constant. Statistical uncertainty, as represented
here by the number $\Omega$ of macroscopically indistinguishable states, seems
by contrast to be much more something to do with our knowledge (or
rather, lack of knowledge) of bodies than with any intrinsic physical
properties possessed by them. That is, it appears to be in the mind.
It is certainly true that the usual entropy $S$ is a peculiar physical
quantity, much harder to grasp than others, since it is not conserved
like mass and energy, but instead has the strange tendency to increase
in spontaneous natural processes. This, of course, parallels the
undoubted fact that our uncertainty about the state of a body will
increase when some uncontrolled change occurs. We will indeed show later
that this type of analogy extends to fine details, in the sense that
exact mathematical correlations exist between the behavioural properties 
of the entropy $S$ and a suitable statistical measure, exemplified here by
the expression $\ln(\Omega)$.

Mathematical analogies alone, however, are not sufficient to guarantee
the identity of the underlying concepts, since it sometimes happens that
similar mathematical theories can apply to quite unrelated phenomena. A
physical postulate is unneccesary in order to link theory and experience
in the present instance. A reasonable assumption is that the 
statistically calculated properties of macroscopic systems will agree
with experiment. This goes far beyond what can be rigorously 
demonstrated by actual calculation and observation; but whenever the
mathematical problems can be solved for a realistic model the assumption
is convincingly verified. To connect with thermodynamics, a relevant
remark is that a great accumulation of experience confirms the utility
of entropy for the correlation and understanding of experimental
properties of systems. 

Some of the mystery surrounding the proposed identification of entropy
and uncertainty disappears when we enquire into their functional
dependence on observed data. An adequate thermodynamic description of a
body is at hand when we know the entropy as a function of the internal
parameters. In particular, we may know the volume $V$ of the system and
the amount of matter in it, to be represented by the number $N$ of
particles, assumed all of one type. Then we require the function
\begin{equation}
S=S[\bar{E}, V, N],
\end{equation}
from which many physical predictions may be obtained by differentiation.

There is a tacit assumption here that the above variables are the only 
relevant ones. But it may happen that the derived predictions do not all
agree with experiment. In that case, we do not abandon thermodynamics,
but take it as an indication that some other relevant parameter has been
overlooked --- for instance, the presence of a magnetic field. We then
seek more data to determine the dependence of $S$ on this new quantity and
so discover many further implied relationships to be checked by
observation. The conclusion is that entropy depends very much on what we
choose to measure, which means that it embodies our knowledge of the system.

The quantity $\Omega$ defining the probabilities of the microcanonical
distribution also depends very much on what we choose to measure and
take into account. The Hamiltonian $\hat{H}$ and energy eigenvalues $E_j$, depend
parametrically on volume $V$ and particle number $N$ and the accessible
states are by definition those lying in an experimentally determined
energy interval. Explicitly:
\begin{eqnarray}
\Omega &=&\Omega[ E \leq E_j(V,N) \leq E + \Delta E] \nonumber \\
&\approx&D[\bar{E}, V, N].\Delta E
\end{eqnarray}
The second, approximate, equality expresses $\Omega$ as the product of the
error interval $\Delta E$ and a function $D$, the density of states, which 
represents the number of accessible states, per unit energy, in the
vicinity of $\bar{E}$ . Clearly, $D[\bar{E}, V, N]$ depends on the same kind of data as
does $S$. The approximate expression for $\Omega$ in terms of the density of
states $D$ is adequate, for all practical purposes, when $\Delta E \ll \bar{E}$. If $\Delta E$ is
fixed, and we write
\begin{equation}
S = k_B\ln{(\Omega)} = k_B[\ln{(D(\bar{E}, V, N))} + \ln{(\Delta E)}],
\end{equation}
as suggested earlier for the isolated system, we see that to predict
physical observables, which are derivatives of $S$ thermodynamically, the
required mechanical quantity is the density of states function $D$. Given
this link between thermal and mechanical concepts we see, incidentally,
that measurement of $S$ gives information on the distribution of energy
levels in the system.

However, independently of this proposed connection between $S$ and $\Omega$ (or
equivalently $D$), physical observables may be estimated directly from
their quantum expectation values in the accessible energy eigenstates,
using the probabilities assigned to such states. Now it may well happen
that the majority of accessible states, as calculated for given $V$ and $N$,
in an interval $\Delta E$ surrounding $\bar{E}$, have macroscopic properties which do
not agree with experimental results. As before, we would not therefore
immediately abandon the calculation, but conclude that a relevant
parameter in the system Hamiltonian had been overlooked, and that only a 
subset of the previously determined states is actually accessible under
the conditions of the experiment. Thus the effective density of states
must be changed, its dependence on the new parameter calculated, and its
predictions again probed by observation. The analogy with the
experimental use of entropy is confirmed.

What emerges very clearly from all this is that the statistical method
is an efficient way of reasoning about the data and its relation to the
underlying physics, but is not in itself a physical theory. We might
even claim that the results of statistical mechanics are most useful
when they do not agree with experiment. For then, assuming we have
confidence in our theoretical models and mathematical techniques, we will
learn something new.
\vfil
\newpage

\section{Probability and uncertainty}
\subsection{What is probability?}

\noindent We have talked rather loosely about the concept of probability and must
now attempt to make the idea sharper. This involves a digression from
physics, but a very necessary one, for we wish to divorce the basic
notions of probability theory from any dependence on experimental facts.
As remarked before, the probabilistic aspects of statistical mechanics
are susceptible of a complete treatment independent of the mechanical
laws underlying the observed phenomena.

From the earliest days of probability there have been two competing
views on its nature. One is that it records the ratio of the number of
times a particular type of event occurs to the total number of trials in
a statistical experiment. This is called the Frequency Theory. The other
is that it encodes by a numerical measure the degree of belief one can
reasonably accord to some assertion on the basis of available evidence.
This is know as the Bayesian Theory, in which the probability calculus 
is regarded as a technique for plausible inference. According to the
first interpretation, probability is an intrinsic measurable property of
physical events independent of what we happen to know. On the other view
it is a quantitative way of thinking about suggested propositions even
when our knowledge is not sufficient to decide definitely on their truth
or falsity. That is, probability is taken to be a method for conjectural 
reasoning, making consistent use of the relations of partial implication
connecting conclusions with given hypotheses.

We hold to the second view as being wider and more generally useful than
the first, for it can be applied in situations where statistical testing
is inappropriate or impossible. To avoid unnecessary considerations of
time ordering, it is also convenient to think of probabilities as being
attached to propositions rather than to events. However, if statistical
evidence on events is in fact available, and relevant, it may still be
incorporated into assessments of probability. Our working hypothesis on
the nature of the theory is that it provides sets of numerical codings 
for the credibilities we may consistently associate with propositions,
given relevant background knowledge and particular evidence. In other
words all probabilities are conditional and a change in the evidence may
well entail a change in the probabilities. Clearly therefore we are not
regarding probability as a physical property of objects or processes but
as a representation of our state of knowledge about the world. 

The main trouble with this interpretation, in the view of many authors,
is that it seems too vague and subjective, while the frequency theory
appears definite and not dependent on the idiosyncrasies of individuals.
The objection does not hold up on further examination. The notion of
probability as an algebra of plausible inference  can be made quite
adequately sharp, as we shall show, by invoking a criterion of internal
consistency. In addition, the apparent subjectivity of degrees of belief
can be removed by adopting consistent rules of numerical coding, so that
when different people assess the same proposition in the light of the
same evidence they arrive at the same number for the probability. The
foundation of a satisfactory theory of probability thus requires the 
solution of two major problems: \hfil

\begin{itemize}
\item{(I)} How to set up a convenient and consistent calculus for
combining and manipulating probability symbols.
\item{(II)} How to associate definite and unambiguous numerical
magnitudes with appropriate symbols of the calculus.
\end{itemize}

The above questions arise on any view of the subject. Let us consider
briefly how they appear in the frequency interpretation. The second one
now looks rather trivial since in that theory the probability values are
defined directly in terms of numbers of instances in a long series of
trials. The catch is that different finite series of trials for the same
type of event usually give somewhat different numbers. Thus we are
forced to postulate that the relative frequency we require will settle
down to a definite ratio in the limit of an infinite number of trials.
The finite experiments that we can in reality perform yield only
estimates of the assumed intrinsic probabilities, and thus some of the
definiteness of the frequency idea begins to evaporate. A formal
solution to the first problem can be constructed by defining sample
spaces of events and associating probability measures with the various
subsets of events in those spaces. Then, by analogy with the observed
behaviour of stable relative frequencies, some quantitative axioms for
the measures are laid down. But one might reasonably enquire whether the
particular axioms chosen are consistent, in the sense that different
ways of doing a calculation will not result in contradictions. It is
also pertinent to ask if other sets of axioms can be found which might
possibly extend the scope of the theory. The frequency model itself now
begins to look decidedly indefinite and arbitrary and we return to our
favoured view.

A much more powerful and persuasive approach to the question of how to 
construct a calculus of probabilities was advanced by R.T.Cox \cite{[Cox46]}, around
the middle of the 20th century, but has not yet become standard material in books. On
the basis of two highly plausible {\it qualitative\/} assumptions, together with 
an axiom of consistency, he proved that there is only one possible
algebra of probable inference. His argument leads to definite
{\it quantitative\/} rules for the combination of probabilities, though their
exact form can be changed by purely mathematical transformations. The
proof is extremely interesting and will be sketched below. It turns out
to be surprisingly easy to understand considering that the main tools
employed are the rather unfamiliar ones of symbolic logic and functional
equations.

The other part of probability theory, how to choose numerical values for
probabilities in a rational way, is not at present a completely solved
problem. When we can reasonably show for some situation that there
exists a set of exhaustive and mutually exclusive inferences and that
our knowledge does not suggest any preferred selections among them, then
clearly a principle of indifference can be formulated. The only rational
choice is equal probability for each inference, as in our example of the
microcanonical description of an isolated body. But, when the conditions
for indifference are not satisfied, it is tempting to believe that no 
general method is possible except to rely on frequency data. If such
data are not available the problem of assigning initial probabilities in
a calculation becomes acute. However, there does exist a certain class
of problems, which includes the questions of statistical mechanics, for
which it {\it is\/} possible to develop a useful theory of probability
assignments, independent of frequencies.

The central concept of this alternative assignment theory originated in 
work of C.E.Shannon \cite{[Shannon]}, also around the middle of
the 20th century, and its significance for the problem of inference was pointed out by E.T.Jaynes
\cite{[Jaynes1]} some ten years later. The fundamental idea is to associate with a probability
distribution over a set of propositions a quantity we shall call the Uncertainty.
Bear in mind that the probability for a single inference
represents our state of partial belief in the truth of the inference,
and so implies some degree of uncertainty. To begin with, the concept of
the total amount of uncertainty implied by the full set of probabilities 
is a little vague, but we try to invent a way of representing, by a 
single number, our overall state of doubt about which proposition is
true. If we believe more strongly in some of the possibilities than in
others we usually feel less uncertain about the true situation than if
our evidence implied that all propositions were equally likely. The
problem is to quantify this kind of feeling by an objective measure. By
making a few plausible qualitative assumptions about how such a measure
of uncertainty should behave, and imposing a condition of self-consistency, 
Shannon \cite{[Shannon]} proved that there exists a simple unique function
of the probabilities which has the desired characteristics. The argument
proceeds by setting up and solving functional equations and the whole
method is similar in principle to that used by Cox \cite{[Cox46]} for deriving the
probability calculus itself. In view of the importance of the result we 
shall outline the proof.

The contribution of Jaynes \cite{[Jaynes1],[Jaynes2]} was to use the Uncertainty function as a tool
for assigning probabilities. He proposed that in the absence of
sufficient information, statistical or otherwise, for direct evaluation
of probabilities, they should be chosen so as to make the Uncertainty an
absolute maximum, subject only to the constraints imposed by the 
evidence we do possess. The underlying idea is to admit that our state
of doubt in any situation is limited solely by the amount of relevant
data we have acquired. The effect of this procedure is to distribute the
total probability as evenly as possible over the allowed inferences, 
without contradicting the given evidence, and the method emerges as a
natural generalisation of the Principle of Indifference. In particular
it gives some weight to every proposition not completely ruled out by
the data. Such a spreading out of the probability, or hedging of bets,
is clearly a desirable feature for any method of inference.

\subsection{Algebra of probability} 

\noindent We now intend to show how to derive a consistent calculus for combining
probabilities without invoking the idea of frequencies in a series of
trials. This is important because the interesting questions in
statistical mechanics are not really statistical at all, but concern
reasonable expectation. Basically we assume that degrees of belief in
various propositions $A$, $B$, etc., or perhaps $\{A_m\}, m=1 \to M$, may be
represented by real numbers on some scale. We also assume that there is
some information or evidence available to be used in assessing the truth
of a proposition. The statement of this background knowledge will be
denoted by $K$. Occasionally the writing of $K$ will be suppressed, but it
must always be considered as present. The primitive notion of the
credibility of an inference $A$, given evidence $K$, will be symbolised briefly by $(A\vert K)$.

The measure to be attached to a credibility $(A\vert K)$ is very vague at this
stage, and clearly any function of such a number will also constitute a 
measure. Instead of using the term probability immediately, let us
assume merely the {\it existence\/} of at least one convenient scale, written as
a numerical function $\alpha (A\vert K)$ of the credibility. This number will be
called the {\it assessment\/} of $A$ given $K$, and we make it a convention that the
stronger the credibility then the larger will be the number $\alpha$. There 
will always be some element of convention in setting up a scale of  
assessment, but it is also necessary that any scheme for combining
assessments should be consistent with the algebra of propositions.

We shall take it that the propositions $A$, $B$, etc., are statements that can 
be either true or false, but are never self-contradictory or devoid
of meaning, and that to assert $A$ is just to say that $A$ is true. We also
introduce below the usual notation for the negation of a proposition and
for two distinct ways of combining statements into compound assertions:-
\begin{equation}
\sim A = {\rm NOT}\ A \ \Rightarrow A \ {\rm is \ not \ true}
\end{equation}
\begin{equation}
A.B = A\ {\rm AND} \ B  \ \Rightarrow {\rm Both} \ A \ {\rm and} \ B \ {\rm are \ true}
\end{equation}
\begin{equation}
A \lor B=A\ {\rm OR}\ B \Rightarrow A \ {\rm is \ true \ or} \ B \ {\rm is \ true \ or \ both \ of \ them\ are \ true}
\end{equation}
In what follows brackets are employed to distinguish compound statements 
which are to be treated as logical units, e.g., $C \equiv (A.B)$ or $B \equiv (\sim A)$.

There are a number of simple rules obeyed by the indicated operations
and it is possible to set up an axiomatic theory for them. We shall not
do this formally since the rules required are obvious transcriptions of
common sense. Thus the symbol $\sim (\sim A)$ just means that it is not true
that $A$ is not true, i.e. $A$ is true, and we easily see that:
\begin{equation}
\sim (\sim A) = A
\end{equation}
Similarly, $(A.A)$ means only that $A$ is asserted twice, which is just to
assert $A$. Hence we obtain the rule:
\begin{equation}
A.A = A
\end{equation}
In the same sort of way a direct consideration of the ordinary meanings
of words enables us, with a little thought, to see the truth of the 
following relations:
\begin{eqnarray}
A.B&=&B.A \\
\sim(A\lor B)&=&(\sim A).(\sim B) \\
A.(B.C)&=&(A.B).C = A.B.C, 
\end{eqnarray}
together with the corresponding equations in which the symbols AND and OR
are interchanged. Where confusion might arise we shall always
spell out the meanings of our equations in full.

Consider now how we could assess the credibility of $(A.B)$ on evidence $K$.
Our belief in the truth of both $A$ and $B$ will presumably be conditioned,
first, on how strongly we believe in $B$ given $K$ and, second, on what 
credibility can be associated with $A$ given that both $B$ and $K$ are true.
In short, the credibility $(A.B\vert K)$ must depend somehow on the 
credibilities $(B\vert K)$ and $(A\vert B.K)$. Expressed mathematically, this means
that any proposed assessment numbers should satisfy the equation:
\begin{equation}
\alpha(A.B \vert K) = F[\alpha(A\vert B.K), \alpha(B\vert K)],
\end{equation}
where $F[x,y]$ is some two-variable function.

This is obviously a very weak assumption, especially when we remember
that no specific scale for the assessments has so far been suggested.
Yet the function $F$ is not completely indeterminate, since we require
internal consistency, which implies that different ways of assessing a
credibility should lead to the same assessment number. A useful
condition on $F$ is obtained by looking at the conjunction $(A.B.C)$ of
three propositions, in the two equivalent forms $(A.B).C$ and $A.(B.C)$.
Suppressing $K$ for the moment, we have from the first form that
\begin{eqnarray}
\alpha[(A.B).C]&=&F[\alpha(A.B\vert C),\alpha(C)] \\
&=&F[F[\alpha(A\vert B.C), \alpha(B\vert C)],\alpha(C)],
\end{eqnarray}
while the same number expressed using the second form leads to
\begin{eqnarray}
\alpha[A.(B.C)]&=&F[\alpha(A\vert B.C),\alpha(B.C)] \\
&=&F[\alpha(A\vert B.C), F[\alpha(B\vert C),\alpha(C)]].
\end{eqnarray}
Hence, if we write $\alpha(A\vert B.C)=x, \ \alpha(B\vert C) = y$ and $\alpha(C)=z$, consistency
requires that
\begin{equation}
F[F(x,y),z)] = F[x,F(y,z)],
\end{equation}
which is a non-trivial functional equation restricting the form of $F$.

It is easily verified that a solution of this equation is expressible by
means of an arbitrary, monotonic, one-variable function $G$ as follows:
\begin{equation}
F(x,y)=G^{-1}[G(x)G(y)],
\end{equation}
where $G^{-1}$ is the function inverse to G. Cox \cite{[Cox46]} has shown, under rather mild
conditions of differentiability and continuity on $F$, that this is the 
most general solution. A more convenient form for the above results is
\begin{equation}
G[F(x,y)]=G(x)G(y),
\end{equation}
This way of writing the solution suggests the definition of another
numerical measure of credibility which is related to, but more useful
than, the originally assumed function that we called the assessment. For
it is now possible to state that whatever scale we may have selected for
the $\alpha$-numbers, provided that it represents consistent reasoning, we can
always find a function of those numbers, given by
\begin{equation}
\beta(A\vert K) = G[\alpha(A\vert K)],
\end{equation}
which obeys a simple multiplicative rule of combination. We shall call
this new measure the {\it believability\/} of $A$ given $K$. Thus, on taking the
function $G$ of both sides of our starting equation (42), we see easily that
\begin{eqnarray}
G[\alpha(A.B\vert K)]&=&G[F[\alpha(A\vert B.K),\alpha(B\vert K)]] \nonumber \\
&=&G[\alpha(A\vert B.K)]G[\alpha(B\vert K)],
\end{eqnarray}
so that the rule for believability of conjoined statements on evidence $K$
reduces to
\begin{equation}
\beta(A.B\vert K) = \beta(A\vert B.K)\beta(B\vert K),
\end{equation}
If we take $G$ to be monotonically increasing, then our convention, that
greater assessment numbers $\alpha$ correspond to greater credibilities,
applies also to the believabilities $\beta$.

The above remarkable result implies a definite numerical value for the
believability to be associated with an inference which is certain on
given evidence. Putting $A=B$, and remembering that $(A.A)=A$, we obtain
\begin{equation}
\beta(A\vert K) = \beta(A\vert A.K)\beta(A\vert K).
\end{equation}
Any proposition $A$ not contradicted by $K$ is clearly certain given that
$(A.K)$ is true; hence we easily deduce the extreme value
\begin{equation}
\beta(\rm Certainty) = 1.
\end{equation}
To complete the calculus we need a rule connecting the believabilities
of a proposition $A$ and its negation $(\sim A)$. Again a qualitative
hypothesis is sufficient, if we begin to believe more strongly in the 
truth of some statement then we usually believe less strongly in its
denial, which is only to say that we feel that their believabilities are
somehow related. In symbols:
\begin{equation}
\beta(\sim A\vert K) = f[\beta(A\vert K)]
\end{equation}
and
\begin{equation}
\beta(A\vert K) = f[\beta(\sim A\vert K)],
\end{equation}
where $f(x)$ is some one-variable function and the second equation follows
from the first on replacing $A$ by $(\sim A)$ and using $\sim(\sim A)=A$. 
By putting the first equation into the second and writing $\beta(A\vert K) = x$ we see that
the function $f$ must satisfy the relation
\begin{equation}
f[f(x)] = x.
\end{equation}
This is not a very restrictive condition and another one must be found
by a further argument based on consistency requirements. One possibility
is to consider two different ways of evaluating the number $\beta((\sim A).B\vert K)$.
We use our hypothesis on negation together with the multiplicative rule for
the believability of conjoined propositions which was derived above.
The symbol $K$ for the background evidence will be dropped temporarily in 
order to simplify the equations, but is still understood to be present.
In particular, we need to assume in the following manipulations that the
propositions and their negations are not impossible given evidence $K$.
With this proviso we may write the sequence of relations:
\begin{eqnarray}
\beta((\sim A).B)&=&\beta(\sim A\vert B)\beta(B) \nonumber \\
&=&f[\beta(A\vert B)]\beta(B) \nonumber \\
&=&f\left [{\beta(A\vert B)\beta(B) \over \beta(B)}\right ]\beta(B) \nonumber \\
&=&f\left [{\beta(A.B)\over \beta(B)} \right ] \beta(B).
\end{eqnarray}
Next consider the expression $B.(\sim(A.B))$. This asserts that $B$ is true
and that simultaneously not both of $A$ and $B$ are true. This is the same
as to say that $(\sim A)$ is true and that $B$ is also true, i.e. to assert
$((\sim A ).B)$. Thus we get the logical equation
\begin{equation}
(\sim A).B = B.(\sim(A.B)),
\end{equation}
from which we obtain the corresponding believability relation
\begin{equation}
\beta((\sim A).B) = \beta(B.(\sim(A.B))).
\end{equation}
Now it is easy to see that the second compound proposition is actually
of the same form as the first, if we think of $(\sim A)$ being replaced by $B$
and $B$ being replaced by $\sim(A.B)$. On repeating the above sequence of
relations with these replacements we soon find that
\begin{equation}
\beta(B.(\sim(A.B))) = f\left [{\beta((\sim B).(\sim(A.B))) \over \beta(\sim(A.B))} \right ]\beta(\sim(A.B))
\end{equation}
But with a little thought, it is clear that to assert $((\sim B).(\sim(A.B))$ is to
assert $(\sim B)$ alone since it states that $B$ is not true and that not both of $A$ 
and $B$ are true, which is always true if $(\sim B)$ is, independently 
of the truth of $A$. In terms of believabilities therefore
\begin{equation} 
\beta((\sim B).(\sim(A.B))) = \beta(\sim B),
\end{equation}
and our chain of equations establishes that
\begin{equation}
f\left[ {\beta(A.B) \over \beta(B)} \right]\beta(B) = f\left[ {\beta(\sim B) \over \beta(\sim(A.B))} \right ]
\beta(\sim(A.B)).
\end{equation}

Finally, we write $\beta(B)=x$, so that $\beta(\sim B)=f(x)$, and $\beta(\sim(A.B))=y$,
implying $\beta(A.B) = f(y)$, to obtain the equation
\begin{equation}
xf\left [ {f(y) \over x} \right ] = yf\left [ {f(x) \over y} \right ] 
\end{equation}
as an interestingly symmetric condition to be satisfied by the function $f$
along with the first consistency relation (57) $f[f(x)]=x$.

Cox \cite{[Cox46]} has proved that the unique general solution of these two equations,
again under mild assumptions of differentiability, is
\begin{equation}
f(x) = [1-x^m]^{1/m},
\end{equation}
where $m$ is a non-zero real number. A more useful form for the result is
\begin{equation}
(f(x))^m = [1 - x^m].
\end{equation}
With this theorem to hand it is now possible to define a particularly convenient
final scale for a measure of credibility by writing
\begin{equation}
p(A\vert K) = \beta^m(A\vert K), \ \ m>0.
\end{equation}
Since our previous equations imply that
\begin{equation}
\beta^m(A.B\vert K) = \beta^m(A\vert B.K)\beta^m(B\vert K)
\end{equation}
and
\begin{equation}
\beta^m(\sim A\vert K) = (f[\beta(A\vert K)])^m = 1 - \beta^m(A\vert K),
\end{equation}
we see that the numbers $p(A\vert K)$, now called {\it probabilities\/}, satisfy the
very simple combination rules
\begin{equation}
p(A.B\vert K) = p(A\vert B.K)p(B\vert K)
\end{equation}
and 
\begin{equation}
p(\sim A\vert K) = 1 - p(A\vert K),
\end{equation}
while the choice $m>0$ ensures, in line with our previous conventions, 
that stronger credibility corresponds to greater probability.

The last two relations constitute one form of the usually accepted rules
of probability calculus and we have demonstrated, independently of any
considerations of frequencies, that they are equivalent to all possible
{\it consistent\/} versions of such a calculus. Numerical values for the $p(A\vert K)$
are fixed, so far, only at the extremes of certainty and impossibility
since $p=\beta^m$ yields
\begin{equation}
p({\rm Certainty}) = 1
\end{equation}
and the probability of an impossible proposition follows easily. For if
$A$ is certain on given evidence then $(\sim A)$ is impossible on the same
evidence and hence, by the second rule shows
\begin{equation}
p({\rm Impossibility}) = 0.
\end{equation}
Notice, however, that the rules tell us nothing about how to assign
probabilities for propositions which are neither certain nor impossible
on the given evidence, except that they should be chosen between zero
and unity to be consistent with our convention that probability is an
increasing function of credibility. It is therefore clear that to
address the problem of {\it assigning\/} probabilities we have to go beyond the
algebraic rules for manipulating and combining them. The assignment
problem of probability theory is indeed hardly mentioned at all in most
accounts of the subject, apart from simple cases where it is plausible
to identify sets of equally probable propositions, or where it seems
reasonable to take an observed relative frequency as the initial value.

Our aim now is to develop a small part of the theory implied by the two
rules we have derived; but this will be sufficient to approach the other
and rather neglected aspect of the subject, which attempts to formulate
rational procedures for choosing explicit numerical values for the
probabilities in at least some fairly general situations. The cases we
have in mind, of course, are more particularly those for which frequency
data are not available and in which there are no obvious sets of
primitive propositions to be taken as equally probable. The method we
present, however, can also be used to justify and extend earlier ideas.

\subsection{Some useful theorems}
\noindent The results derived in the last subsection give rise to an extensive 
mathematical theory of probability. In this subsection we collect together
only the few theorems that we shall need subsequently. Most of them are
straightforward deductions  from the basic rules, but are so useful that
they should be regarded as working rules in their own right. Summarising
the content of our consistent theory of probable inference, we have
shown that the degree of belief in a proposition $A$, given evidence $K$,
may be represented by a number $p(A\vert K)$, satisfying the restrictions
\begin{equation}
0 \leq p(A\vert K) \leq 1,
\end{equation}
in which the upper and lower limits denote certainty and impossibility,
respectively, and that the necessary computational rules for combining
such probability numbers can be reduced to the elementary forms:
\begin{equation}
{\rm Rule \ 1} \ \ \ \ \ \ \ \ \ \ p(A.B \vert K) = p(A\vert B.K)p(B \vert K)
\end{equation}
and
\begin{equation}
{\rm Rule \ 2} \ \ \ \ \ \ \ \ \ \ p(A \vert K) + p(\sim A\vert K)=1
\end{equation}
Our first deduction follows at once from Rule(1) by using the logical
relation $A.B=B.A$, so that
\begin{equation}
p(A.B\vert K)=p(B.A\vert K).
\end{equation}
Interchanging propositions $A$ and $B$ in Eq.(75) gives
\begin{equation}
p(B.A\vert K) = p(B\vert A.K)p(A\vert K),
\end{equation}
and on combining these equations with Rule(1) we find
\begin{equation}
{\rm Rule \ 3} \ \ \ \ \ \ \ \ \ \ p(A \vert B.K)p(B\vert K) = p(B\vert A.K)p(A \vert K)
\end{equation}
This simple result is known as Bayes' Theorem and it is one of the most 
useful in probability theory since it codifies the process of learning
from experience. To see this, suppose we are considering the probability 
of proposition $A$ on evidence $K$, and then acquire some new evidence $B$.
Rewriting Eq.(79) in the form
\begin{equation}
p(A\vert B.K)=\left [ {p(B\vert A.K)\over p(B\vert K)} \right ] p(A\vert K)
\end{equation}
we see that the {\it prior\/} probability $p(A\vert K)$, appropriate when we know only
$K$, gets transformed into the {\it posterior\/} probability $p(A\vert B.K)$ as a result
of the new evidence $B$. The transforming factor is the ratio of the
probability of the truth of $B$, supposing that both $A$ and $K$ are given, to
the probability that $B$ is implied by $K$ alone. This ratio is sometimes
called the {\it likelihood\/} and may be either greater than or less than unity.
Hence the new evidence will strengthen or weaken our belief in the truth
of $A$ in a calculable way.

The next result is a generalisation of Rule(2). Writing Rule(1) first as
it is, and then with $A$ replaced by $(\sim A)$, we obtain the equivalent pair
\begin{equation}
p(A.B \vert K) = p(A\vert B.K)p(B\vert K),
\end{equation}
and
\begin{equation}
p((\sim A).B \vert K) = p(\sim A\vert B.K)p(B\vert K),
\end{equation}
Addition of these two equations yields
\begin{eqnarray}
p(A.B \vert K) + p((\sim A).B \vert K)&=&[p(A \vert B.K) + p(\sim A \vert B.K)]p(B\vert K) \nonumber \\
&=&p(B\vert K),\ \ {\rm on \ using \ Rule(2)},
\end{eqnarray}
Thus
\begin{equation}
{\rm Rule \ 4} \ \ \ \ \ \ \ \ \ \ p(A.B \vert K) + p((\sim A).B \vert K) = p(B\vert K).
\end{equation}
It is easily seen that Rule(2) is a special case of this last equation,
arising when $B$ is a proposition certain on evidence $K$.

As a final general theorem we deduce a formula for the probability of
the compound statement $(A \lor B)$, called the disjunction of two
propositions, which asserts that either $A$ is true or $B$ is true or both
are true. We recall that the operation of disjunction is related to the
conjunction (AND) operation by
\begin{equation}
\sim (A \lor B) = (\sim A).(\sim B).
\end{equation}
We now drop $K$ temporarily and write Rule(4) in two slightly different
but equivalent forms as follows
\begin{equation}
p((\sim A).B) + p((\sim A).(\sim B)) = p(\sim A),
\end{equation}
\begin{equation}
p(A.B) + p((\sim A).B) = p(B).
\end{equation}
The next step is to apply Rule(2) with $A$ replaced by $(A \lor B)$:
\begin{eqnarray}
p(A \lor B)&=&1 - p(\sim (A \lor B))\ \ \ \ {\rm from\ Rule(2)} \nonumber \\
&=&1 - p((\sim A).(\sim B))\ \ \ \ {\rm from\ Eq.(85)} \nonumber \\
&=&1 - p(\sim A) + p((\sim A).B)\ \ \ \ {\rm from\ Eq.(86)}   \nonumber \\
&=&p(A) + p((\sim A).B)\ \ \ \ {\rm from\ Rule(2)} \nonumber \\
&=&p(A) + p(B) - p(A.B)\ \ \ \ {\rm from\ Eq.(87)}  .
\end{eqnarray}
On restoring $K$ we obtain the pleasingly symmetric theorem
\begin{equation}
{\rm Rule \ 5} \ \ \ \ \ \ \ \ \ \ p((A\lor B) \vert K) + p(A.B \vert K) = p(A\vert K) + p(B\vert K).
\end{equation}
From this point on we specialise to propositions of the type exemplified
in statistical mechanics. Consider first a whole set of inferences such
that our background knowledge $K$ implies the truth of at {\it least\/} one of the
possibilities. This kind of set is said to be {\it exhaustive\/} on $K$. In the
notation of symbolic logic, a set of propositions $\{ A_m \}, \ m = 1 \to M$, which
is exhaustive on $K$, can be characterised by saying that the compound statement
\begin{equation}
C = (A_1\lor A_2\lor \ldots\ldots\lor A_M)
\end{equation}
is certain given $K$. But we know that the probability of a certain
proposition is assigned the value unity, i.e.,
\begin{equation}
{\rm Rule \ 6} \ \ \ \ \ \ \ \ \ \ p(A_1\lor A_2\lor \ldots\ldots\lor A_M) = 1.
\end{equation}
On the other hand, if the evidence $K$ implies only that at {\it most\/} one of
the set of propositions $\{ A_m \}, \ m = 1 \to M$, can be true, then every
conjunction $(A_m.A_n), \ m \not= n$, must be impossible given $K$. Such a set is
said to be {\it exclusive\/} on $K$ and the condition is symbolised by
\begin{equation}
p(A_m.A_n\vert K) = 0, \ \ m \not= n.
\end{equation}
Hence we have, by Rule(5), that in this case
\begin{eqnarray}
p(A_1\lor A_2\lor \ldots\ldots\lor A_m\vert K)&=&p(A_1\lor( A_2\lor A_3\lor \ldots\ldots\lor A_M)\vert K) \nonumber \\
 =p(A_1\vert K) + p(A_2\lor A_3\lor \ldots\ldots\lor A_M\vert K) &-& p(A_1.(A_2\lor A_3\lor \ldots\ldots\lor A_M)\vert K),
\end{eqnarray}
The last term on the right hand side is zero by the condition of 
exclusion, since if $(A_1.A_m), \ m > 1$ is impossible, then so is the 
proposition $A_1.(A_2\lor A_3\lor \ldots\ldots\lor A_M)$. Repetition of this process by separating
off $A_2$, then $A_3$, etc., soon shows that for a set of mutually exclusive
propositions
\begin{equation}
{\rm Rule \ 7} \ \ \ \ \ \ \ \ \ \ p(A_1\lor A_2\lor \ldots\ldots\lor A_M \vert K) = \sum_m p(A_m\vert K)
\end{equation}
Finally, if the set of inferences $\{ A_m \}$ is both exhaustive (at least one
is true) and exclusive (at most one is true) on evidence $K$, then it is
clear that exactly one of them is true, though in general we do not know
which. Thus from Rules (6) and (7) we derive the special theorem, valid for a set of
$M$ exhaustive and exclusive propositions,
\begin{equation}
{\rm Rule \ 8} \ \ \ \ \ \ \ \ \ \ \sum_m p(A_m\vert K) = 1
\end{equation}
 
This is the appropriate place to discuss the Principle of Indifference
once again, though now in a more abstract form than before. If the
conditions  of Rule(8) are satisfied, but the evidence $K$ provides no
grounds for believing in any one of the propositions more strongly than
in any other, then there is clearly an element of symmetry present. We
know from Rule(8) that the probabilities must add up to unity, but that
is {\it all\/} we know. A rational way to proceed is to appeal to the now
familiar idea of consistency.

Suppose that some non-uniform distribution of the probabilities over the
labels $m$ is suggested, i.e., in general  $p(A_m\vert K) \not= p(A_n\vert K)$ for $m\not= n$. But
then, by permuting the propositions and relabelling them, say with the
new subscript $j = 1 \to M$, we can set up a differently ordered but
equivalent collection $\{ A_j \}$ with each $A_j$  equal to an $A_m$ of the previous
labelling. If all inferences carry with them the originally suggested
probabilities, then in the new scheme the distribution over the ordering
labelled by $j$ will look quite different to the earlier distribution over
the ordering specified by $m$. Yet the new arrangement is entirely
equivalent to the first with respect to the given evidence and should
for consistency appear to have the same distribution over labels. The
only choice of probabilities consistent with permutation invariance is
to take them all equal, so for the given conditions we should have
\begin{equation}
{\rm Rule \ 9} \ \ \ \ \ \ \ \ \ \ p(A_m\vert K) = p(A_n\vert K), \ \ \forall \ m, n
\end{equation}
Combining Rule(8) and Rule(9) we obtain, at last, some definite
numerical values for the probabilities of propositions which are neither
certain nor impossible on the given information, but only in a very
special situation. The simple result is
\begin{equation}
{\rm Rule \ 10} \ \ \ \ \ \ \ \ \ \ p(A_m\vert K) = {1 \over M}, \ \ \forall \ m
\end{equation}
for a set of $M$ inferences $\{ A_m \}$ known only to be exhaustive and mutually 
exclusive on given premisses $K$.

To conclude this subsection we discuss briefly the expected value of a 
function $X$ taking on numerical values $X_m$ for the inferences of an 
exhaustive and exclusive set $\{ A_m \}$. It may be necessary, as in 
statistical  mechanics, to estimate a single value for $X$, for comparison
with, or prediction of, experimental quantities. One possibility is just
to take the value of $X$ associated with the most probable inference. This 
number is called the {\it mode\/}. However, there may not be a single most
probable inference or perhaps there are several inferences with very
similar probabilities to which we would like to give some weight.  There
are no hard and fast rules available here, so we must rely on common
sense. In many situations, especially when the appreciable probabilities
are for a subset of inferences attached to similar values of $X_m$, a very
reasonable guess for the expected value is the {\it mean\/}, defined by
\begin{equation}
\bar{X} = \langle X \rangle = \sum_m p(A_m\vert K) X_m,
\end{equation}
As a test of whether $\bar{X}$ is a reliable estimate it is also reasonable to
consider the mean of the squared deviations of $X_m$, denoted by
\begin{equation}
(\Delta X)^2  = \sum_m p(A_m\vert K) (X_m - \bar{X})^2,
\end{equation}
and to demand that the {\it root mean square deviation\/} $\Delta X$ is, in some sense,
small. It is an easy exercise to show from the definitions, and from
Rule(8), that 
\begin{equation}
(\Delta X)^2  = \langle X^2 \rangle - \langle X \rangle^2.
\end{equation}
It is also easy to show that the mean square deviation of $X$ away from
any estimate $\tilde{X}$ of the expected value of $X$ is a minimum for $\tilde{X} = \bar{X}$. In the
development of the formalism of statistical mechanics we shall mainly be
concerned with the calculation of mean values of observable quantities, 
but it is frequently necessary to check that our average values are in 
fact {\it useful\/} predictions of the expected results of experiment. For this
purpose, it is usually sufficient to calculate also the root mean square
deviation from the average and verify that it is small compared to any
likely measurement error. If it should, nevertheless, turn out that the
observed discrepancy of the calculated mean from experiment is  greater
than the estimated experimental errors, then the most likely conclusion
is that the model used for calculation is deficient.

\subsection{Theory of uncertainty}

\noindent For nearly three centuries the only known method for assigning initial
probabilities in a calculation, apart from their approximate assessment
by finite statistical experiments, was to appeal to the Principle of
Indifference or Equiprobability. As we have seen, this requires the
identification of a set of exhaustive and exclusive propositions which
are symmetrical with respect to the available background knowledge.

In this subsection we investigate what can be done when it is clear from
the evidence $K$ that the propositions of an exhaustive and exclusive set
are not all equally probable. It may happen, for example, that direct
observations of a quantity associated with the possible inferences yield
values systematically different from the average calculated on the basis
of equal probabilities. Or one may have a strong theoretical reason for
believing in the existence of a sharp mean value, as in the case of the
energy of an isolated body, and want to choose probabilities consistent
with such an imposed average. Almost always, however, the amount of
information at our disposal is not enough to define a unique set of
probabilities, and we have to select, among the many distributions which
satisfy the known constraints, that which is, in some sense, the best or
the most honest one. Our only guide is the strong feeling that a truly
impartial choice of distribution should somehow imply the greatest
possible uncertainty, about the actual situation, as is compatible with
our special knowledge.

Let us therefore proceed by trying to formulate a measure of the state
of doubt, or uncertainty, implied by a given probability distribution
$p(A_m\vert K), \ m = 1 \to M$, over a set of exhaustive and exclusive propositions
$\{A_m\}$. For such probabilities we condense the notation by writing
\begin{equation}
p_m = p(A_m\vert K)
\end{equation}
with
\begin{equation}
\sum_m p_m =  \sum_m p(A_m\vert K) = 1,
\end{equation}
by Rule(8) of the previous subsection.

On being given numerical values for different special sets of
probabilities $\{p_m\}$, it is often easy to see at a glance that they
represent quite markedly different states of knowledge. For instance,
one set may attribute similar probabilities to many possibilities, so
that a choice between them is difficult. Another may indicate several
conceivable inferences, but with only one or two of them at all likely.
The first case gives a greater impression of uncertainty than the second
and it is only a short step to the idea that this kind of judgement can
be quantified. However, there is no guarantee that a {\it unique\/} measure of
uncertainty can be found or even that a {\it single\/} number will adequately
summarise our feelings of doubt.

We nevertheless postulate the existence of some numerical function of
the probabilities $\{p_m\}$, to be denoted by
\begin{equation}
U_m(p_1, \ldots , p_m, \ldots p_M) = U_M(\{p_m\}),
\end{equation}
which represents by its values the overall amount of uncertainty to be
associated with a given probability distribution. It is a surprising
theorem that this uncertainty function is determined up to an arbitrary
multiplicative constant by requiring it to satisfy two innocuous seeming
but natural conditions and a simple composition rule arrived at by
plausible considerations of consistency. These conditions are meant as a 
precise formulation of our intuitive ideas about how a measure of
``Amount of Uncertainty'' should behave. In particular, the notion of 
amount or quantity of something suggests that uncertainties should be
combined by some form of addition rule. The three conditions are: \hfil

\begin{itemize}
\item{(i)} The quantity $U_M(\{p_m\})$ should  be a smooth continuous function of
its arguments. Otherwise, small adjustments of the probabilities
may lead to large changes in the uncertainty, which would seem
highly unreasonable.
\item{(ii)} If all $p_m=1/M$, so that the function $U_M$ has $M$ equal arguments
and depends only on $M$, i.e.,
\begin{equation}
U_M(1/M, \ldots \ldots , 1/M) = f(M),
\end{equation}
then $f(M)$ should be a monotonically increasing function of $M$.
Clearly, the larger the number of equally probable possibilities
the greater should be our uncertainty about which one is true.
\item{(iii)} Different ways of evaluating the numerical value of $U_M(\{p_m\})$
should give the same answer. This condition of consistency can be
turned into a definite statement only by specifying alternative
routes for arriving at the answer.
\end{itemize}

Instead of giving the probabilities $\{p_m\}$ for the individual propositions
$\{A_m\}$ directly, we might provide equivalent information in two stages.
Suppose we group the first $r$ propositions into a compound block
\begin{equation}
B_1 = (A_1 \lor A_2 \lor, \ldots , \lor A_r),
\end{equation}
the next $s$ propositions into the block statement
\begin{equation}
B_2 = (A_{r+1} \lor A_{r+2} \lor, \ldots , \lor A_{r+s}),
\end{equation}
and so on, until we have $N$ blocks $\{B_i\}, \ i=1 \to N$, altogether and all
the $\{A_m\}, \ m = 1 \to M$ are used up. Then, information equivalent to the
$\{p_m\}$ values is specified by giving the set of probabilities for the
blocks, $\{P_i  = p(B_i\vert K)\}$, followed by all the conditional probabilities
$\{p(A_m\vert B_i.K)\}$ for an $A_m$ to be the true proposition, given that it occurs
in a block $B_i$.

From Rule(7), and the definition of the $\{B_i\}$, we have
\begin{equation}
P_1 =  \sum_1^rp_m, \ \ \ P_2 = \sum_{r+1}^{r+s}p_m, \ \ \ {\rm etc.},
\end{equation}
and it is clear that the $\{P_i\}$ satisfy the condition 
\begin{equation}
\sum_{i=1}^NP_i = \sum_{m=1}^M p_m = 1.
\end{equation}
Thus the probabilities of the $N$ exhaustive and exclusive block
statements $\{B_i\}$ contribute to an amount of uncertainty
\begin{equation}
U_N(P_1, P_2, \ldots , P_N) = U_N(\{P_i\}).
\end{equation}
To see how the specified conditional probabilities $\{p(A_m\vert  B_i.K)\}$, within
the blocks, are related to the original probabilities $\{p_m\}$ consider in
particular the conjunction $(A_m.B_1)$. This asserts that $A_m$ is true and
simultaneously that one of the propositions in the first block is true.
Thus the probability $\{p(A_m.B_1 \vert K)\}$ can be non-zero only for propositions in
the first block, and we have easily that
\begin{eqnarray}
p(A_m.B_1,\vert K) &=& p_m, \ \ \ 1 \leq m \leq r \nonumber \\
&=& 0, \ \ \ r+1 \leq m \leq M,
\end{eqnarray}
But by Rule(1) of the previous subsection we know that
\begin{equation}
p(A_m.B_1 \vert K) = p(A_m\vert B_1.K)p(B_1 \vert K)
\end{equation}
and, since $p(B_1 \vert K) = P_1$ is just the probability of block $B_1$, the
required conditional probabilities are expressed by
\begin{eqnarray}
p(A_m\vert B_1.K)&=&{\displaystyle {p_m \over P_1}}, \ \ \ 1\leq m \leq r \nonumber \\
p(A_m\vert B_2.K)&=&{\displaystyle {p_m \over P_2}}, \ \ \ r+1 \leq m \leq r+s
\end{eqnarray}
and similarly for all $N$ blocks. Now, on evidence $(B_1.K)$, the $r$
statements $\{A_m\}, \ 1 \leq m \leq r$, of the first block, form an exhaustive and
exclusive set with
\begin{equation} 
\sum_{m=1}^r p(A_m \vert B_1.K) = \sum_{m=1}^r {p_m \over P_1} = 1,
\end{equation}
so we see that the internal uncertainties of the blocks can be
represented by expressions of the type
\begin{equation}
U_r\left ( {p_1\over P_1} , \ldots , {p_r\over P_1} \right ),
U_s\left ( {p_{r+1}\over P_2} , \ldots , {p_{r+s}\over P_2} \right ),\ \ \ \ {\rm etc.}
\end{equation}
To  compute the final answer, we note that each of these uncertainties is
encountered only with the associated block probability; hence they
should be weighted with those probabilities and added to the uncertainty
$U_N(\{P_i\})$ arising from the choice of blocks. An alternative computational
path has now been specified and the result, for consistency, should be
the same as before, i.e.,
\begin{equation}
U_M(p_1, \ldots ,p_M) = U_N(P_1, \ldots ,P_N) +
P_1U_r\left ( {p_1\over P_1} , \ldots , {p_r\over P_1} \right ) +
P_2U_s\left ( {p_{r+1}\over P_2} , \ldots , {p_{r+s}\over P_2} \right ) + \ldots
\end{equation}
This form of composition rule is required to hold for all possible
regroupings of the original propositions, and in each case the two sides
of the equation will contain exactly the same probability information,
though differently arranged. Our formulation of the consistency
condition is now complete. However, it must be admitted that the
argument leading to the above formula is not entirely compelling. It
would seem more natural to equate the total uncertainty to the direct
sum of external and internal block uncertainties rather than to weight
the latter quantities with the corresponding block probabilities before
summing. Such a scheme does not lead to a useful result and we must 
therefore conclude that the proper and convincing form of
words to justify exactly the given rule has not yet been found, In spite
of this, we observe that the exhibited composition formula is still a
very reasonable embodiment of consistency.

It is not difficult to verify that, for any constant $\alpha > 0$, all three of
our conditions are satisfied by the function
\begin{equation}
U_M(\{p_m\}) = -\alpha\sum_{m=1}^M p_m\ln{(p_m)},
\end{equation}
and Shannon\cite{[Shannon]} has shown that it is unique. This expression neatly captures
the essence of uncertainty and has remarkable mathematical properties
that we shall indicate and use later on. It was originally developed by 
Shannon in connection with work on the efficient encoding and
transmission of messages, but it has turned out to have significance for
many other problems involving probabilities. What it is important to
observe is that the derivation makes no reference to physics, though the
result can in fact be applied very effectively in that special field. 
Indeed, our interest in the uncertainty function arises precisely
because it may be used as the basic tool for a rational choice of
probability distributions in statistical mechanics.

\subsection{Assignment of probabilities}
\noindent The uncertainty is sometimes called the {\it missing information\/} of a
probability distribution since it can be thought of as representing the
extra knowledge required to reduce all doubts to zero, i.e. to determine
the true proposition. It should be clear, however, that any distribution
of probabilities conditional on evidence $K$ always embodies {\it some\/}
knowledge about the situation, even if only, as in the special cases we
have considered, that we are dealing with propositions which are
exhaustive and mutually exclusive. When it comes to assigning
probabilities, we want to be sure that the selected distribution is a
true reflection of the knowledge we really have, and does not have
unwarranted assumptions built into it. A reasonable proposal, therefore,
is that the most honest probabilities compatible with the known data are
the ones for which our uncertainty, or missing information, is still as
large as possible. The definite measure established for this lack of
knowledge can then be used as a predictive instrument, as follows. The
least committal probability distribution among all those that conform
with the actual data, is that which maximises the missing information
$U_M(\{p_m\})$. This solves the initial assignment problem in many interesting 
cases.

The above suggestion has been called by Jaynes the Principle of Maximum 
Entropy\cite{[Jaynes_MaxEnt]}, since the measure of uncertainty was originally called entropy
by Shannon\cite{[Shannon]}. Notice particularly that it is {\it not \/} claimed that the
probabilities thus produced are the final word or are in any sense the
{\it real\/} ones. There are no such things as real physical probabilities, only
codings of our current state of knowledge, and the algorithm yields
those codings which represent, in a well-defined impartial way, just our
actually existing information, and nothing else. The power of the
principle becomes apparent only when the maximum entropy probabilities
are used to estimate a quantity before it is observed. It is interesting
if the prediction is successful, but even more interesting if it is not,
since that indicates the presence of some previously unknown constraint.

If the constant in the formula for the uncertainty is chosen to have the
value $\alpha = 1$ the resulting function will be written as $\sigma(\{p_m\})$ and
called the {\it information entropy\/} for the probabilities of $M$ exhaustive and
exclusive propositions. Explicitly
\begin{equation}
\sigma(\{p_m\}) = -\sum_m p_m\ln{(p_m)}
\end{equation}
with 
\begin{equation}
\sum_m p_m = 1 \ \ {\rm always}
\end{equation}
and we can state the formal rule to be known as: \hfil

\begin{itemize}
\item{\bf Jaynes' Principle:}\ Probabilities should be chosen so as to render
$\sigma(\{p_m\})$ an absolute maximum, subject only to 
the constraints imposed by the data $K$\cite{[Jaynes1],[Jaynes2],[Jaynes4]}.
\end{itemize}

Pending further investigation we shall assume that a distribution found
in this way is unique, so that the maximising set $\{p_m(K)\}$ depends solely
on the evidence $K$. In some problems it is also convenient to allow $M$ to
approach infinity, provided that the constraints ensure convergence of $\sigma(\{p_m\})$.

Before applying Jaynes' Principle, we consider a few mathematical 
properties of the expression for Information Entropy, the most obvious
being that $\sigma > 0$ if $0 < p_m < 1$ for all $m$. One potential source
of difficulty can be cleared up quickly. If some particular $A_m$ is
impossible on the evidence, then $p_m=0$, and the corresponding term in
the entropy seems ill defined.  But as $x \to 0, \ x\ln{(x)} \to 0$; so we make the
convention that $-p_m\ln{(p_m)} = 0$ when $p_m=0$. This is reasonable since
there is no uncertainty associated with an impossible inference --- we
know it is not true. The convention also implies a reasonable result
when one of the $A_m$ is certain, i.e. $p_m=1$, which entails that $p_n=0$
for $n \not= m$. The contribution of the vanishing probabilities is zero and
the entropy reduces to $\sigma = -\ln{(1)} = 0$. Clearly there should be no
uncertainty associated with a certain proposition either. Thus we have
established that $\sigma(\{p_m\}) \geq 0$ for any distribution and that $\sigma$ reaches its
lower limit of zero when one of the inferences is certain.

A more interesting inequality arises if we take any pair of
probabilities, say $p_1$ and $p_2$, and  make them more nearly equal while
keeping their sum constant. It is straightforward to prove that this
increases the entropy. For a constant sum $p_1 + p_2 = C$, write $p_1=p$
and $p_2 = C-p$, and consider the varying part of $\sigma(\{p_m\})$, namely
\begin{equation}
f(p) = -p\ln{(p)} - (C-p)\ln{(C-p)},
\end{equation}
as a function of $p$. its derivative is $df/dp = \ln{[(C-p)/p]}$, and it
follows easily that $df/dp = 0$ for $p=C/2$ and that $df/dp$ is greater than
or less than zero, respectively, according as $p$ is less than or greater
than $C/2$. Hence any change towards equality of probabilities, while
retaining the condition $\sum p_m = 1$, increases $\sigma$.

This brings out a very attractive feature of the Entropy Principle. The
process of maximising $\sigma$ consists of adjusting the $\{p_m\}$, without
violating any constraints, to make their differences as small as
possible. The effect, as mentioned earlier, is to spread out the
probability as evenly as is consistent with the data, which agrees
nicely with the idea of honest assessment. Naturally, in the presence of
several constraints, it will not in general be possible to make {\it all\/} the
probabilities equal to each other. When, however, the condition $\sum p_m = 1$
is the {\it only\/} constraint, the above result shows immediately that the 
absolute maximum of $\sigma$ {\it is\/} achieved by taking all the probabilities equal,
for then no further adjustment can increase the entropy. Hence we must
have $p_m = 1/M$ for all $m$, and we have recovered the Principle of Indifference
as a special case of the entropy algorithm. The 
corresponding maximised value of the information entropy is found to be
\begin{equation}
\sigma ( \{1/M\}) = \ln{(M)}.
\end{equation}
Any additional data beyond $\sum p_m=1$ may still be consistent with equal
probabilities, but is more likely to imply some differences between
them; so, under further constraints, the entropy maximum must in general
be less than the above value. Thus extra knowledge usually reduces the 
uncertainty, as we would expect.

The interpretation of entropy as missing information also looks very
natural in the light of the above formula. For we have $\ln{(M)} \approx \log_2(M)$,
which is a good estimate of how many questions, with yes/no answers, are
necessary to be sure of locating the true proposition among $M$ equally
likely possibilities. If for example $M=32$, then $\log_2(32) = 5$ questions
of the type ``Is it in the first sixteen?'' and so on, 
which halve the 
number of possibilities at each stage, will always ensure success. For
other given probability distributions the expected number of necessary
questions is similarly proportional to the value of $\sigma(\{p_m\})$.

Several other properties of the uncertainty function will be derived in
the next section; we have said enough, for the moment, to justify its
use as a tool for assigning probabilities by the Maximum Entropy Principle.
Alternative derivations of the method have been given in the last few years,
which may be more appealing to some tastes,
but they tend to be more technical than the simple argument
presented here. Before leaving the topic, however, we should mention
briefly one completely independent approach to the maximum entropy idea.
This is to base the method firmly on the Principle of Indifference, to
bypass entirely any general discussion of uncertainty, and to consider
instead how one might {\it generate\/} a probability distribution satisfying
given constraints. In other words we set out to analyse the likely
behaviour of a device for producing such distributions automatically.

Imagine a row of $M$ boxes into which $N$ similar small objects are somehow
placed, preliminary observations having indicated no reason for
supposing that any box is favoured as a receptacle. All we know is that
every object finishes up in some box. By the Principle of Indifference,
each of the $M^N$ ways of distributing the objects is equally likely; but,
in general, a final result consisting only of the set $\{n_m\}$, where $n_m$ is
the number of objects in the m'th box, can be achieved in many ways,
differing merely in the order of placing the objects. The numbers $\{n_m\}$
must, by hypothesis, satisfy the condition $\sum_m (n_m/N) = 1$, and by counting
the number of ways of choosing $N$ things $n_m$ at a time, without regard for
order, we find that the probability for generating the set $\{n_m\}$ is
\begin{equation}
P(\{n_m\})  = \left [ { N! \over \prod_m n_m!} \right ] M^{-N}. 
\end{equation}
Now the ratios $(n_m/N)$ will serve as probability values $p_m$ and they can
be made as fine as we like by taking $N$ indefinitely large. The above
expression then becomes the probability of a mechanically generated
probability distribution $\{p_m = (n_m/N)\}$ satisfying $\sum_m p_m=1$. Hence on
using Stirling's Theorem in the simplified form 
\begin{equation}
x! \approx x^x
\end{equation}
which is valid for indefinitely large factorials, we soon find that
\begin{equation}
P(\{p_m\}) \approx \left [ {\exp{ [\sigma(\{p_m\})] } \over M} \right ]^N,
\end{equation}
where
\begin{equation}
\sigma(\{p_m\})  = -\sum_m p_m\ln{(p_m)}
\end{equation}
is the by now familiar entropy function, which we already know satisfies
$\sigma \leq \ln{(M)}$. If we choose to consider only the distributions $(\{p_m\})$
which satisfy some specified constraints, then clearly the most probable
distribution generated in this way is the one which maximises $\sigma$ subject
to the constraints.

Furthermore, the ratio of probabilities for the maximising distribution
$(\{p_m\})_{\rm max}$ and any other set $(\{\tilde{p}_m\})$ consistent with the same data, is given by
\begin{equation}
{P_{\rm max} \over \tilde{P} }  = \exp{ [N(\sigma_{\rm max} - \tilde{\sigma})]},
\end{equation}
which increases rapidly with $N$. So, for large $N$, the maximum entropy
distribution is overwhelmingly the most likely to be generated. We may
therefore conclude that, in a very definite sense, the methods we have
explained yield the most probable probability distribution conforming to
our knowledge of the situation.
\vfil
\newpage

\section{Uncertainty and entropy}
\subsection{Isolated equilibria} 

\noindent The material of the second section was intended as a simple preliminary
discussion of the main problem to be solved before the properties of a
body at equilibrium can be estimated. A system with observables constant
in time is reasonably described in terms of energy eigenstates and we
require to know which eigenstate occurs. Since it is {\it unreasonable\/} to
suppose that the exact eigenstate occurring can be determined, the best
we can hope to do is to assign probabilities to the possible states and
to estimate the measurable quantities by averaging. The ensuing account
of probability theory in the third section was purposely kept rather
abstract in order to emphasise that the {\it methods\/} used for assigning the
probabilities do not, in themselves, depend in any way on the mechanics
of bodies; the physics  of the situation enters only when the general
method is applied to find the eigenstate probabilities consistent with
our background knowledge.

The earlier discussion was based on the assumption that an isolated
body, with known Hamiltonian $\hat{H}$, was in some normalised state $\vert\Psi(t)\rangle$,
which developed in time according to the Schr\"odinger equation
\begin{equation}
\hat{H}\vert \Psi_{\alpha}(t) \rangle = i\hbar {\partial \vert \Psi_{\alpha}(t) \rangle \over \partial t},
\end{equation}
where we have now attached to $\vert\Psi\rangle$ a subscript $\alpha$ to denote all the labels
required to specify the state.

As soon as we realise that the available data $K$ are never sufficient to
determine $\vert\Psi_{\alpha}\rangle$ uniquely, it is clear that the previous treatment has to
be generalised. To do this we suppose that $\vert\Psi_{\alpha}(t)\rangle$ is one member of a
complete set of states $\{\vert\Psi_{\alpha}(t)\rangle, \vert\Psi_{\beta}(t)\rangle, \ldots\}$, orthonormal
at $t=0$. Thus, for any $\alpha$ and $\beta$,
\begin{equation}
\langle\Psi_{\alpha}(0) \vert \Psi_{\beta}(0)\rangle = \delta_{\alpha\beta}.
\end{equation}
It then follows easily from the wave equation that this orthonormalisation
is preserved at all later times $t>0$. Hence the states $\{\vert\Psi_{\alpha}(t)\rangle,\ldots\}$ 
form an exhaustive and exclusive set of possibilities for
description of the system and, with any evidence $K$, each will occur with
some probability $p(\alpha\vert K)$, where $\sum_{\alpha}p(\alpha\vert K)=1$.

The discussion of what happens, on the hypothesis that the state is
$\vert\Psi_{\alpha}(t)\rangle$, is essentially the same as before. If we wait long enough, the
time dependence of reproducible macroscopic properties will effectively
damp out and our estimates of energy and other quantities in state $\vert\Psi_{\alpha}\rangle$
take on the constant equilibrium forms
\begin{equation}
\langle E \rangle_{\alpha} = \sum_j \langle\psi_j\vert\hat{H}\vert\psi_j\rangle p(j\vert\alpha)
\end{equation}
and 
\begin{equation}
\langle Q \rangle_{\alpha} = \sum_j \langle\psi_j\vert\hat{Q}\vert\psi_j\rangle p(j\vert\alpha)
\end{equation}
In these expressions the matrix elements are expectation values with
respect to energy eigenstates $\vert\psi_j\rangle$, defined by $\hat{H}\vert\psi_j\rangle = E_j\vert\psi_j\rangle$;
and, from the rules of quantum mechanics, the probabilities associated with the 
energy eigenstates are given by $p(j\vert\alpha) = \vert a_j^{\alpha}\vert^2$, where the amplitudes
$a_j^{\alpha}$ are the constant coefficients of the expansion of the system state
in energy eigenfunctions, i.e.,
\begin{equation}
\vert\Psi_{\alpha}(t) \rangle = \sum_j a_j^{\alpha}\vert\psi_j\rangle\exp{(-E_jt/\hbar)}.
\end{equation}
The probabilities $\{p(j\vert\alpha)\}$ are thus evaluated on the explicit assumption
that the actual state is $\vert\Psi_{\alpha}(t)\rangle$. However, in order to obtain useful
estimates of the system observables we must also take specific note of
the probabilities $\{p(\alpha\vert K)\}$ for the occurrence of the states of the set
$\{\vert\Psi_{\alpha}(t)\rangle\}$. A very reasonable plan is to construct the mean values
\begin{equation}
\bar{E} = \sum_{\alpha}\langle E\rangle_{\alpha}p(\alpha\vert K )
\end{equation}
and 
\begin{equation}
\bar{Q} = \sum_{\alpha}\langle Q \rangle_{\alpha}p(\alpha\vert K ),
\end{equation}
from which, on remembering that $\langle \psi_j\vert\hat{H}\vert\psi_j\rangle=E_j$, we find the mean energy
\begin{equation}
\bar{E} = \sum_{\alpha,j} E_j p(j\vert\alpha)p(\alpha\vert K),
\end{equation}
while for the mean values of other quantities we finish up with
\begin{equation}
\bar{Q} = \sum_{\alpha,j} \langle\psi_j\vert\hat{Q}\vert\psi_j\rangle p(j\vert\alpha)p(\alpha\vert K).
\end{equation}
Now, by Bayes' Theorem, Rule(3) of subsection 3.3, we have that 
\begin{equation}
p(j\vert\alpha)p(\alpha\vert K) = p(j.\alpha\vert K) = p(\alpha\vert j.K)p(j\vert K),
\end{equation}
which leads immediately to the conclusion that
\begin{equation}
\sum_{\alpha}p(j\vert\alpha)p(\alpha\vert K) = p(j\vert K)\sum_{\alpha}p(\alpha\vert j.K).
\end{equation}
Moreover, we have by hypothesis that the states $\{\Psi_{\alpha}(t)\rangle, \ldots \}$ form an
exhaustive and exclusive set of possibilities, on any evidence. Hence
$\sum_{\alpha}p(\alpha\vert j.K)=1$ and we see that the mean values reduce to
\begin{equation}
\bar{E} = \sum_j E_jp(j\vert K)
\end{equation}
and 
\begin{equation}
\bar{Q} = \sum_j \langle\psi_j\vert \hat{Q}\vert\psi_j\rangle p(j\vert K).
\end{equation}
The upshot of this more careful discussion is, that when equilibrium is
reached, it is not necessary to consider what the actual state of the 
system might be, but only to assign probabilities for the various energy
eigenstates on the basis of the data $K$. Clearly, the energy states
$\{\vert\psi_j\rangle\ldots\}$ are also exhaustive and exclusive on any evidence $K$, since a
measurement of energy must always result in some particular eigenstate.
Thus the finally required probabilities $\{ p(j\vert K)\}$ must satisfy the
condition $\sum_jp(j\vert K) = 1$.

We are now faced with exactly the type of problem envisaged in Jaynes' 
Principle\cite{[Jaynes1],[Jaynes2],[Jaynes4]}.
There are $M$ possibilities, where $M$ may be indefinitely large,
or even denumerably infinite, to be considered in the light of given 
data $K$. The most impartial assignment of the respective probabilities is
achieved by maximising the information entropy
\begin{equation}
\sigma(\{p_j\}) = -\sum_j p_j\ln{(p_j)},
\end{equation} 
subject  to probability normalisation and the constraints implied by $K$.
Assuming that there is a unique maximising distribution, which needs to
be checked in each case, the resulting probabilities $p(j\vert K)$ will depend
on the data $K$ and may be used to construct the mean values of
observables for comparison with experiment. The actual implementation of
this programme, for the most commonly considered types of assumed prior
knowledge, will be spelt out in the next section. Notice, however, that
once the probabilities have been decided on there appears to be no
further use for the uncertainty function $\sigma$. While this is, strictly
speaking, correct, since everything of experimental interest can be
calculated from the probabilities, it is not entirely idle to
investigate the physical meaning of the maximised uncertainty.

The most important remark is that the existence of a unique maximising 
distribution $\{p(j\vert K)\}$ implies that the corresponding maximal value of
the information entropy
\begin{equation} 
\sigma_{\rm max}(\{p(j\vert K)\}) = \sigma_{\rm max}(K)
\end{equation}
is a function of the assumed data. We shall show that the function
$\sigma_{\rm max}(K)$ possesses all the properties of the thermodynamic entropy $S$. In
fact we shall try to make plausible the definite identification
\begin{equation}
S=k_B\sigma_{\rm max}(K) + S_c,
\end{equation}
in which $k_B$ is a constant determined by the experimental definition of
$S$, and $S_c$ is an arbitrary constant. Such a demonstration amounts to a
detailed statistical interpretation of the thermodynamical entropy.

The theoretical formula for entropy as maximal uncertainty has a
definite mathematical expression for each specific ``system plus relevant
data''. It depends both on the structure of the body and on what we have
chosen to measure, and this is also true of the experimental entropy $S$.
It is clearly very useful to be able to identify $\sigma_{\rm max}(K)$ with $S(K)$, in 
suitable units, for then all the well-developed machinery of
thermodynamics can be applied to the statistical expression. This enables rapid
derivations of results in statistical mechanics which would be
tedious to construct directly.

\subsection{Thermodynamic entropy}

\noindent We proceed by examining some of the well-known properties of thermodynamic
entropy $S(K)$, with the ultimate aim of showing mathematically 
that $\sigma_{\rm max}(K)$ has the same properties. To connect the mathematics with
the physics we shall assume throughout that statistically calculated
mean values $\bar{Q}$ are accurate estimates of the corresponding observed
quantities; it will then become highly plausible that $S$ and $\sigma_{\rm max}$ should
be closely related. We note first that in classical thermodynamics
nothing is assumed about the microscopic structure of bodies, but that a
large accumulation of experience may be understood by postulating the
existence of an entropy function $S$ with the following properties: \hfil

\begin{itemize}
\item{(i)} It is defined for equilibrium conditions,
\item{(ii)} is a function of the ``state'' of the system,
\item{(iii)} is not decreased by an adiabatic process,
\item{(iv)} is an extensive or additive variable, and
\item{(v)} is related experimentally to heat and temperature.
\end{itemize}

Each of these items is also characteristic of our maximised uncertainty
function. The first two are included for the sake of completeness and
can be disposed of relatively quickly. To show that $\sigma_{\rm max}$ possesses the
next two properties is more complicated, but the proofs depend only on
general considerations. The last item alludes to the definite
experimental procedure for measuring entropy which underlies the
formulation of the previously listed properties. We shall defer the
demonstration, that maximised uncertainty is determined empirically by
exactly the same procedure, until the appropriate probability
distribution has been derived. In this section we concentrate attention
on the first four listed properties of $S$ and $\sigma_{\rm max}$ .

\noindent{\bf Equilibrium Restrictions:} The entropy $S$ is traditionally introduced as a
quantity which has meaning only for the macroscopically static condition
of equilibrium. Its application to processes arises from the possibility
of comparing values of $S$ in initial and final equilibrium states. Such
comparisons may be used, for example, to indicate what final state a
system will reach after internal constraints have been removed; but
nothing is predicted about rates of evolution or relaxation times. The
experimental entropy is not usually even defined for the intermediate
time-varying states of a process unless the changes are so gradual that
the body is in effective equilibrium at all stages. Processes fulfilling
the last condition are said to be quasi-static and are considered
precisely for the purpose of measuring or computing entropy differences.

The uncertainty $\sigma$ is also introduced primarily as a means of describing
equilibrium conditions, though in a more microscopic and theoretical way
than $S$. The probabilities for the possible energy eigenstates are found
by maximising $\sigma(\{p_j\})$ subject to appropriate macroscopically static
constraints. Hence $\sigma_{\rm max}(K)$ can be defined for equilibrium data $K$ and, by
obvious extension, for the intermediate states of a quasi-static
process. We shall soon show that  knowledge of the function $\sigma_{\rm max}(K)$
provides the same sort of information about processes as that given by
$S(K)$. More generally, and by using data on rates of change, it is
possible to formulate an uncertainty function which applies even in non-static
intermediate situations. This as least suggests that the concept of
entropy need not be restricted solely to equilibrium conditions, but
discussion of such a far-reaching generalisation of ordinary thermodynamics
and statistical mechanics will not be attempted here.

\noindent{\bf State Functions:} That the entropy $S$ and uncertainty $\sigma$ are both functions
of the equilibrium data $K$ is expressed conventionally by saying that
they are state-functions. But the word ``state'' has also been used to
denote the quantum wave-vector of the system. The two meanings of the 
word are quite distinct and it should always be clear from the context
whether we are referring to the quantum state of the body or to the
available data at equilibrium. It is perhaps worth remarking that, whatever
the actual physical conditions in the body, the thermodynamic state
is defined entirely and solely in terms of the measured data.

The statistical quantity $\sigma_{\rm max}(K)$ is a state-function by construction,
but the discovery of that special function of equilibrium data which is
denoted by $S(K)$ was the central result of classical thermodynamics. The
point is that thermal processes are described experimentally in terms of
the non-integrable differential\ $\dbar q$ of heat transfer, so it was certainly
of fundamental interest that there exists a universal integrating factor
which converts\ $\dbar q$ into the perfect differential $dS$. A more extensive
discussion of this will be given in the next section when we describe
exactly how $S$ or $\sigma_{\rm max}$ may be measured. For the moment we shall just
indicate in a general way how $S$ depends on the state.

The required equilibrium data must include an estimate of the mean
energy $\bar{E}$ of the system and the values of selected external and internal
parameters, typically the volume $V$ and the number of atoms $N$. These are
meant only as examples and it is quite possible that other controllable
parameters will eventually have to be specified in order to understand
the full range of behaviour of the system. With this proviso, the
entropy may be written as a function of experimental data in the form
\begin{equation}
S = S_{\rm exp}(K,K_0) + S_0,
\end{equation}
where $K \equiv (\bar{E}, V, N)$ and $K_0 \equiv (\bar{E}_0, V_0, N_0)$ are corresponding data for an
equilibrium reference state with arbitrarily assigned entropy $S_0$. The
entropy $S_{\rm exp}(K,K_0)$ is determined empirically from the results of certain
specific measurements of temperature and heat transfer, as the system is
caused to undergo a quasi-static change from the arbitrary reference
state $K_0$ to the final condition with data $K$. When the function has been 
mapped out over a sufficient domain of data, physical predictions about
the system follow by differentiation or by comparing values of $S$ in the
different states of equilibrium. In all such applications of the concept
of entropy the final results are independent of the arbitrary constant
$S_0$ and the initial reference state $K_0$.

The microscopic description of equilibrium in terms of energy eigenstates
also makes essential use of the data $K$. In particular, the energy
operator $\hat{H}$, which determines the eigenstates $\{\vert\psi_j\rangle\}$, depends on various
parameters such as $V$ and $N$, and the corresponding energy eigenvalues are
the functions $E_j(V,N)$. It is usually assumed in addition that the system
possesses a definite mean energy $\bar{E}$. Hence a normalised probability
distribution for the eigenstates must satisfy, at least, the constraints
\begin{equation}
\sum_j p_j = 1 \ \ \ {\rm and} \ \ \ \sum_j E_j(V,N)p_j = \bar{E}.
\end{equation}
The unique maximal uncertainty consistent with these restrictions will
thus depend on the same data, $K \equiv (\bar{E}, V, N)$, as is required for definition
of $S$, i.e., we have $\sigma_{\rm max} = \sigma_{\rm max}(K)$.

No mention is made here of a reference state $K_0$, or of any arbitrary
additive constant, because the maximised uncertainty is an absolute
quantity and has a natural zero in which the exact eigenstate of the
body is known. The reason for this difference is that $\sigma_{\rm max}(K)$  has so far
been formulated as a theoretical expression whose explicit construction
depends on a detailed knowledge of the eigenvalue spectrum of the system
Hamiltonian as well as on the given data. Technically, it is a function of $K$
and a functional of the eigenspectrum.  To prove the basic identity of
the functions $\sigma_{\rm max}$ and $S$ we will certainly need to show that the quantum
information can be replaced to some extent by a knowledge of just those
thermodynamic measurements which are needed to evaluate $S$. It will then
become plain that experimental determinations of $\sigma_{\rm max}(K)$ are necessarily
relative to an initial equilibrium reference state $K_0$ and that they
leave undetermined an arbitrary constant representing $\sigma_{\rm max}(K_0)$.

On the other hand it is sometimes feasible to obtain direct experimental
information on the energy spectrum, from which the function $\sigma_{\rm max}(K)$ may
be calculated as an absolute quantity. Comparison of its variation with
$K$ to that of the function $S(K)$ then gives objective support to the idea
that the two concepts are intimately related. Several interesting
insights into thermal behaviour have been obtained as by-products. For
instance, the theoretical formula leads at once to a statistical
interpretation of the Nernst\cite{[Nernst]} Postulate on entropy, sometimes called
the Third Law of Thermodynamics. Before discussing such relatively minor
matters, however, we shall show in the next subsection how to prove
the probabilistic analogue, for $\sigma_{\rm max}(K)$, of the Second Law of Thermodynamics
governing the behaviour of $S(K)$.

\subsection{Adiabatic processes}

\noindent When the properties of an otherwise isolated body are changed by
altering the configuration of the confining walls, or by modifying the
applied fields, or by adjusting the mechanical stresses acting on it, in
a controllable way, the process is said to be adiabatic. The effect is
that the system Hamiltonian is varied in a definite knowable manner and,
in general, work is done on the system. Bodies may also be subjected to
processes in which no obvious mechanical work is done on them, as when
they are exposed to a flame, or are placed in an oven or a refrigerator.
Such non-adiabatic processes always involve the presence of another body
which can exchange energy with the given system in a mechanically
uncontrollable way. In essence, the effective Hamiltonian of the given
body suffers unpredictable or random perturbations which are not macroscopically 
recognisable as work. This kind of energy transfer is called 
heating (or cooling), and for a pure heating process the Hamiltonian of 
the body at the end is the same as at the start of the operation. For an
adiabatic process the system Hamiltonian at the end may be the same 
or may be different from the initial one, but in either case the energy
gained or lost by the body can be calculated from external mechanical
measurements. In a heating process this is not possible and the final
energy transfer can only be determined by direct measurement of the
difference between the initial and final energies of the body. The most
general type of process involves both heating and working.

The heating of a body takes place by radiation, conduction or convection 
of energy to or from the outside world. In practice, it is often
possible to surround the system with heat-proof walls which, to a good 
approximation, isolate it from these microscopically random influences.
The fundamental law of entropy change to be discussed now applies only
under such conditions of adiabatic isolation. This rule is one version
of the Second Law of Thermodynamics and it states that the entropy of a 
body never decreases as the result of a reproducible adiabatic process.
It is assumed there that the system starts and finishes in a condition of
complete thermodynamic equilibrium, with the initial and final entropies
$S(\bar{E}_I, V_I)$ and $S(\bar{E}_F, V_F)$ determined entirely by the corresponding
equilibrium data. We have suppressed the dependence of $S$ on the number $N$
of atoms in the body, on the understanding that $N$ is constant for an
isolated system. The mathematical expression of this version of the
Second Law is that, for any adiabatic transformation,
\begin{equation}
S(\bar{E}_F, V_F) \geq S(\bar{E}_I, V_I).
\end{equation}
It follows immediately from this relation that any equilibrium condition
with entropy less than that of the initial state of the body is
adiabatically inaccessible. The existence of equilibria which can not be
reached from a given initial condition by the performance of work alone
is a primitive experimental observation. Indeed, Carath\'eodory\cite{[Carath]} was able
to base thermodynamics precisely on the general principle that, in the
neighbourhood of any equilibrium state, there are adiabatically
unattainable states. From this he deduced mathematically the existence
of an entropy function $S$ obeying the above inequality. However, we do
not propose to discuss here the various axiomatic foundations of thermodynamics;
we shall just take it as well-confirmed that there is a 
function $S$, depending on experimental data, which satisfies the law we have 
stated. Our concern is to show that the maximised uncertainty 
function $\sigma_{\rm max} (\bar{E},V)$ has the same property.

Consider, therefore, an isolated body which is in equilibrium at initial 
time $t=0$. If the initial Hamiltonian is $\hat{H}_I$ and the data are $K_I = (\bar{E}_I, V_I)$,
then by maximising the uncertainty function, subject to $K_I$, we derive
the probabilities $\{p[j(I)\vert K_I] = p_j(I)\}$ for the occurrence of the orthonormalised
energy eigenstates $\{ \vert \psi_j^I \rangle = \vert \Psi_j^I(t=0) \rangle \}$ which satisfy
\begin{equation}
\hat{H}_I \vert\Psi_j^I(0)\rangle = E_j^I\vert\Psi_j^I(0)\rangle.
\end{equation}
The corresponding initial information entropy is represented by
\begin{equation}
\sigma_{\rm max}(\bar{E}_I, V_I) = -\sum_j p_j(I) \ln{(p_j(I))}.
\end{equation}
Now suppose the system is caused to undergo an adiabatic change
involving controlled work done on it, or by it. This is described
theoretically by means of the time-dependent Hamiltonian
\begin{equation}
\hat{H}(t) = \hat{H}_I + \hat{W}(t).
\end{equation}
where $\hat{W}(t)$ is the potential energy operator of the forces acting on the
body. Eventually the work process will cease, say at $t=t_c$, so that the
final potential energy operator, $\hat{W}_F = \hat{W}(t_c)$, is constant for all $t \geq t_c$.
The final Hamiltonian of the body is thus given by
\begin{equation}
\hat{H}_F = \hat{H}_I + \hat{W}_F, \ \ \ \ t\geq t_c.
\end{equation}
The system is then allowed to come to equilibrium once more, with the
Hamiltonian $\hat{H}_F$ and freshly measured equilibrium data $K_F = (\bar{E}_F, V_F)$. This
final equilibrium condition will naturally be described, via the maximisation
of uncertainty, by the new probabilities $\{q(m(F) \vert K_F) = q_m(F)\}$,
which refer to the occurrence of energy eigenstates $\{\vert \varphi_m^F\rangle\}$ defined by
\begin{equation}
\hat{H}_F\vert\varphi_m^F\rangle = E_m^F\vert\varphi_m^F\rangle.
\end{equation}
The appropriate final information entropy at the end of the adiabatic
process, when equilibrium has again been reached, takes the form
\begin{equation}
\sigma_{\rm max}(\bar{E}_F, V_F) = -\sum_m q_m(F)\ln{(q_m(F))},
\end{equation}
and we require to show that, necessarily,
\begin{equation}
\sigma_{\rm max}(\bar{E}_F, V_F) \geq \sigma_{\rm max}(\bar{E}_I, V_I),
\end{equation}
i.e. that the uncertainty can never decrease in this type of change. The
demonstration has some subtle points, so we shall develop it slowly. It
proceeds in two main stages, with the hardest, but essential, step being
the proof of an intermediate inequality. We shall also need to discuss
carefully an often repeated objection to the derivation, which we shall
show is based on a common misinterpretation of the formalism.

We observe first that the varying Hamiltonian $\hat{H}(t)$ is supposed known.
Accordingly, it is possible in principle to follow the evolution of the
system by integrating the wave equation
\begin{equation}
\hat{H}(t)\vert\Psi(t)\rangle = i\hbar {\partial \vert\Psi(t)\rangle \over \partial t},
\end{equation}
from $t=0$ to $t>t_c$, with the initial condition that $\vert\Psi(0)\rangle$ is one or
other of the eigenstates $\{\vert\Psi_j^I(0)\rangle\}$, of $\hat{H}_I$, that we started with. Thus
by the time $t=t_F>t_c$, that the system has come to equilibrium again, 
the functions $\{\vert\Psi_j^I(0)\rangle\}$ will have evolved into the predictable states
$\{\vert\Psi_j^I(t_F)\rangle\}$, which remain orthonormal, but are in general no longer
eigenstates of either $\hat{H}_I$ or $\hat{H}_F$. Furthermore, each of them will occur 
with the initial probability $p(j(I)\vert K_I) = p_j(I)$ since no information is
lost by integrating the Schr\"odinger equation. This is true because the
equation is of first order in time, implying that the final state is 
determined uniquely by $\hat{H}(t)$ and the initial state. Hence, for any such
reproducible process, we should be able to predict the values of the
observables at the final equilibrium by using estimates of the form
\begin{equation}
\bar{Q}_F = \sum_j \langle \Psi_j^I(t_F)\vert \hat{Q} \vert \Psi_j^I(t_F) \rangle p(j(I)\vert K_I),
\end{equation}
which we assume would agree with direct experimental measurements on the
system. This last assumption is actually the crux of the matter, as we
shall see; but is really saying no more than that we believe in the 
correctness of quantum theory.

From the given postulate it follows that a knowledge of the initial set
of probabilities, and of the time-developed wave functions originating
from the eigenstates of $\hat{H}_I$, is sufficient to deduce the properties of
the final situation. In a sense, therefore, the information entropy at
the end of the process, as calculated from the constant probabilities,
is exactly the same as the maximised uncertainty $\sigma_{\rm max}(I)$ of the initial
set-up. In other words, our initial uncertainty about the actual state 
of the system will not be increased if the final state-vectors are
uniquely and controllably determined from the initial ones as is implied
by the deterministic  wave equation of quantum mechanics.

The above result is very well known and appears to {\it contradict\/} the
observation that adiabatic changes generally lead to an increase of
entropy. The obvious conclusion, that information entropy can never be
identified with thermodynamic entropy, is, however, totally erroneous.
The correct conclusion is exactly the opposite. The demonstrated
constancy of the uncertainty under the given conditions, far from contradicting
the Second Law, is all that is required to {\it prove\/} an
analogous law for {\it maximised\/} uncertainties.

This very surprising theorem is due to E.T.Jaynes\cite{[Jaynes1]}. The {\it point\/} is that to
define an uncertainty function identifiable with thermodynamic entropy
we have to reject detailed knowledge of the intermediate non-equilibrium
history of the body and use {\it only\/} the initial or final equilibrium
measurements.\footnote{The one exception to the remarks of this paragraph occurs when the
change is performed without friction and so slowly that the system
remains effectively in equilibrium at all times. Any such process, and
not only an adiabatic one, is called quasi-static and reversible.
However, if the process {\it is\/} adiabatic, as well as quasi-static and reversible,
then we are able to follow the intermediate evolution of the
system by means of reproducible macroscopic measurements of, for
example, the pressure. This extra data is sufficient to ensure that the
final maximised uncertainty $\sigma_{\rm max}(F)$ is equal to the initial uncertainty
$\sigma_{\rm max}(I)$. The details of this important {\it isentropic\/} process will be
elaborated in the next section, as part of the discussion of entropy 
change in an arbitrary quasi-static reversible process, one which may
involve heat transfer in addition to work.}
In particular, it is the maximised uncertainty determined by the 
final equilibrium data, {\it alone\/}, which represents the final thermodynamic
entropy. The qualitative reason for the theorem is now easy to
see. Indeed, it even seems trivial to remark that the {\it maximal\/}
uncertainty consistent with the finally observed properties can not be
less than any other information entropy compatible with the same data,
and in general it must be greater. However, the remark is not entirely
trivial because the final maximising probabilities refer to states which
are different from the non-stationary evolved states $\{\vert\Psi_j^I(t_F)\rangle\}$. Hence
the associated uncertainties are not directly comparable.

To see how this difficulty may be overcome we recall that the 
description of the final equilibrium should be given in terms of the
eigenstates $\{\vert\varphi_m^F\rangle\}$ of $\hat{H}_F$. We remember also that at equilibrium 
only the constant part of the expectation value of a reproducible observable
effectively survives. So, if $t_F$ is the time at which equilibrium has
been reached, then for all $t \geq t_F$ 
\begin{equation}
\langle \Psi_j^I(t)\vert \hat{Q} \vert\Psi_j^I(t)\rangle \approx 
\sum_m \langle\varphi_m^F \vert \hat{Q} \vert \varphi_m^F\rangle 
p(m(F) \vert j(I)).
\end{equation}
Here, $p(m(F) \vert j(I))$ is the probability of finding the state $\vert\varphi_m^F\rangle$ when
the system is known to be in the state $\vert\Psi_j^I(t)\rangle$, the latter 
having evolved, under the action of $\hat{H}(t)$, from the initial state $\vert\Psi_j^I(0)\rangle$. For
$t\geq t_F$, $\hat{H}(t)$ will by hypothesis have become the constant operator $\hat{H}_F$ and
at such times the state vector $\vert\Psi_j^I(t)\rangle$ will satisfy the wave equation
\begin{equation}
\hat{H}_F \vert\Psi_j^I(t)\rangle  = i\hbar {\partial \vert\Psi_j^I(t)\rangle \over \partial t}.
\end{equation}
It follows that the $a$-coefficients in the expansion
\begin{equation}
\vert\Psi_j^I(t\geq t_F)\rangle =\sum_m a[m(F)\vert j(I)] \vert\varphi_m^F\rangle\exp{(-iE_m^Ft/\hbar)}
\end{equation}
are constants, and that the required quantum probabilities are
\begin{equation}
p(m(F) \vert j(I)) = \left \vert a[m(F) \vert j(I) ] \right \vert^2.
\end{equation}
Summation of these probabilities over $m$ must yield unity since the
states $\{\vert\varphi_m^F\rangle\}$ are taken to be complete and orthonormal. Also, the
square modulus of the $a$-coefficient can be interpreted in quantum theory 
as the probability $p(j(I)\vert m(F) )$ of finding $\vert\Psi_j^I(t\geq t_F)\rangle$ when the state is
known to be $\vert\varphi_m^F\rangle\exp{(-iE_m^Ft/\hbar)}$. Hence, by completeness and orthonormality
the sum of the probabilities $p(m(F)\vert j(I)) = p(j(I)\vert m(F) )$ over $j$ must be
unity as well. In abbreviated notation, therefore, these normalization 
conditions are
\begin{equation}
\sum_m p(m\vert j) = 1 = \sum_j p(m\vert j).
\end{equation}

To complete the calculation of mean value estimates of observables we
take into account that the evolved system states $\{\vert\Psi_j^I(t)\rangle\}$ occur only
with probabilities $p(j\vert K_I)$. Thus at final equilibrium, when $t \geq t_F$,
\begin{eqnarray}
\bar{Q}_F &=& \sum_j \langle \Psi_j^I(t)\vert \hat{Q} \vert\Psi_j^I(t)\rangle p(j\vert K_I) \nonumber \\
&\approx& \sum_{m,j} \langle\varphi_m^F \vert \hat{Q} \vert \varphi_m^F\rangle p(m\vert j)p(j\vert K_I).
\end{eqnarray}
The above argument is clearly just a modification of the development in
subsection (4.1) and the compounded probabilities in the expression for $\bar{Q}_F$
may be treated by the same technique as given there. By Bayes' theorem
\begin{equation}
p(m\vert j)p(j\vert K_I) = p(m.j\vert K_I) = p(j\vert m.K_I)p(m\vert K_I),
\end{equation}
and we derive immediately, by summation, the result
\begin{equation}
\sum_j p(m\vert j)p(j\vert K_I) = p(m\vert K_I)\sum_j p(j\vert m.K_I).
\end{equation}
Since the states $\{\vert\Psi_j^I(t)\rangle\}$ are complete and orthonormal it follows at
once that $\sum_j p(j\vert m.K_I) = 1$. Consequently, the mean value formula becomes
\begin{equation}
\bar{Q}_F  = \sum_m \langle\varphi_m^F \vert \hat{Q} \vert \varphi_m^F\rangle p(m\vert K_I),
\end{equation}
where
\begin{equation}
p(m\vert K_I) = \sum_j p(m\vert j)p(j\vert K_I)
\end{equation}
may be interpreted as the probability of the occurrence of the final
eigenstate $\vert\varphi_m^F\rangle$ given the probabilities of the initial eigenstates
$\{\vert\Psi_j^I(0)\rangle\}$ and the details of the adiabatic process. Since $\sum_m p(m\vert j)=1$
and $\sum_j p(j\vert K_I)=1$, it is easy to see also that $\sum_m p(m\vert K_I)=1$.

We have now shown how to calculate the final equilibrium probability
distribution over the {\it appropriate\/} energy eigenstates and can consider
the implied information entropy. This is
\begin{equation}
\sigma(F\vert I) = -\sum_m p(m\vert K_I)\ln{(p(m\vert K_I))},
\end{equation}
which in general is {\it not\/} the same as the maximised initial uncertainty given in
terms of initial state probabilities by
\begin{equation}
\sigma_{\rm max}(I) = -\sum_j p(j\vert K_I)\ln{(p(j\vert K_I))}.
\end{equation}
The difference between them arises because of the quantum uncertainty
inherent in the decomposition of the time-developed states $\{\vert\Psi_j^I(t\geq t_F)\rangle\}$
into the final eigenstates $\{\vert\varphi_m^F\rangle\}$ of $\hat{H}_F$. A reasonable guess is that the
initial uncertainty can not be decreased by this unavoidable quantum 
transition effect; but a formal proof is required. In summary, we have
succeeded so far in identifying the correct {\it mechanically calculable\/}
information entropy of the final equilibrium and now wish to show that
\begin{equation}
\sigma(F\vert I) \geq \sigma_{\rm max}(I).
\end{equation}

The proof of this basic result depends on the elementary inequality
\begin{equation}
x - 1 \geq \ln{(x)},
\end{equation}
valid for $x>0$, and with equality only at $x=1$. The simplest way to
see the truth of this relation is to draw the graphs of $y=\ln{(x)}$ and
its tangent $y = x-1$ at $x=1$, but of course an analytic derivation is 
easily constructed. Replacing $x$ by $1/x$ in the above equation we find
quickly a further inequality which, written along with the first, yields
\begin{equation}
x-1 \geq \ln{(x)} \geq 1 - 1/x,
\end{equation}
again with the equality signs holding only for $x=1$. These simple
relations form the basis for many proofs in information theory.

Using the definition of $p(m\vert K_I)$, and the probability normalization
conditions, we can write $\sigma(F\vert I)$ and $\sigma_{\rm max}(I)$ in the forms
\begin{equation}
\sigma(F\vert I) = -\sum_{m,j} p(m\vert j)p(j\vert K_I)\ln{(p(m\vert K_I))}
\end{equation}
and 
\begin{equation}
\sigma_{\rm max}(I) = -\sum_{m,j} p(m\vert j)p(j\vert K_I)\ln{(p(j\vert K_I))},
\end{equation}
so that their difference becomes
\begin{eqnarray}
\sigma(F\vert I) - \sigma_{\rm max}(I) &=& \sum_{m,j} p(m\vert j)p(j\vert K_I)[\ln{(p(j\vert K_I))} - \ln{(p(m\vert K_I))}] \nonumber \\
 &=& \sum_{m,j} p(m\vert j)p(j\vert K_I)\ln{  \left [ {p(j\vert K_I) \over p(m\vert K_I)} \right ] } \nonumber \\
 &\geq & \sum_{m,j} p(m\vert j)p(j\vert K_I)\left [ 1 - {p(m\vert K_I) \over p(j\vert K_I)} \right ] \nonumber \\
 &=& \sum_{m,j} p(m\vert j)p(j\vert K_I) - \sum_{m,j} p(m\vert j)p(m\vert K_I) \nonumber \\
 &=& \sum_j p(j\vert K_I) - \sum_m p(m\vert K_I) = 0, \nonumber \\
 &\Rightarrow& \sigma(F\vert I) \geq \sigma_{\rm max}(I),
\end{eqnarray}
and we have proved the difficult intermediate result. The equalities in
the above reasoning are either obvious mathematical transformations or
follow from previously displayed normalization conditions, while the
fundamental inequality is an instance of $\ln{(x)} \geq 1 - 1/x$. This possible
increase of uncertainty for a system with known evolution is a pure
quantum effect and is usually negligible for macroscopic bodies.

To complete the proof and thus identify the origin of observable entropy
increase we must show that $\sigma_{\rm max}(F) \geq \sigma(F\vert I)$. But now the trivial remark
made earlier applies in its full triviality. The reason for this is that
we have calculated the final mean values in the form
\begin{equation}
\bar{Q}_F  = \sum_m \langle\varphi_m^F \vert \hat{Q} \vert \varphi_m^F\rangle p(m\vert K_I),
\end{equation}
with associated entropy represented by 
\begin{equation}
\sigma(F\vert I) = -\sum_m p(m\vert K_I)\ln{(p(m\vert K_I))},
\end{equation}
and have assumed that these computed $\bar{Q}_F$ values would agree with direct
experimental measurements, as summarised by the data $K_F$ of the final
equilibrium condition. On the other hand, a direct assignment of probabilities
for the occurrence of the eigenstates $\{\vert\varphi_m^F\rangle\}$, based on the
same data values, would be done by maximising the uncertainty expression
$\sigma = -\sum_m q(m)\ln{(q(m))}$ subject to the constraints
\begin{equation}
\bar{Q}_F  = \sum_m \langle\varphi_m^F \vert \hat{Q} \vert \varphi_m^F\rangle q(m).
\end{equation}

By construction, the maximising distribution $\{q(m\vert K_F)\}$ yields a value of
the uncertainty, namely
\begin{equation}
\sigma_{\rm max}(F) = -\sum_m q(m\vert K_F)\ln{(q(m\vert K_F))},
\end{equation}
which can not be less than the information entropy for any other
distribution, such as $\{p(m\vert K_I)\}$, which satisfies the same constraints.
Therefore, we have proved quite generally that
\begin{equation}
\sigma_{\rm max}(F) \geq \sigma(F\vert I) \geq \sigma_{\rm max}(I) ,
\end{equation}
for an adiabatic process connecting initial and final conditions of
complete thermodynamic equilibrium.

This result is so important, and so central to a proper understanding of
the statistical foundations of thermodynamics, that it is worthwhile to
summarise the steps in the argument and the assumptions underlying them.
The really fundamental hypothesis is that solution of the time-dependent
Schr\"odinger equation will, in principle, enable accurate prediction of
the future observables of a system when we know the probabilities of the
various possible initial states. Those probabilities were found by
maximizing the uncertainty function subject to the constraints implied
by the initial Hamiltonian and the values of measured observables. It
was then assumed that the evolution of each initial eigenstate could be
calculated up to and beyond the time $t_F$ at which equilibrium would again
be established. The probabilities for these evolving states remain
constant, but will presumably yield the experimentally observed final
data. The initial information entropy thus also remains constant; in
general, however, it does not correspond to the maximized uncertainty
calculated using only the final Hamiltonian and measured data. The 
latter entropy would be relative to the final eigenstates, so we first
show how to calculate an intermediate entropy greater than the initial
one as a result of quantum transitions to final eigenstates. It then
becomes obvious that this intermediate entropy must be less than the 
maximum uncertainty constructible from the same final data.

\subsection{Composite systems}

\noindent Two separate isolated bodies, each at equilibrium, may be regarded as a
single equilibrated system with internal adiabatic and mechanical constraints,
and it may be verified experimentally that the total energy is
the sum of the energies of the subsystems. The bodies will also have
separately measurable entropies. Now the total entropy $S$ can certainly
be {\it defined\/} by the expression
\begin{equation}
S = S_1 + S_2,
\end{equation}
but it is important to realise that total entropy $S$ can not be given a
useful operational meaning independent of this definition so long as the
bodies remain insulated from each other.

Additivity of thermodynamic entropy becomes an experimental proposition
only when the adiabatic barrier is lifted and the bodies are allowed to
exchange energy by heat transfer. At the same time the internal
mechanical constraints may be wholly or partially removed so that one
body can do macroscopically measurable work on the other. Under these
conditions the energies and volumes of the subsystems will in general
change; but, when {\it mutual\/} equilibrium has been achieved, it is usually
observed that the bodies remain separately in equilibrium if contact
between them is broken. It is thus reasonable to assume that bodies in a 
state of mutual equilibrium possess individually reproducible
properties. The entropy of the compound system can then be calculated,
as above, by adding the entropies of the constituent bodies. This is now
an experimentally verifiable statement, and not just a definition, 
because the entropy of the total system can also be determined by global
measurements and compared with the separately measurable entropies of
the subsystems. The point of this discussion is that the principle of
increase of thermodynamic entropy is used as a tool for {\it predicting\/} the
final equilibrium state of a compound body whose parts can exchange
energy. The eventual final state is taken to be the one of maximum total
entropy as calculated from the sum of the entropies of the parts.

It is definitely implied in this account that we may attribute sharp
intrinsic properties to the subsystems making up the composite body. We
are postulating further that the total energy and volume are themselves
well represented experimentally by the sums
\begin{equation}
\bar{E} = \bar{E}_1 + \bar{E}_2 \ \ \ {\rm and} \ \ \ V = V_1 + V_2,
\end{equation}
which is equivalent to assuming that the coupling interaction between 
the bodies, and the region over which it operates, are very small
compared to the individual energies and volumes, respectively. The
approximation is valid to a high degree of accuracy for macroscopic
bodies interacting by surface contact. Hence we are taking it as an
observed fact that the empirical entropy of a compound system formed
from two bodies in mutual equilibrium is given by
\begin{equation}
S(\bar{E} = \bar{E}_1 + \bar{E}_2, V = V_1 + V_2) = S_1(\bar{E}_1, V_1) + S_2(\bar{E}_2, V_2).
\end{equation}
A similar result holds if the two bodies can exchange matter as well as
energy and volume. In the simplest case both subsystems are made of the
same kind of atoms and the total number of atoms is given by $N=N_1+N_2$.
When the composite system is in a state of thermal, mechanical and
diffusive equilibrium it is found that
\begin{equation}
S(\bar{E}, V, N) = S_1(\bar{E}_1, V_1, N_1) + S_2(\bar{E}_2, V_2, N_2),
\end{equation}
with an obvious generalisation if several kinds of atoms are present.
It is also known empirically that two bodies separately in equilibrium
with a third will be in equilibrium with each other. Thus the basic
addition rule may be extended easily to a collection of more than two
bodies, all in mutual equilibrium.

The above generalisations support the proposition that the additive
property of entropy holds even for a single macroscopic body, thought of
as conceptually divided into a large number of macroscopic subsystems
labelled by $r=1\to R$. All the subsystems are individually in
equilibrium and the body could, in principle, be physically separated
into these parts by the insertion of suitable barriers, assuming that
the operation does not appreciably disturb the equilibrium conditions.
Explicitly, if the total entropy $S=S(\bar{E}, V, N)$ is determined by the global
data, and the entropies of the parts are given by $S_r=S_r(\bar{E}_r, V_r, N_r)$, in
terms of the data for the subsystems, then we can write
\begin{eqnarray}
S &=& \sum_r S_r, \ \ \ \ \ \ \ \bar{E} = \sum_r\bar{E}_r, \nonumber \\
V &=& \sum_r V_r \ \ \ \ {\rm and} \ \ \ N = \sum_r N_r.
\end{eqnarray}
Now in order to compare entropy with the concept of maximal uncertainty,
as applied to physical systems, the crucial point to grasp is that the
thermodynamic entropy $S_r$ of a subsystem is determined by the intrinsic
data $K_r\equiv (\bar{E}_r, V_r, N_r)$ of that subsystem. It does not depend on the data $K_s$,
$s\not= r$, of any other subsystem. The data sets $K_r, r=1\to R$, are not themselves
independent of each other, since for fixed global data $K\equiv (\bar{E}, V, N)$
we have the above sum-relations; but all inferences about the internal
microscopic states of a given subsystem are to be assessed only on the
corresponding data $K_r$, and knowledge of any $K_s$ with $s\not=r$ is irrelevant.
Thus we are free to consider each subsystem separately, with its own
Hamiltonian, energy levels and data, and to construct an expression for
the total uncertainty associated with a collection of such internally
independent sub-problems.

In our physical applications the uncertainty is a function of the probabilities
for the occurrence of possible energy eigenstates at
equilibrium. We shall show generally that the additivity property of the
uncertainty function is a purely mathematical result which is valid
whenever the question at issue can be resolved into a set of independent
statistical problems. This is not quite enough, however, to constitute a
complete analogy with thermodynamic entropy, so we shall also need to
discuss the conditions under which physical subsystems are in mutual
equilibrium, and which ensure that the total uncertainty can be written
as a function of the global summed data. These various aspects of 
additivity will be treated for the simple case of a composite body with
two components, but the ideas are straightforwardly extendable. This
apparently rather intricate treatment of additivity is necessary to
establish firmly the parallels between uncertainty and entropy.

Consider first the uncertainties associated with two independent, 
problems, one having $M$ possibilities $\{A_m\}$, data $K_A$, and uncertainty
maximising probabilities $\{p(A_m\vert K_A)\}$, the other with $N$ propositions $\{B_n\}$,
data $K_B$, and corresponding probabilities $\{q(B_n\vert K_B)\}$. In a mechanical
application the propositions refer to energy eigenfunctions, e.g.,
\begin{equation}
A_m \equiv \ ({\rm The\ eigenstate\ \vert\varphi_m^A\rangle \ of\ Hamiltonian\ \hat{H}_A\ occurs}).
\end{equation}
Assuming that each set of possibilities is internally exhaustive and
mutually exclusive the assigned probabilities must satisfy the normalization
conditions $\sum_m p_m = 1 = \sum_n q_n$, and the individual maximised
uncertainties are given by
\begin{equation}
\sigma_{\rm max}(K_A) = -\sum_m p_m\ln{(p_m)}\ \ {\rm and} \ \ \sigma_{\rm max}(K_B) = -\sum_n q_n\ln{(q_n)}.
\end{equation}
Now construct a compound problem by forming all conjunctions $(A_m.B_n)$
considered on evidence $(K_A.K_B)$. Such compound propositions are also
exhaustive and mutually exclusive if the evidence and inferences in one
sub-problem are irrelevant to inferences in the other. This constitutes
the condition of independence. The joint probabilities are
\begin{equation}
P(A_m.B_n\vert K_A.K_ B) = P_{mn}, \ \ {\rm with} \ \ \sum_{m,n} P_{mn} = 1,
\end{equation}
and the total information entropy takes the form
\begin{equation}
\sigma(K_A.K_B) = -\sum_{m,n} P_{mn}\ln{(P_{mn})}.
\end{equation}
But, by Rule(1) of subsection (3.3) we know that
\begin{equation}
P(A_m.B_n\vert K_A.K_B) = P(A_m\vert B_n.K_A.K_B)P(B_n\vert K_A.K_B),
\end{equation}
which, from the stated condition of independence, reduces to
\begin{equation}
P_{mn} = p(A_m\vert K_A)q(B_n\vert K_B) = p_mq_n.
\end{equation}
It is now easy to see that we may derive the simple result
\begin{eqnarray}
\sigma(K_A.K_B) &=& -\sum_{m,n} p_mq_n\ln{(p_mq_n)} \nonumber \\
&=& -\sum_{m,n} p_mq_n [ \ln{(p_m)} + \ln{(q_n)} ] \nonumber \\
&=& -\sum_m p_m\ln{(p_m)} - \sum_n q_n\ln{(q_n)} \nonumber \\
&=& \sigma_{\rm max}(K_A) + \sigma_{\rm max}(K_B).
\end{eqnarray}
All we have shown so far is that the total uncertainty $\sigma(K_A.K_B)$
associated with two arbitrary bodies, each in a condition of internal
equilibrium, is represented by the sum of the maximised uncertainties
$\sigma_{\rm max}(K_A)$ and $\sigma_{\rm max}(K_B)$ of the separate isolated systems. There is no
implication that $\sigma(K_A.K_B)$ is the same as the maximised uncertainty
$\sigma_{\rm max}(K)$ which would follow from a knowledge of the summed data $K=[\bar{E}, V]$,
where $\bar{E} = \bar{E}_A + \bar{E}_B$ and $V=V_A+V_B$, and indeed this would not be true in general.

The situation is different, however, if the bodies are in thermal and
mechanical contact and have come into {\it mutual\/} equilibrium, but are otherwise
isolated from external influences. The resulting composite body may
be regarded as a single system, in total internal equilibrium, having a 
maximal uncertainty determined by the summed data. By construction, this
maximised uncertainty can not be less than any other uncertainty which
has been calculated from alternative probabilities consistent with the
same summed data. The implication now is that the two subsystems will
take up energies and volumes consistent with the known sums, but also
constrained by the condition of mutual equilibrium. If they were not in
equilibrium when first placed in contact then the small coupling
interaction between them would initiate an adiabatic process inside the 
composite body which would continue until equilibrium {\it was\/} attained. By
the results of the last subsection the final maximised uncertainty can not
be less than the initial uncertainty and will in general be greater.
Since, moreover, we have assumed that the subsystems will still have
independent uncertainties determined by the final equilibrium data $K_A$
and $K_B$, we see that, for consistency, the maximised total uncertainty at
{\it mutual equilibrium\/} must be represented by
\begin{equation}
\sigma_{\rm max}(K) = \sigma_{\rm max}(K_A) + \sigma_{\rm max}(K_B).
\end{equation}

It is clear from the above discussion that the additivity of {\it maximised\/}
information entropy, under the given conditions, is a direct deduction
from the basic dynamical result that maximal uncertainty is not
decreased by an adiabatic process. We have also made essential use of
the qualitative knowledge that the composite body is made from two subsystems
with independently constructible uncertainties. The maximal
uncertainty implied by the summed data is attained when the subsystems
reach mutual equilibrium with definite partitioning of the total energy
and volume. We see immediately that the addition rule for {\it maximised\/} 
uncertainties becomes a powerful tool for {\it predicting\/} the conditions of
mutual equilibrium between bodies.

Suppose, for example, that we know the total energy $\bar{E}=\bar{E}_A+\bar{E}_B$ and total
volume $V=V_A+V_B$ of two isolated bodies, each at equilibrium, and also
their maximised uncertainties, $\sigma_{\rm max}(\bar{E}_A, V_A)$ and $\sigma_{\rm max}(\bar{E}_B, V_B)$, as functions
of the respective data. If the bodies remain in equilibrium when placed
in thermo-mechanical contact then the uncertainty of the resulting 
composite system, namely
\begin{equation}
\sigma(\bar{E}, V) = \sigma_{\rm max}(\bar{E}_A, V_A) + \sigma_{\rm max}(\bar{E}_B, V_B),
\end{equation}
must also be maximal for the summed data. Assuming that $\bar{E}=\bar{E}_A+\bar{E}_B$ and
$V=V_A+V_B$ are fixed at given values it follows at once that conditions for
mutual equilibrium may be expressed by the maximising relations
\begin{equation}
{\partial \sigma_{\rm max}^A \over \partial\bar{E}_A} = {\partial \sigma_{\rm max}^B \over \partial\bar{E}_B} \ \ \ \
{\rm and} \ \ \ \ {\partial \sigma_{\rm max}^A \over \partial V_A} = {\partial \sigma_{\rm max}^B \over \partial V_B}.
\end{equation}
We have here a natural extension of the principle of maximum information
entropy which shows how $\sigma_{\rm max}(K)$ yields definite criteria for equilibrium
between interacting bodies. Since the empirically determined entropy $S$
is used in exactly the same way for prediction of the values of macroscopic
observables of bodies in mutual equilibrium, the analogy between
these theoretical and experimental quantities is greatly strengthened.
The corresponding relations for $S$ are
\begin{equation}
{\partial S_A \over \partial\bar{E}_A} = {\partial S_B \over \partial\bar{E}_B} \ \ \ \ {\rm and}
\ \ \ \ {\partial S_A \over \partial V_A} = {\partial S_B \over \partial V_B},
\end{equation}
which are well known to express, respectively, equality of temperature
and pressure in the two systems. Thus a successful identification of
$\sigma_{\rm max}(K)$ with $S$ opens up the possibility of a statistical interpretation of 
temperature.

\subsection{Meaning of entropy}

\noindent The results of the last few subsections establish the existence of a
theoretical quantity, $\sigma_{\rm max}(\bar{E}, V, N)$, which has some of the characteristic
properties of thermodynamic entropy $S$. It depends explicitly on the data
at equilibrium and on the Hamiltonian of the system, and is a direct
measure of our uncertainty about the exact state of the body. The really
vital assumption in the demonstrations was that calculated mean values
of observables would agree with experimental measurements of the
observables. This postulate provided the necessary link between the
mathematical properties of $\sigma_{\rm max}$ and possible processes in physical
systems. The empirical entropy, $S(\bar{E}, V, N)$, on the other hand, is found,
up to an arbitrary additive constant, entirely from observations of
energy and temperature changes in quasi-static, reversible processes,
and requires no knowledge of the internal structure of the body. Hence,
for logical completeness, we should show that the function $\sigma_{\rm max}(\bar{E}, V, N)$
can be determined in the same way, apart from a corresponding arbitrary
constant. However, it is convenient to defer this clinching step in the
identification of $S$ with $\sigma$ to the next section.

It is, nevertheless, already plausible that the thermodynamic and
statistical functions are closely related and that the theoretical
interpretation gives a definite meaning to the usual remark that entropy
is a measure of disorder in the system. More precisely, the disorder is
not in the system, which is always in some particular quantum state, but  
in our knowledge about the system. The available data are insufficient to
specify the actual state and we are forced to use probabilistic
inference. The uncertainty function then provides a means for estimating
probabilities and its maximised value is a natural measure of our state
of doubt. The uncertainty can only be made smaller by acquiring more
data. The truly surprising thing is that just a little information,
supplemented by efficient guesswork, can lead to accurate prediction of
reproducible observations.

Although a full justification has not yet been given, we now make the
working hypothesis that thermodynamic entropy and maximised uncertainty
are essentially equivalent representations of our lack of knowledge
about the actual state of a body. Writing $\sigma_{\rm max} = \sigma$ for brevity, we assume
\begin{equation}
S = f(\sigma),
\end{equation}
where $f(\sigma)$ must be a monotonic increasing function of $\sigma$ since both $S$
and $\sigma_{\rm max}$ are non-decreasing in adiabatic processes. Also, because $S$
and $\sigma_{\rm max}$ are both additive for interacting but effectively independent systems at
mutual equilibrium, we must have, for two such bodies, that
\begin{equation}
S=S_1 + S_2 \ \ \ \ {\rm and} \ \ \  \sigma = \sigma_1 + \sigma_2,
\end{equation}
which leads immediately to the functional equation
\begin{equation}
f(\sigma_1 + \sigma_2) = f(\sigma_1) + f(\sigma_2),
\end{equation}
to be satisfied by $f(\sigma)$. Differentiating successively with respect to $\sigma_1$,
and $\sigma_2$, we find quickly that 
\begin{equation}
f^{\prime}(\sigma_1 + \sigma_2) = f^{\prime}(\sigma_1) = f^{\prime}(\sigma_2) = k_B,
\end{equation}
where $k_B$ is a positive constant. Integration then gives
\begin{equation}
S = f(\sigma) = k_B\sigma_{\rm max}(\bar{E}, V, N) + S_c,
\end{equation}
with $S_c$ an arbitrary constant. 

This result should be compared with a slightly expanded version of the
formula already given for the expression of thermodynamic entropy in
terms of experimental data, namely
\begin{equation}
S = S_{\rm exp}[(\bar{E}, V, N), (\bar{E}_0, V_0, N_0)] + S_0
\end{equation}
in which $S_0$ is an arbitrary constant.

The quantity $S_{\rm exp}(K, K_0)$, where $K$ and $K_0$ stand for the data sets, is
determined from definitely prescribed experimental operations on the
system, which take it from a reference condition $K_0$ to the final
equilibrium $K$. If we now write our theoretical interpretation of $S$ in
the rearranged form
\begin{equation}
S = k_B[\sigma_{\rm max}(K) - \sigma_{\rm max}(K_0)] + [S_c + k_B\sigma_{\rm max}(K_0)],
\end{equation}
and choose the arbitrary constants $S_0$ and $S_c$ such that
\begin{equation}
S_0 = S_c + k_B\sigma_{\rm max}(K_0),
\end{equation}
then the experimental entropy relative to $K_0$ may be identified as
\begin{equation}
S_{\rm exp}(K, K_0) = k_B[\sigma_{\rm max}(K) - \sigma_{\rm max}(K_0)].
\end{equation}
The one remaining gap in our arguments is that we have not shown that
the differential relations underlying the experimental procedures for
measuring $S_{\rm exp}$  are the same as those describing the changes of $k_B\sigma_{\rm max}$
during such operations. The missing demonstration necessarily involves 
an explanation of the statistical meaning of temperature, together with
an identification of $k_B$ as Boltzmann's constant. When this has been 
achieved the derivation of thermodynamics from statistical mechanics
will be essentially complete. The proof will be given after the 
appropriate probability distributions have been derived.

We conclude this subsection by remarking that the experimental entropy $S_{\rm exp}$
would have an absolute significance if we could measure it relative to a
condition $K_0$ for which the maximised uncertainty is either zero or a
characteristic structural constant for the given system. The uncertainty
could be zero only if the data $K_0$ are such that we know the exact state
of the body, but we have argued that this is generally impossible. An
exception would occur, however, if the measured energy, identified with
the theoretical mean value $\bar{E}_0$, were equal to the ground state energy $E_G$
of the system. The body must then be in its ground state and, if the
state is non-degenerate, the associated uncertainty would vanish. It is
{\it believed\/} that the ground states of all real physical systems are in fact
non-degenerate; but, if the lowest level should happen to comprise $\Omega_0$
degenerate states, as it does in some simplified theoretical models, the
uncertainty would by $\ln{(\Omega_0)}$. This latter value is almost always found to 
be independent of macroscopic mechanical parameters such as the volume. 
Thus, provided the body remains in its ground level, externally imposed
alterations of the system configuration will not change the uncertainty.
The previous probabilistic considerations would then be unnecessary. The
body may be treated as a pure mechanism whose properties are calculable,
in principle, by the standard techniques of quantum theory.

For any other equilibrium condition specified by data $K\equiv (\bar{E}, V, N)$ it now
seems reasonable to represent the thermodynamic entropy as an absolute
quantity by writing
\begin{equation}
S(\bar{E}, V, N) = k_B\sigma_{\rm max}(\bar{E}, V, N),
\end{equation}
with the implication that, as $\bar{E}$ approaches the ground state energy $E_G$
the entropy $S$ tends to zero or to a constant which is independent of the
mechanical parameters. This is the Third Law of Thermodynamics. It will
be shown later that $\bar{E}$ can equal $E_G$ only at the absolute zero of
temperature, which is in practice unattainable, but can be approached
very closely. Hence a good approximation to absolute entropy can often
be determined by starting with the system at a very low temperature. The
value of $S(\bar{E}, V, N)$ found subsequently, from thermal experiments, may then
sometimes be compared with an independent measurement of $k_B\sigma_{\rm max}(\bar{E}, V, N)$
using structural data obtained spectroscopically. Such comparisons can
yield valuable information.

Development of the implications of the Third Law in physics and physical
chemistry and discussions of its experimental status are well beyond our
scope. The whole topic of physics at very low temperatures is extremely
intricate and involves delicate decisions about whether or not the body
being investigated can be considered to be in equilibrium. Some degrees
of freedom may be frozen into metastable states whose evolution towards
equilibrium is hampered by potential energy barriers. The application of
thermodynamics to such systems is problematical. It is sometimes claimed
that the Third Law explains the {\it observed\/} decrease of quantities such as 
thermal heat capacities at the low temperatures reachable by experiment,
but modern analyses require more information on $S(\bar{E})$  as $\bar{E}$ approaches $E_G$.
\vfil
\newpage

\section{Entropy and temperature}
\subsection{Constrained maximization}

\noindent We have now shown, pending an investigation of how uncertainty can be
measured directly, that $\sigma_{\rm max}(\bar{E}, V, N)$ has all the {\it vital\/} properties of the
thermodynamic entropy $S(\bar{E}, V, N)$. Hence, on using the physical principle
that statistical estimates of reproducible observables will agree with
experimental measurements, the two quantities may be related by writing
\begin{equation}
S(\bar{E}, V, N) = k_B\sigma_{\rm max}(\bar{E}, V, N),
\end{equation}
where $k_B$ is a constant depending on the measurement units.

This proposed identification of the theoretical quantity $\sigma_{\rm max}$ and the
phenomenological $S$ greatly simplifies the conceptual development of
statistical mechanics. In particular, the absolute temperature $T$ can be
introduced into the formalism by using the traditional thermodynamical
arguments involving the Second Law. Such treatments show that $1/T$ is an
integrating factor for the differential of heat transfer and is equal to
the partial derivative $\partial S/\partial \bar{E}$. The above formula then indicates that $T$ is
derived from theory by means of the relation $\partial \sigma_{\rm max}/\partial\bar{E} = 1/(k_BT)$. Further
discussion of these matters requires an explanation of how to calculate
the absolute maximum of the information entropy $\sigma = -\sum_j p_j\ln{(p_j)}$, and the
corresponding probability distribution $\{ p_j \}$, in terms of the assumed
data $K = (\bar{E}, V, N)$. The result will be an explicit theoretical expression
for the thermodynamic entropy, built up from a knowledge of the microscopic
constitution of the system and the values of the measured data.

We start on this by considering the general problem of constrained maximisation,
which may be formulated as follows. Given that there are $M$ 
variables $\{ x_j \}, \ j=1 \to M$, it is required to find the absolute maximum
of a function
\begin{equation}
F(x_1, \ldots , x_j, \ldots ,x_M),
\end{equation}
and the corresponding set of maximising values of the variables $\{ x_j \}_{\rm max}$,
when the allowed $x$'s must always satisfy the $n$ constraint equations
\begin{equation}
G_{\alpha}(x_1, \ldots , x_j, \ldots ,x_M) = g_{\alpha},
\end{equation}
with $\alpha = 1\to n,\ n<M$ and the $g_{\alpha}$'s a set of fixed numbers.

Now for a fully satisfactory solution of the problem, as it arises in
statistical physics, we must certainly do two things: \hfil

\begin{itemize}
\item{1)} somehow derive a proposed solution, and
\item{2)} prove that it {\it is\/} the absolute maximum.
\end{itemize}

The second part of the demonstration is omitted in most accounts of the
calculation, but is really necessary for logical completeness since the
usual variational techniques do not ensure that the absolute maximum has
been attained. It appears in our particular problem that the variational
method, using Lagrange Multipliers, does in fact provide an elegant and
correct solution; but it is still of some interest that a relatively
simple proof of this fact can be given which exploits the special form
of the uncertainty function. The idea of this proof will be explained in
the next subsection after the canonical probability distribution of thermal
physics has been derived.

To return to the general problem set out above we proceed by considering
small variations $\delta x_j$ in the $x_j$'s and calculate the associated variation
$\delta F$ of the function $F$. This variation must certainly vanish at any smooth
maximum of $F$ and thus a necessary condition to be satisfied by $F$ at such
a point, for small, but to a large extent, arbitrary $\delta x_j$, is given by
\begin{equation}
\delta F = \sum_j \left ( {\partial F \over \partial x_j } \right ) \delta x_j = 0.
\end{equation}
If there were no subsidiary equations constraining the allowable
variations of the $\{ x_j\}$ then we could argue as follows. Since the $\delta x_j$ may
be chosen independently and arbitrarily, we can set all but one of them,
say $\delta x_k$, equal to zero and hence derive the condition $(\partial F/\partial x_k)\delta x_k = 0$ to
be satisfied for any small value of $\delta x_k$. Clearly this can hold only if
the partial derivative is zero. Extension of the argument to all the $\delta x_j$
then shows that at a smooth maximum of $F$ or, indeed, at any smooth
extreme point of the function, we must have
\begin{equation}
{\partial F \over \partial x_j } = 0, \ \ j=1\to M.
\end{equation}
Investigation of the exact nature of the extrema thus determined usually
requires computation of the second derivatives also, but even when this
is done it may still prove difficult to pick out the {\it absolute\/} maximum.

The argument just given breaks down when the constraint conditions are
imposed since the variations $\delta x_j$ can no longer be taken to be arbitrary,
but must be related by the $n$ equations $(\alpha = 1\to n)$
\begin{equation}
\delta G_{\alpha} = \sum_j \left ( {\partial G_{\alpha} \over \partial x_j } \right ) \delta x_j = 0.
\end{equation}
We could at this point just use these equations to eliminate $n$ of the
$\delta x_j$ from the expression for $\delta F$ and then, by the same type of process as
above, vary the others arbitrarily so as to obtain $(M-n)$ equations
connecting the variables at an extremum. Taking into account the $n$
constraints $G_{\alpha} = g_{\alpha}$ themselves, we would thus have enough equations to
determine the extremising values of the $\{ x_j \}$, at least in principle. It
is, however, mathematically more elegant and, physically, much more convenient
to introduce instead certain auxiliary constants $\lambda_{\alpha}$ known as the 
Lagrange Multipliers, one for each of the constraint equations. Every
such condition is multiplied by a corresponding $\lambda_{\alpha}$ and the results added
together and to the variation $\delta F$ to yield the equation:
\begin{equation}
\delta F + \sum_{\alpha} \lambda_{\alpha}\delta G_{\alpha} = \sum_j \left [ {\partial F\over \partial x_j}
+ \sum_{\alpha} \lambda_{\alpha}\left ( {\partial G_{\alpha}\over \partial x_j } \right ) \right ]\delta x_j = 0.
\end{equation}
Now the so far undetermined constants $\lambda_{\alpha}$ are at our disposal and we are 
free to choose them so that the coefficients 
$(\partial F/ \partial x_j)+ \sum_{\alpha} \lambda_{\alpha} (\partial G_{\alpha}/ \partial x_j)$
of $n$ of the $\delta x_j$ are zero. The remaining $(M-n)$ of the $\delta x_j$ may then be varied independently
and arbitrarily as in the process for unconstrained extremisation given
above. We could, for example, choose them all zero except for $\delta x_k$, say,
so that we arrive at the variational condition
\begin{equation}
\left [ {\partial F\over \partial x_k}
+ \sum_{\alpha} \lambda_{\alpha}\left ( {\partial G_{\alpha}\over \partial x_k } \right ) \right ]\delta x_k = 0,
\end{equation}
which can be satisfied for arbitrary $\delta x_k$ only by having the coefficient
of $\delta x_k$ equal to zero. The result of a systematic application of this new
method is the set of $(M+n)$ equations
\begin{eqnarray}
{\partial F\over \partial x_j}
+ \sum_{\alpha} \lambda_{\alpha}\left ( {\partial G_{\alpha}\over \partial x_j} \right ) &=& 0 , \ \ \ j=1\to M, \\
G_{\alpha}(\{ x_j \}) - g_{\alpha} &=& 0, \ \ \ \alpha = 1\to n,
\end{eqnarray}
which determines the $(M+n)$ unknowns $x_1, \ldots,x_M; \lambda_1 \ldots ,\lambda_n$.

This analytical method is very powerful when applied in statistical
mechanics, as we shall see, but we must also take note of the potential
problems. There will in general be many different solutions to the above
equations other than the one corresponding to the greatest maximum that
we seek; they will give rise to lesser maxima, or to minima, or to complicated
saddle points of the function $F$. The differential technique may
even miss the absolute maximum altogether if it occurs at a point where
$F$  has some cusp-like shape with no well defined derivatives. Thus when a
``solution'' has been generated it is certainly necessary to convince oneself
that it corresponds to the highest value of $F$.

To conclude this subsection we apply the above mentioned method to the simple case of
an isolated body which has $N$ particles in volume $V$ and an energy $\bar{E}$ known
to be  in the interval from $E$ to $E + \Delta E$. As before, we suppose that the
Hamiltonian $\hat{H}(V, N)$ can be constructed and that the eigenstates $\psi_j(V,N)$,
and corresponding energy eigenvalues $E_j(V, N)$, of the body can be found.
We can thus calculate the number $\Omega = \Omega(\bar{E}, V, N)$ of {\it accessible\/} states whose
energies obey the condition $E < E_j < E + \Delta E$ and we wish to make an honest
assessment of the probabilities $\{ p_j \} $ with which these states occur. This
is precisely the problem that we solved by common sense in section 2 and 
by an invariance argument in section 3; it is to be hoped that the third 
method will give the same answer, namely, that the probabilities for the
accessible states are all equal to $1/\Omega$. We also wish to prove, of
course, that this solution corresponds to the absolute maximum of the
uncertainty or information entropy function.

Assuming that the $\Omega$ accessible energy eigenstates are orthogonal and are
labelled by $j=1 \to \Omega$ we require to find the probabilities $\{ p_j \}$ of these
states by maximising the uncertainty function
\begin{equation}
\sigma = -\sum_j p_j\ln{(p_j)},
\end{equation}
subject to the constraint
\begin{equation}
\sum_j p_j = 1,
\end{equation}
which expresses that the accessible states form a set of exhaustive and
mutually exclusive possibilities, i.e., that exactly one of them occurs.

Varying the $\{ p_j \}$, and setting the induced variation of $\sigma$ to zero, gives
\begin{equation}
-\delta\sigma = \sum_j [1 + \ln{(p_j)}]\delta p_j = 0,
\end{equation}
in which the allowed variations are restricted by the condition
\begin{equation}
\sum_j \delta p_j = 0.
\end{equation}
Multiplying the latter variational equation by a constant $\alpha$ and adding
it to the former one now yields the combined equation
\begin{equation}
\sum_j [\alpha + 1 + \ln{(p_j)}]\delta p_j = 0,
\end{equation}
in which the coefficients of the $\{\delta p_j \}$ can, by the arguments above, be
all equated to zero. Hence for all $j$ we find easily that $p_j = \exp{(-\alpha-1)}$,
and application of the probability normalisation constraint on the $\Omega$
possible states then leads to the previous result
\begin{equation}
p_j = {1 \over \Omega}, \ \ j=1\to\Omega.
\end{equation}
We already know from subsection 3.5 that if, consistently with
the constraints, we can make any pair of probabilities more equal, then
the value of the uncertainty function will increase. Since, in the above
problem, the probabilities are {\it derived\/} to be all equal, the absolute
maximum of the $\sigma$-function must have been attained, namely, $\ln{[\Omega(\bar{E}, V, N)]}$.

\subsection{Canonical probabilities}

\noindent 
We are now in a position to derive the probability distribution over
energy eigenstates for a body subjected to the experimental conditions
most commonly encountered in thermal physics. Since these conditions are
overwhelmingly the most usual ones, the corresponding probabilities are
called {\it canonical\/}. The simple final formula for the probability of
occurrence of some particular energy eigenstate may be regarded as the
central result in the standard form of statistical mechanics. Once this
formula has been established, all the preceding framework of probability
and uncertainty theory can essentially be disregarded; work can then be
concentrated on understanding, and modelling, the mechanics of various
individual systems and attempting to evaluate the sums over states that
arise from the canonical treatment. Most modern research and teaching in
statistical physics does in fact take this line, to all intents assuming
the probability formula as an axiom; but much greater understanding, and
confidence in the results, can be achieved by making some effort to
appreciate the purely inferential nature of the fundamental formula. It
is, of course, essential to do this before generalisations of the theory
can safely be undertaken. We will eventually derive a minor extension of
the canonical method, mainly for computational reasons; but the point is
that this is easy to do when the probabilistic background is understood.
It is not necessary to invent a new axiom. The underlying probability 
methods may also be applied to give formalisms appropriate for other
static experimental conditions and even for non-equilibrium situations.

The canonical form of equilibrium statistical mechanics is designed to
describe the properties of a system for which, not the energy, but the
{\it temperature\/} is known. The concept of temperature arises in a natural way
from the reproducible observation of spontaneous energy transfer between
bodies placed in loose energy contact. Such a process occurs in general
even when the two bodies are themselves both in equilibrium initially. 
Systems in thermal contact eventually come to a mutual equilibrium and
then each remains in equilibrium when separated from the other. Hence it
is possible to start with two separately equilibrated systems, allow
them to exchange energy by thermal contact, and finally to separate them
again as two mutually and internally equilibrated bodies. The body which
loses energy in this type of process is said to be {\it hotter\/} than the other
and two bodies which remain in their initial equilibrium states when
placed in thermal contact are said to be equally hot. The primitive
notion is the idea of equal hotness and it is desired experimentally to
construct a reproducible scale for measuring the hotness of bodies so as 
to determine easily if two or more systems will be in mutual equilibrium
when placed in thermal contact. The usual procedure is to choose some
standard body, with a readily measurable, hotness-dependent property, to
act as a thermometer. If the thermometer gives the same readings when in
thermal equilibrium with two separate bodies, then this may be taken to
imply that the two measured systems would also be in mutual equilibrium
if placed in thermal contact. The systems are now said to be at the same
temperature. There are many physical properties suitable as thermometric
indicators for which no ambiguities arise in the assignment of equal
hotness to bodies.

It is also found experimentally that the readings of thermometers may be
used to judge the relative hotness of bodies; which is to say that these
readings may be ordered so that, for example, the values observed for
the temperatures of two different bodies can be correlated unambiguously
with the direction in which net energy transfer occurs when the bodies
are allowed to equilibrate with each other. The development of accurate
thermometers is a highly technical subject which we will not go into. We
wish only to draw attention to the fact that there is a reproducibly
measurable property of systems, the temperature, which controls the net
spontaneous exchange of energy between coupled systems, each initially
in equilibrium. The real nature of this quantity is somewhat obscure in
macroscopic thermodynamics,, but since it is connected with {\it net\/} energy
transfers there are strong indications that it is basically statistical.
It was eventually discovered to be possible to set up a universal or
absolute scale of temperature, denoted by $T$, which is independent of the
properties of particular substances, such that greater hotness always 
corresponds to greater $T$. This will be discussed in more detail in the 
next subsection, but it is worth remarking here that the really important
feature of $T$  is that all thermodynamic relations between measurable
properties of systems are valid only when the absolute scale is used.

Historically and logically in the development of thermal physics the
concept of absolute temperature was bound up inextricably with the idea
of entropy. Our purpose now is to show that application of the formalism
of maximal uncertainty also implies a universal scale for a parameter of
systems which orders the hotness of isolated  bodies, each in a state of
internal equilibrium. That is, we shall show for such systems that there
exists a quantity, calculable in principle from mechanics, corresponding
to absolute temperature. Although for most bodies the parameter has a 
very complicated expression in terms of the mechanical specifications,
its value is the same for two systems which remain in equilibrium when
brought into thermal contact. Further, its values, for two bodies which
do not remain in equilibrium on establishment of thermal contact, enable
the direction of spontaneous energy exchange to be inferred. Thus it has
the operational features of temperature and its importance lies in the
fact that it may be found experimentally without a detailed knowledge of
internal dynamics. The possibility of such direct measurement depends in
an essential way on how uncertainty may be determined experimentally
from macroscopic information. This will be explained later.

We turn now to the derivation of canonical probabilities for occurrence
of energy eigenstates in a body known to be in equilibrium, on the basic
assumption that the system possesses a reproducibly measurable average
energy. A constant mean energy $\bar{E}$ certainly exists for any isolated body,
even if it is not in equilibrium, as we showed in section 2, but is also
observed for an equilibrated body which is thermally coupled to another,
much larger system (or heat bath), itself in equilibrium. In the latter
situation it is found that although repeated measurements of the energy
of the body may show small fluctuations, there is a very sharply defined
mean value which may also be taken as an estimate of the energy when the
body is removed from contact with the heat bath. The important point is
that the body does not show any systematic tendency to exchange energy
with the bath. Such considerations, of course, underlie our preliminary
discussion of the temperature concept and in practice what is really
measured is the temperature of the body; but it is not possible at this
stage to introduce that concept directly. We shall see later, however,
that for any specific body a given average energy $\bar{E}$ implies a definite
temperature and vice versa. Still, the mean energy is a measurable thing
and it is certainly allowable to investigate the form of the eigenstate
probabilities on the basis of given mean system energy $\bar{E}$, together with
specification of the volume $V$, the number of particles $N$ and the system
Hamiltonian. The connection between energy and temperature then emerges
naturally from the formalism and the subsequent  description of how the 
maximised uncertainty can be found by experiment (this gives the usual
thermodynamic entropy) also indicates how the temperature parameter of
the body can be measured and inserted directly into the final formulae.

Thus we now consider a body, either isolated, or in loose energy contact
with a heat bath, to which can be attributed a quite sharply defined
mean energy $\bar{E}$. We do not assume the artificial condition of the microcanonical
method that the energy is definitely known to be in a small
interval from $E$ to $E + \Delta E$ and hence we can no longer claim that there is
only a finite number of possibilities for the exact eigenstate. Instead,
we must now envisage that any energy eigenstate has some probability of
occurrence. This implies that at the end of every particular application
of the new method we should certainly check that the expected deviations
from the assigned mean energy are indeed small, i.e., that the energy of
the system {\it is\/} a sharply defined and therefore reproducible quantity in
accord with experiment.

We assume as before that the Hamiltonian $\hat{H}(V, N)$ of the system is known 
and that the eigenstates $\psi_j(V, N)$ belonging to eigenvalues $E_j(V, N)$ can be
constructed. The index $j$ represents all the quantum numbers required to
distinguish the energy eigenstates from each other, so that states which
are degenerate in energy are also separately labelled. Given the mean
energy $\bar{E}$ of the body at equilibrium, we can now straightforwardly assign
probabilities for the occurrence of the energy eigenstates by maximising 
the uncertainty function
\begin{equation}
\sigma = -\sum_j p_j\ln{(p_j)},
\end{equation}
subject to the constraint equations
\begin{equation}
\sum_j p_j = 1 \ \ \ {\rm and} \ \ \ \sum_j p_jE_j(V, N) = \bar{E},
\end{equation}
which symbolise that the states form a set of exhaustive and mutually
exclusive possibilities and that the average energy is specified. 
Varying the $\{ p_j \}$, and setting the implied variation of $\sigma$ to zero, yields
\begin{equation}
-\delta\sigma = \sum_j [1 + \ln{(p_j)}]\delta p_j = 0,
\end{equation}
to be satisfied under the variational restrictions
\begin{equation}
\sum_j \delta p_j = 0 \ \ \ {\rm and} \ \ \ \sum_j \delta p_j E_j(V, N) = 0.
\end{equation}
Multiplying the two last equations respectively by $\alpha$ and $\beta$ and adding them to the
previous one results in the combined equation
\begin{equation}
\sum_j [\alpha + 1 + \beta E_j(V, N) + \ln{(p_j)}]\delta p_j = 0,
\end{equation}
in which, by previous arguments, the bracketed coefficients of the $\{\delta p_j \}$
can all be equated to zero. Hence we deduce that, for all $j$,
\begin{equation}
p_j = \exp{(-\alpha - 1)}\exp{[-\beta E_j(V, N)]}.
\end{equation}
We see therefore that the probability of occurrence of the state $\psi_j(V, N)$
depends only on the energy eigenvalue $E_j(V, N)$ of that state and on the 
fixed, but so far undetermined, parameters $\alpha$ and $\beta$.

The factor $\exp{(-\alpha - 1)}$ is quickly reducible to a useful form by inserting the last equation
into the normalisation condition, which give us at once that
\begin{equation}
\sum_j p_j = 1 = \exp{(-\alpha - 1)}\sum_j \exp{[-\beta E_j(V, N)]}.
\end{equation}
Equation (224) can thus be rewritten as 
\begin{equation}
p_j = {\exp{[-\beta E_j(V, N)]} \over Z(\beta, V, N) },
\end{equation}
where
\begin{equation}
Z(\beta, V, N) = \sum_j \exp{[-\beta E_j(V, N)]}.
\end{equation}
By this manipulation the probability normalisation condition is made manifest.
The, at first sight, trivial normalisation factor $Z$ above 
actually plays a central r\^ole in all subsequent developments of
the canonical method and is called the {\it partition function\/} or, sometimes,
the {\it sum-over-states\/}. It is a function of $\beta$ and mechanical parameters.

Having eliminated the explicit appearance of $\alpha$ in the formalism, we may
attempt a similar exercise for $\beta$. However, it is not possible to do this
in general; nor is it even desirable, since $\beta$ proves to be important in
its own right. It is a statistical parameter that is closely related to
the mean energy $\bar{E}$, itself an intrinsically statistical quantity, though
determinable, of course, by measurement. The required connection is
easily derived by inserting the probabilities above into the second form of the normalisation condition;
this leads to the sequence of equalities
\begin{eqnarray} \bar{E} &=& \sum_j p_jE_j = {\sum_j \exp{[-\beta E_j(V, N)]}E_j(V, N) \over Z(\beta, V, N)} \nonumber \\
&=& -\left ( {1\over Z} \right ){\partial \sum_j \exp{[-\beta E_j]} \over \partial \beta} = 
-\left ( {1\over Z} \right ){\partial Z\over \partial\beta},
\end{eqnarray}
from which we deduce the crucially important equation
\begin{equation}
\bar{E} = -{\partial \ln{[Z(\beta, V, N)]} \over \partial\beta}.
\end{equation}
Regarded as an equation for determining $\beta$ from assumed information about
$\bar{E}$, the above Eq.(229) does not seem at all satisfactory, since it generally
implies a complicated transcendental relation for $\beta$ as a function of $\bar{E}$.
But in every application it is much more convenient to have $\bar{E} = \bar{E}(\beta)$, as
in Eq.(229), the reason being that $\beta$ is a simple function of the absolute
temperature of the system and is much easier to measure. The theoretical
importance of this equation is to show that knowledge of $\beta$ is equivalent to a
specification of $\bar{E}$ and may be used to replace that assumed input for the
assignment of probabilities. The extremely pleasant outcome is that the
very accurately measurable temperature of a body enters, via its simple and
unique relation to $\beta$, directly into the formulae (226) and (227) for the
canonical probabilities. One of the beautiful and surprising features of
the variational method is that quantities originally introduced just for
mathematical convenience frequently assume direct physical significance.
We will soon show that $\beta$ is in fact proportional to the reciprocal of 
the absolute temperature $T$ so that Eq.(226) can be construed as giving the 
state probabilities implied by knowledge of $T$, rather than of $\bar{E}$, as we
assumed at the start. But before moving on to this we give the simple
proof that, under the given constraints, the variationally derived
probabilities do yield the absolute maximum of the uncertainty function.

We use Eq.(226) for the canonical probabilities and recall the inequality
employed in section 4, namely $\ln{(x)} \leq x-1$, with equality holding only
when $x=1$. Consider now any other probability distribution $\{ q_j\}$ for the
energy eigenstates, which satisfies the same constraint equations, i.e.,
\begin{equation}
\sum_j q_j = 1 \ \ \ {\rm and} \ \ \ \sum_j q_jE_j(V, N) = \bar{E}.
\end{equation}
Thus the $\{ q_j\}$ are compatible with the given data and we have also that
\begin{equation}
\ln{(p_j/q_j)} \leq (p_j/q_j) - 1,
\end{equation}
with equality only if $p_j=q_j$, taking the $\{ p_j\}$ as normed probabilities.

Multiplying this relation by $q_j$ and summing over all eigenstates gives
\begin{equation}
\sum_j q_j\ln{(p_j/q_j)} \leq \sum_j q_j(p_j/q_j - 1),
\end{equation}
i.e.
\begin{equation}
\sum_j q_j\ln{(p_j)} - \sum_j q_j\ln{(q_j)} \leq \sum_j p_j - \sum_j q_j = 0,
\end{equation}
on making use of the probability normalisation conditions. Hence we find
\begin{equation}
\sigma(\{q_j \}) = -\sum_jq_j\ln{(q_j)} \leq -\sum_j q_j\ln{(p_j)},
\end{equation}
with equality only when {\it all\/} $q_j = p_j$.

Now substitute Eq.(226) for the canonical $\{p_j\}$ into Eq.(234) to obtain
\begin{equation}
\sigma(\{q_j \}) \leq -\sum_jq_j[-\beta E_j - \ln{(Z)}],
\end{equation}
which by use of the constraint equations (230) becomes
\begin{equation}
\sigma(\{q_j \}) \leq \beta\bar{E} + \ln{(Z)} = \sigma(\{p_j\})
\end{equation}
the last equality following easily from Eqs.(219) and (226).

The basic point to observe about the relation (236) is that it applies to
any arbitrary set of probabilities $\{q_j\}$ which happens to satisfy the few
constraints imposed by experiment.
The quantity $\beta\bar{E} + \ln{[Z(\beta, V, N)]}$ in Eq.(236) depends only on the given
data $(\bar{E}, V, N)$ since $\beta$ is determined from that same data by Eq.(229). Hence
it remains constant as the probabilities $\{q_j\}$ are varied (consistently
with the data constraint equations). Therefore, for all sets $\{q_j\}$ which
agree with the data, i.e., which satisfy $\sum_jq_j = 1$ and $\sum_jq_jE_j = \bar{E}$, we have
\begin{equation}
\sigma(\{q_j\}) \leq \sigma(\{p_j\}) = \beta\bar{E} + \ln{[Z(\beta, V, N]}
\end{equation} 
We conclude that $\sigma(\{p_j\})$ is the {\it absolute maximum\/} of $\sigma$ under the given
constraints since equality in Eq.(237) is attained only when {\it all\/}
\begin{equation}
q_j = p_j = {\exp{[-\beta E_j(V, N)]} \over Z(\beta, V, N) }.
\end{equation}

To finish this subsection we gather together the basic results. Under the
conditions $\sum_j p_j = 1$ and $\sum_j p_jE_j = \bar{E}$ the uncertainty function $\sigma = -\sum_j p_j\ln{(p_j)}$ is
maximised by the canonical probabilities
\begin{equation}
p_j = {\exp{[-\beta E_j(V, N)]} \over Z(\beta, V, N) },
\end{equation}
in which
\begin{equation}
Z(\beta, V, N) = \sum_j \exp{[-\beta E_j(V, N)]},
\end{equation}
and the Lagrange multiplier $\beta$ us related to $\bar{E}$ by the equation
\begin{equation}
\bar{E} = -{\partial \ln{[Z(\beta, V, N)]} \over \partial \beta }.
\end{equation}
The maximised uncertainty function itself then takes the value
\begin{equation}
\sigma_{\rm max}(\{p_j\}) = \beta\bar{E} + \ln{[Z(\beta, V, N)]}.
\end{equation}
We emphasise that these formulae apply to any system whatever that is
in equilibrium under isothermal (constant $T$) conditions. They constitute in
fact the fundamental machinery of statistical mechanics and in principle
require no further development except to incorporate additional relevant
macroscopic variables for specific systems. In the last subsection of this section
we will extend the  formalism itself in a minor way in order to
ease the computational problems for quantum gases and to treat systems
for which only the mean particle number is specified, but no essentially
new ideas are involved.

\subsection{Meaning of temperature}

\noindent The purpose of this subsection is to explain further the significance of
the parameter $\beta$ which arises naturally in the canonical formalism. We
assume throughout the principle of conservation of energy or, as it is
called in thermal physics, the First Law of Thermodynamics.

We show first that the value of $\beta$ for a given system depends only on the
differences of the energy eigenvalues, and assumed mean energy, from an
arbitrary reference point. This is a necessary step in the demonstration
that $\beta$ has an absolute physical significance since energy is definable 
only up to an additive constant. Thus all the energy eigenvalues $\{E_i\}$ of
a particular body must be referred to a constant energy of freely chosen
value $E_0$, and they may be written as
\begin{equation}
E_j = \epsilon_j + E_0,
\end{equation}
in which only the differences $\{\epsilon_j\}$ are physically relevant.

It follows easily from the normalisation of the eigenstate probabilities
that the specified mean energy $\bar{E}$ of the system also contains the same
arbitrary constant $E_0$ since we have
\begin{equation}
\bar{E} = \sum_j p_jE_j = \sum_j p_j(\epsilon_j + E_0) = \bar{\epsilon} + E_0,
\end{equation}
so that the physically relevant quantity is the mean energy difference
\begin{equation}
\bar{\epsilon} = \sum_j p_j\epsilon_j = \sum_j p_j(E_j - E_0).
\end{equation}
The demonstration that the canonical probabilities $\{p_j\}$ do not depend on
$E_0$ follows simply from the  Eqs.(239) and (240) of the previous subsection. On
substituting into them Eq.(243) above we find easily that the distribution
no longer contains the constant $E_0$ explicitly, but is expressed in terms
of the set of energy differences $\{\epsilon_j\}$. At the same time, the parameter $\beta$
is clearly determined by the relation Eq.(245), just given, which involves $\bar{\epsilon}$.
This result is already a strong indication that the value of $\beta$ can be
given an absolute significance, in contrast to energy values. We will,
however, retain the previous notation for energies on the understanding
that the canonical formalism really depends only on energy differences.
It is often convenient to take the $\{E_j\}$ as the {\it excitation\/} energies of a
system above its ground state and, with this convention, it is not very
difficult to see what various possible extreme values for $\beta$ might imply.

Take the ground state energy to be $E_0=0$ and suppose first that $\beta \to \infty$.
Then, since all excited states of the system have $E_j > 0$, $\exp{(-\beta E_j)} \to 0$;
from which it follows that the corresponding $p_j \to 0$, while states with
the ground energy retain finite equal probabilities. On the other hand,
if we suppose that $\beta \to +0$, i.e., approaches zero from above, then all $p_j$
clearly tend towards equality, though states with lower energies always
have greater probabilities. When the system has no finite upper bound to
its possible energy eigenvalues, the limiting value $\beta = 0$ would not be
sensible, since the state probabilities for finite energy would vanish.
Hence $\beta > 0$ is the usual situation in thermal physics; but it is
sometimes possible to isolate effectively, in a real body, a subsystem
which {\it does\/} have a finite number of energy eigenstates. In this case the
value $\beta = 0$ leads at once to equal non-zero probabilities for all eigenstates
and it  even becomes sensible to consider what happens when $\beta$ is
taken to be negative. Inspection of the expression for $p_j$ now shows that
the states with {\it greater\/} energy have larger probability and that $p_j \to 0$
as $\beta \to -\infty$, excepting only the $p_j$ for states having the maximum allowed
energy eigenvalue, which become equal. To summarise, for all real bodies
in complete thermal equilibrium, the physically possible values of $\beta$ lie
in the range $0 < \beta \leq \infty$ while for subsystems with a maximum energy, which
can also be isolated long enough to come to internal equilibrium, values
of $\beta$ in the range $-\infty \leq \beta \leq 0$ may be possible in addition.
 
Subsystems described by negative $\beta$ are said to be inverted, since their
state probabilities $\{p_j\}$ increase with increasing $E_j$, in sharp contrast
to the situation for all bodies in complete equilibrium. We shall soon
see that subsystems with $\beta \leq 0$ are always hotter than any body having 
positive $\beta$. However, it must be said that these negative temperature
systems are of extremely specialised interest and we will touch on them
only briefly in the remainder of this article. In what follows, therefore,
we will restrict ourselves mainly to consideration of bodies in complete
internal equilibrium among all degrees of freedom and take it that the
possible values of $\beta$ are somewhere in the range $0 < \beta \leq \infty$. We are led to
the adoption of strictly positive $\beta$ in the formalism by the discussion
above and the observation that physical systems have no definite upper
limit to their energy spectra when all their mechanical variables are
taken into account. Attainment of the extreme value $\beta = \infty$ is believed to
be impossible for any real body. That limit would correspond, as we saw,
to the system being brought to its ground state; but this does not seem
to be achievable by macroscopically feasible processes. The matter will
not be discussed further here, except to note that a formal statement of
such effective impossibility is very often taken as an axiom and called
the Third Law of Thermodynamics.

The next step is to apply the canonical method to a system composed of
two bodies, each in complete internal equilibrium and also in a state of
{\it mutual\/} equilibrium with each other. The composite body can be thought of
as either isolated or in equilibrated energy contact with another, much
larger body, which acts as a source or sink for heat transfer. Under the 
stated conditions, the composite system may be ascribed a definite mean
total energy $\bar{E}$, and the component bodies will also possess well defined
average energies $\bar{E}_1$ and $\bar{E}_2$, such that $\bar{E} = \bar{E}_1 + \bar{E}_2$.
We assume that the interaction coupling the two bodies is sufficiently large that energy
equilibrium is established easily, but at the same time is essentially
negligible compared with the individual Hamiltonians. Hence to excellent
approximation the total Hamiltonian $\hat{H}$ may be written in terms of the two
subsystem energy operators as $\hat{H} = \hat{H}_1 + \hat{H}_2$. The Schr\"odinger equation for
the energy eigenstates and eigenvalues of the compound body has the form
\begin{equation}
\hat{H}\vert\psi_i\rangle = (\hat{H}_1 + \hat{H}_2)\vert\psi_i\rangle = E_i\vert\psi_i\rangle,
\end{equation}
and it is easy to see that the eigenstates are products of eigenstates
of the individual bodies while the eigenvalues are sums of corresponding
separate eigen-energies. Thus, on writing the compound state labels $(i)$
as $(i) = (jk)$, where the sets of quantum numbers $(j)$ and $(k)$ refer to
the states of bodies 1 and 2, respectively, we have that
\begin{equation}
\vert\psi_i\rangle = \vert\varphi_j\rangle\vert\chi_k\rangle \ \ {\rm and} \ \ E_i = E_j + E_k,
\end{equation}
in which the states $\vert\varphi_j\rangle$ and $\vert\chi_k\rangle$ of bodies 1 and 2 are solutions of
\begin{equation}
\hat{H}_1\vert\varphi_j\rangle = E_j\vert\varphi_j\rangle\ \ {\rm and} \ \ 
\hat{H}_2\vert\chi_k\rangle = E_k\vert\chi_k\rangle.
\end{equation}
Equilibrium probabilities $\{p_i\}$, for the occurrence of the eigenstates
$\vert\psi_i\rangle$ of the composite body, are assigned by maximising the uncertainty
\begin{equation}
\sigma(\{p_i\}) = -\sum_i p_i\ln{(p_i)},
\end{equation}
subject to the assumed constraints
\begin{equation}
\sum_i p_i = 1 \ \ {\rm and} \ \ \sum_i p_iE_i = \bar{E}.
\end{equation}
By the arguments of the last subsection this process leads to the result
\begin{equation}
p_i = {\exp{(-\beta E_i)} \over Z(\beta) },
\end{equation}
where
\begin{equation}
Z(\beta) = \sum_i \exp{(-\beta E_i)},
\end{equation}
and the parameter $\beta$ is related to the total mean energy value by
\begin{equation}
\bar{E} = -{\partial \ln{(Z(\beta)} \over \partial \beta }.
\end{equation}
The maximised uncertainty itself is then expressed by
\begin{equation}
\sigma_{\rm max} = \beta\bar{E} + \ln{[Z(\beta)]}.
\end{equation}
By replacing each compound eigenvalue $E_i$ by the corresponding sum of the
eigenvalues $E_j$ and $E_k$, as indicated in Eq.(247), we see immediately, from
the elementary properties of the exponential function, that we may write
\begin{equation}
p_i = p_jp_k,
\end{equation}
with
\begin{equation}
p_j = {\exp{(-\beta E_j)} \over Z_1(\beta) } \ \ {\rm and} \ \ p_k = {\exp{(-\beta E_k)} \over Z_2(\beta) },
\end{equation}
where
\begin{equation}
Z_1(\beta) = \sum_j \exp{(-\beta E_j)} \ \ {\rm and} \ \ Z_2(\beta) = \sum_k \exp{(-\beta E_k)},
\end{equation}
and the same value of $\beta$ as above is used in the separated equations.

It is also clear from Eqs.(247), (250) and (255) that $\bar{E}$ may be written as
\begin{equation}
\bar{E} = \bar{E}_1 + \bar{E}_2,
\end{equation}
in which
\begin{equation}
\bar{E}_1 = \sum_j p_jE_j \ \ {\rm and} \ \ \bar{E}_2 = \sum_k p_kE_k.
\end{equation}
Alternatively, these relations may be derived easily from Eq.(253) since
it is obvious that $Z(\beta) = Z_1(\beta)Z_2(\beta)$. Inspection of Eqs.(256), (257) and
(259) now reveals that the expressions for the sets of numbers $\{p_j\}$ and
$\{p_k\}$ are exactly of the form of canonical probabilities for the separate
systems 1 and 2, respectively, and would in fact be numerically equal to
such probabilities if the mean energies of those systems were known to
have the values $\bar{E}_1$ and $\bar{E}_2$. The uncertainties associated with the bodies,
separately considered, would then have the maximum values for that input
information. Thus from Eq.(254) we would have for the composite body that
\begin{eqnarray}
\sigma_{\rm max} &=& \beta(\bar{E}_1 + \bar{E}_2) + \ln{[Z_1(\beta)Z_2(\beta)]} \nonumber \\
&=& \{ \beta\bar{E}_1 + \ln{[Z_1(\beta)]} \} + \{ \beta\bar{E}_2 + \ln{[Z_2(\beta)]} \} \nonumber \\
&=& \sigma_{\rm max}(1) + \sigma_{\rm max}(2).
\end{eqnarray}
We conclude therefore, as a consequence of our uncertainty maximisation
principle, that Eqs.(256)--(259) {\it determine\/} the average energies $\bar{E}_1$ and $\bar{E}_2$
possessed by two bodies, in equilibrated thermal contact, which together
comprise a single compound system of given mean energy $\bar{E}$. It is clear in
addition that each of the components must have the same value for its $\beta$
parameter as is required in the description of the composite body.

Yet another way to derive this equality of $\beta$ values is to consider small
changes in the mean energies $\bar{E}_1$ and $\bar{E}_2$ consistent with the total average
energy $\bar{E} = \bar{E}_1 + \bar{E}_2$ remaining constant. For the total uncertainty to be a
maximum such variations must give
\begin{equation}
{\partial \sigma_{\rm max} \over \partial \bar{E}_1 } = {\partial \sigma_{\rm max}(1) \over \partial \bar{E}_1 }
+ {\partial \sigma_{\rm max}(2) \over \partial \bar{E}_1 } = 0,
\end{equation}
i.e.,
\begin{equation}
{\partial \sigma_{\rm max}(1) \over \partial \bar{E}_1 } = {\partial \sigma_{\rm max}(2) \over \partial \bar{E}_2 }
\end{equation}
But from the general relations Eq.(241) and (242) of the last subsection we get
\begin{eqnarray}
{\partial \sigma_{\rm max} \over \partial \bar{E} } 
&=& { \partial [\beta\bar{E} + \ln{Z(\beta)}] \over \partial\bar{E}} \nonumber \\
&=& \beta + \left [\bar{E} + {\partial \ln{Z(\beta)}\over \partial \beta} \right ]{\partial\beta\over\partial\bar{E} } \nonumber \\
&=& \beta.
\end{eqnarray}
Hence, on applying this to Eq.(260), and using also Eq.(261) and (262), we find that
\begin{equation}
\beta = \beta_1 = \beta_2.
\end{equation}
This shows that the uncertainty maximisation conditions for description
of mutual equilibrium, mentioned at the end of subsection 4.4
are equivalent to setting the parameters $\beta$ of canonical distributions to
be equal. We see, therefore, that $\beta$ has at least one of the operational
characteristics of empirical temperature, namely having equal values for
bodies in mutual equilibrium. For any particular system, the quantity $\beta$
may, in theory, be obtained from a knowledge of the energy excitation
spectrum, the degeneracies of the quantum states and the specified mean
energy. It is moreover, independent of the arbitrarily chosen reference
point with respect to which the energies are reckoned. But eventually we
will need to show that $\beta$ can also be determined empirically without such
detailed microscopic information. The macroscopic measurement of $\beta$ will
not, by itself, enable us to infer the quantal energy spectrum or even
the total energy $\bar{E}$ of the system. Measurement of $\bar{E}$ requires the further
techniques of calorimetry for estimating energy transfers. In practice,
one of the fundamental concerns of thermal physics is the dependence of
energy on temperature, since the observed relationships throw much light
on the internal constitution of bodies.

It now remains to show that we can correlate the values of $\beta$ for two
bodies, each initially in equilibrium, with the direction in which a
spontaneous net energy transfer occurs when the bodies come into thermal
contact. We assume that matters are arranged so that the bodies exchange
energy only by essentially microscopic interactions occurring at the
interface and not by any processes recognisable macroscopically as work.
This corresponds experimentally to heat transfer.

Consider, therefore, two systems, each with a fixed volume and number of
particles, one with mean energy $\bar{E}_1$, and corresponding $\beta_1$, the other with
mean energy $\bar{E}_2$ and appropriate $\beta_2$, such that $\beta_1 < \beta_2$. If they are now placed
in thermal contact, through a wall allowing only spontaneous transfer of
heat energy, then it is a reproducible fact of observation that energy
will indeed be exchanged between them, until thermal equilibrium is once
again established. We assume that the volumes of the bodies are held at
their initial values throughout this in general irreversible process and
that the composite system is also isolated from external influences so
that the total energy stays constant at the initial value of $\bar{E}_1 + \bar{E}_2$. At
the final condition of complete mutual equilibrium both bodies, and the
conjoined system, may all be described by canonical distributions, over
corresponding eigenstates, with the {\it same\/} parameter value $\beta$. The separate
energies of the two bodies will now have assumed some new values $\bar{E}_1^{\prime}$ and
$\bar{E}_2^{\prime}$ satisfying $\bar{E}_1^{\prime} + \bar{E}_2^{\prime} 
= \bar{E}_1 + \bar{E}_2$.
Writing the energy transfer from the first body to the second as $\delta\bar{E}$, 
the final energies are represented by
\begin{equation}
\bar{E}_1^{\prime} = \bar{E}_1 - \delta\bar{E} \ \ {\rm and} \ \ 
\bar{E}_2^{\prime} = \bar{E}_2 + \delta\bar{E}.
\end{equation}
We will now demonstrate that, under the given conditions, the body with
the lower initial value of $\beta$ will lose energy to the other body and that 
the final $\beta$ value at equilibrium will lie between the separate initial
values. That is, if $\beta_1 < \beta_2$ as assumed above, then the energy transfer
$\delta\bar{E} > 0$ and $\beta_1 < \beta < \beta_2$. The simple proof depends only on a consideration
of the rate of change of mean energy $\bar{E}$ with $\beta$ at constant volume. If the
volume and particle number of a system are maintained at fixed values,
then the Hamiltonian, energy eigenstates and eigenvalues are unchanging
and a variation of the mean energy $\bar{E} = \sum_j E_jp_j$ to a new equilibrium value
is described entirely by altering the value of $\beta$ entering the expression
for the canonical probabilities, namely, $p_j  =\exp{(-\beta E_j)}/Z(\beta)$. Therefore,
\begin{eqnarray}
{\partial \bar{E} \over \partial \beta} &=& \sum_j E_j{\partial p_j \over \partial \beta} =
\sum_j E_jp_j{\partial \ln{(p_j)} \over \partial\beta} \nonumber \\
&=& -\sum_j E_jp_j\left [ E_j + {\partial\ln{Z(\beta)}\over \partial\beta} \right ] \nonumber \\
&=& -\sum_j E_j(E_j - \bar{E})p_j,
\end{eqnarray}
in which we have used the relation $\bar{E} = -\partial\ln{Z(\beta)}/\partial\beta$ derived earlier.

It is now easy to see, from the definition of $\bar{E}$ and the normalisation of
the probabilities, that this result may be cast in the more useful form
\begin{equation}
{\partial \bar{E} \over \partial \beta} = -\sum_j(E_j - \bar{E})^2p_j = -(\Delta E)^2
\end{equation}
where it is plain that the $(\Delta E)^2$, the expected square deviation
of the system energy from the mean, is strictly positive for all finite
values of $\beta$. A further simple deduction is that $\partial\bar{E}/\partial\beta$ approaches zero
(from below) only as $\beta$ moves towards infinity, where the only surviving
probabilities are those for the lowest, or highest, levels. We conclude
from Eq.(267) that for all attainable $\beta$ the mean energy $\bar{E}$ at equilibrium
is a strictly decreasing function of $\beta$.

Application of this result to the case of two bodies in thermal contact,
as set out above, is straightforward. The final, common $\beta$ value for the
systems can not be less than the smaller initial value because then both
bodies would have gained energy, which is impossible if energy is to be
conserved overall. Similarly, the final $\beta$ can not be bigger than the
larger of the initial values; hence it must take on a value in the range
$\beta_1 < \beta < \beta_2$. It follows, from the proved decrease of the expected energy
with increasing $\beta$, that the first body loses energy while the second one
gains an equal amount. Thus in Eq.(265) we have that the quantity $\delta\bar{E} > 0$.
We may also remark that nowhere in the argument has it been necessary to
assume that the values of $\beta$ be positive. A corollary of the basic result
is, therefore, that a system with $\beta < 0$ is expected to lose energy when
put into thermal contact with a body having $\beta > 0$. Thus systems assigned
negative temperatures are hotter than all ordinary bodies possessing the
usual positive temperature.

This completes our demonstration that the relative hotness of bodies may
be ordered in terms of their canonical distribution parameters $\beta$. All
the supporting arguments are, of course, really statistical in character
since we have employed throughout the probabilistic notions of {\it expected\/}
values of energy and of mean square deviations from the average. We have
also made a quite extensive use of our theorems on the maximisation of
uncertainty and the physical assumptions underlying them. To finish this
subsection, we discuss how $\beta$ can be found experimentally and related to the
conventional absolute temperature $T$ of thermodynamics.

In order to determine the value of $\beta$ for a body in equilibrium it should
be entirely sufficient to place it in thermal contact with some suitable
subsidiary body which can act as a thermometer. It is assumed of course
that mutual equilibrium will soon be attained, that the presence of the
thermometer does not greatly perturb the original equilibrium state of
the investigated body, and that the substances used for measurement have
some hotness dependent properties whose relation to the common $\beta$ value
can be analysed completely. The possibility of constructing a reliable
{\it empirical\/} temperature scale does not in fact depend on the last proviso.
All that is required for an empirical scale is that the chosen indicator
properties can be reproducibly ordered, so that they can serve as easily
recordable criteria for equal or relative hotness of bodies. There are
many materials and instruments with properties suitable for establishing
such practical measures. Examples include the length of a mercury column
in a glass capillary tube, the electrical resistance of platinum and the
potential difference between junctions in a closed circuit formed by two
different metallic wires. Electrical devices, in particular, can be made
very rapidly acting and with such small thermal capacities that their
perturbing effects are almost negligible.

The drawback of most of the practical scales is that it is troublesome
to calibrate them in terms of the universal standard provided by the $\beta$
parameter of the statistical treatment, or by the equivalent absolute
temperature $T$  of thermodynamics. It is certainly necessary to undertake
this task of calibration before any theoretical understanding of thermal
behaviour can be achieved and, in low temperature physics especially, it
is often forced by the absence of more direct ways for assigning the $\beta$ 
values. Fortunately, the subsequent development of thermodynamics shows
that many properties of substances, with their corresponding empirically
defined temperature scales, can be exploited experimentally to yield
accurate assessment of absolute temperature. But on a rather lower level
than this, there are several practical scales which have been found to
possess a reasonably accurate linear relationship to absolute $T$ values.
These are usually based on the easily reproducible ice and steam points
of water, under one atmosphere of pressure. For a Centigrade scale these
fixed points are called $0^{\circ}$C and $100^{\circ}$C, respectively, and some readily
measurable property of a convenient substance is chosen to interpolate
between them or to extrapolate outside the given range. A simple example
is the well-known mercury in glass thermometer which is of great utility
when high accuracy is not required and only a very limited variation of
temperature needs to be monitored. The standard Centigrade scale can be
set up by using the properties of gases. Under accessible experimental
conditions it is found that all gases lead to exactly the same scale and
it is this which is most nearly a linear function of the thermodynamic
temperature. The absolute zero of temperature, at which all bodies reach
their ground states, is then represented approximately by the Centigrade
value $-273^{\circ}$C. For our purposes it is not necessary to examine minutely
the technical aspects of thermometry, but only to describe in sufficient
detail how the gas properties alluded to above may be used to measure $\beta$
values directly. The procedure allows a useful definition of temperature
requiring only one reproducible fixed point to set the numerical scale.

The main problem, as outlined above, is to establish a temperature scale
which is not tied irrevocably to accidental properties of an arbitrarily
chosen substance or to the quirks of a particular instrument. But at the
same time the measurement method should be relatively easy to implement
over a wide range of conditions. By great food fortune, Nature has given
us a system which neatly fulfils all our requirements. This involves the
concept of the ideal gas temperature, which proves on the one hand to be 
identical to the quite differently defined thermodynamic temperature and
on the other hand can be found to extremely good approximation from data
on real gases. The purely thermodynamic definition of temperature will 
be discussed later; our present interest is to indicate in a preliminary 
way how an analysis of the ideal gas system enables direct access to the
statistical $\beta$ values. The crucial consideration from both experimental
and theoretical viewpoints is that all gases, at low enough density, can
be shown to obey a law which involves macroscopic observables and the $\beta$
value in a particularly simple combination. This is called the ideal gas
law. For some gases, under the same condition of low density, there also
exists a simple relation between $\beta$ and the energy of a given mass of the
gas; but this is of only secondary importance. More thermodynamically
relevant is the observation that the energies of all sufficiently dilute
bodies of gas are nearly independent of volume or pressure, over fairly
large ranges, and are controlled solely by the temperature, though the
functional relationships may be quite complicated.

The possibility of establishing the standard gas scale depends, firstly,
on the existence of reproducible equilibrium states for the systems used
to provide constant temperature environments for a gas and, secondly, on
two experimental observations for real gases which can be extrapolated
confidently to the ideal conditions. The results are summarised in the
statement that, as the pressure of a real gas tends to zero, at constant
temperature, then the internal energy and the product $PV$ of pressure and
volume both become independent of $P$ and $V$. These two empirical facts are
known, respectively, as Joule's Law and Boyle's Law. The second gives at 
once a way of defining an experimental temperature scale by the relation
\begin{equation}
PV = R\Theta,
\end{equation}
in which, for a given mass of gas, $R$ is a numerical constant determined
from the measurement units and the value chosen for $\Theta$ at the fixed point
used as a reference temperature.

Application of the canonical formalism to a dilute 
gas of $N$ molecules in volume $V$ implies the ideal gas equation
\begin{equation}
\bar{P}V = N/\beta,
\end{equation}
where $\bar{P}$ is the statistically estimated pressure, assumed equal to $P$. The
same analysis also leads to Joule's Law, but indicates that the equation
of state holds independently of the internal structure of the molecules.
Comparison with Eq.(268) reveals that the empirical gas scale is related
to $\beta$ values by the simple formula $\Theta = N/R\beta$. In the next subsection we prove
that $1/\beta$ for any system is proportional to the thermodynamic temperature
$T$ of the absolute Kelvin scale, which is defined traditionally by means
of Carnot cycles. We deduce that the empirical gas scales can be read as
identical to the Kelvin scale by appropriate choice of units and a fixed
reference state. This result can also be derived by purely thermodynamic 
methods starting from the perfect gas laws. Thus, anticipating slightly,
we write $\Theta = T$ and set $R/N=k_B$ so that $\beta$ values for bodies may be found
directly from readings of a gas thermometer by using the relations
\begin{equation}
{\bar{P}V\over N} = k_BT = {1\over\beta}.
\end{equation}

\subsection{Measurement of uncertainty}

\noindent Having now seen that the statistical $\beta$ parameter of a system is directly
measurable, we are equipped to understand how the uncertainty function
may be determined by experiment. Such measurements will generally yield 
only the value of this quantity relative to the unknown uncertainty of
some arbitrary reference states, though the difficulty can be overcome to
a large extent if the starting conditions correspond to the body being
near to its ground level. For many purposes, however, it does not matter
that this empirical uncertainty includes an undetermined constant, since
we are often interested only in its {\it variation\/} from state to state. From
that knowledge it is possible to deduce many interrelations between the
temperature and other observable properties of a system at equilibrium.
The procedures required for an exploration of changes in the uncertainty
are in all respects equivalent to those needed for entropy measurements, 
and our demonstration of this fact completes, at last, the long chain of
reasoning by which the thermodynamic entropy becomes identified with the 
statistical uncertainty. The argument is not altogether straightforward.

In what follows we retain the notation $\beta$ for the temperature parameter,
since this is traditional in the statistical theory, but keep always in
mind that it is related to the ideal gas scale temperature by $\beta = 1/(k_B\Theta)$,
where $k_B$ is Boltzmann's constant. Further, it is useful at this stage to 
accept that $\Theta = T$, in which $T$ denotes the absolute or Kelvin temperature
defined independently of particular substances, though we have not yet
proved this equality. For an understanding of our current argument it is
really only necessary to grasp that $\beta$  is accessible from experiments and
that detailed knowledge of microstructure is not presupposed. We assume,
therefore, that a system can be prepared in some reproducible reference
state with definite values of volume $V$ and number of particles $N$, and
with a temperature determined by the reading of a thermometer placed in
close thermal contact. It is not possible to deduce the internal energy 
of the body from this meagre information; the numerical values of energy
are anyway arbitrary unless some fixed zero point has been specified.
Neither can anything be said about the underlying quantum energy levels.

It is clear that our initial knowledge about the body falls far short of
enabling us to calculate the initial value of the maximised uncertainty,
denoted by $\sigma_I$. The canonical formalism does however imply some definite
value for this quantity and also a definite magnitude of the expected or
mean energy $\bar{E}_I$, both depending on the observed temperature parameter $\beta_I$.
All we can hope to learn further about the system from gross macroscopic
manipulations is how the internal energy changes as the body goes from
one equilibrium state to another. We assume that it is also possible to
monitor the corresponding changes in temperature, volume and pressure.

The link between processes in thermal physics and the equilibrium states
central to ordinary thermodynamics lies in the idea of quasi-static and
reversible change. The concept envisages the possibility of making small
modifications in the surroundings of a body, and in the forces acting on
it, so that the system remains effectively in equilibrium at all stages
of the process. This defines the quasi-static nature of the postulated 
change; reversibility just means that an application of exactly the same
small modifications in the opposite direction will return the system to
its original condition. The sorts of change considered include addition
or extraction of energy by spontaneous transfer of heat, using external
reservoirs with temperatures slightly changed from that of the body, and
the performance of work on or by the body, induced by alterations in the 
applied pressure. By such means the system can be brought by slow stages 
from the initial state to a final equilibrium state. The information on
changes of energy, temperature and entropy, gained in this way, is vital
for an understanding of the many processes which can connect equilibrium
states through intermediate conditions arbitrarily far from equilibrium.

The concept of quasi-static and reversible change is, of course, only an
idealisation, and in practice the process must be approximated by finite
steps. We shall denote such a small, but finite, increment in a quantity
$x$ by $\delta x$ and assume that the variations can be extrapolated to a limit in
which it is sensible to construct experimental versions of differentials
and integrals. We also need to know that changes in the internal energy
of a body between two states can be found by direct measurement of work
done adiabatically. This work is not itself required to be quasi-static,
but it is in general possible to perform it in one direction only, since
states adiabatically accessible from a given starting point can not have
a smaller uncertainty. In many practical implementations of quasi-static
processes the energy increments are found just from the recorded changes
in temperature, using previous data on the thermal capacity of the body,
i.e., the value of $\partial\bar{E}/\partial\beta$, measured under appropriate constraints.

After this brief sketch of the experimental background, stripped of most
technical details, we can consider how the maximised uncertainty changes
between neighbouring equilibrium states. We denote the initial value by
$\sigma_I$ and recall from Eq.(242) the maximal uncertainty formula
\begin{equation}
\sigma_{\rm max} = \beta\bar{E} + \ln{[Z(\beta, V, N)]},
\end{equation}
the numerical value of which could, in principle, be calculated from a 
knowledge of the microstructure and the measured $\beta$ since the mean energy
$\bar{E}$ is determined by Eq.(241) i.e.,
\begin{equation}
\bar{E} = -{\partial \ln{[Z(\beta, V, N)]} \over \partial \beta}.
\end{equation}
It is also convenient to recall here the expressions for the partition
function $Z$ and the probabilities $p_j$ (Eqs.(240) and (239)):
\begin{equation}
Z(\beta, V, N) = \sum_j \exp{[-\beta E_j(V, N)]},
\end{equation}
and 
\begin{equation}
p_j = {\exp{[-\beta E_j(V, N)]} \over Z(\beta, V, N) }.
\end{equation}
To simplify the notation we shall from now on drop the suffix (max) from
the uncertainty $\sigma$ and suppress the dependence on $N$, assuming for present
purposes that $N$ is fixed. This restriction is easily lifted if required.
We should further bear in mind that the volume $V$, which enters into the 
definition of the Hamiltonian, is taken only as a representative of many
such parameters that could influence the energy eigenvalues. They define, 
the externally controllable environment of the system and modifying them
corresponds to the performance of macroscopically recognisable work. An
increment $\delta V$ in the volume $V$, for example, results in the energy change
$\delta E_j = (\partial E_j/\partial V)\delta V$, thus defining  the effective pressure as
$P_j = -(\partial E_j/\partial V)$, and the work done {\it on\/} the body is related to the
mean pressure $\bar{P}$ by
\begin{equation}
\delta W = \sum_j p_j\delta E_j = -(\sum_j p_jP_j)\delta V = -\bar{P}\delta V.
\end{equation}
The change in the maximised uncertainty of a system, as it proceeds from 
any equilibrium state to one near by, follows easily from Eq.(271) and, to
first order in small finite quantities, is expressed by
\begin{eqnarray}
\delta\sigma &=& \beta\delta\bar{E} + \bar{E}\delta\beta + \delta(\ln{Z}) \nonumber \\
&=& \beta\delta\bar{E} + \bar{E}\delta\beta + \left ( {\partial\ln{Z} \over \partial\beta} \right)\delta\beta
+ \left ( {\partial\ln{Z} \over \partial V} \right)\delta V \nonumber \\
&=& \beta\delta\bar{E} + \left ( {\partial\ln{Z} \over \partial V} \right)\delta V,
\end{eqnarray}
on making use of the general relation between $\bar{E}$ and $\ln{Z}$ given in Eq.(272).

Now, from Eqs.(273) and (274), we quickly find that
\begin{eqnarray}
{\partial\ln{Z} \over \partial V} &=& \left ({1 \over Z} \right ) {\partial Z \over \partial V} \nonumber \\
&=& \beta\sum_j \left ( - {\partial E_j \over \partial V} \right){\exp{(-\beta E_j)}\over Z} \nonumber \\
&=& \beta \sum_j p_jP_j = \beta\bar{P},
\end{eqnarray}
in which $P_j (=-\partial E_j/\partial V)$ is the pressure exerted by the body when it is in
the energy eigenstate $\psi_j$. Thus from Eqs.(275), (276) and (277) we see that
\begin{equation}
\delta\sigma = \beta [\delta\bar{E} + \bar{P}\delta V] = \beta [\delta\bar{E} - \delta W].
\end{equation}
This important result indicates that the change in maximised uncertainty
between neighbouring equilibrium states can be determined experimentally
by multiplying the increase $\delta\bar{E}$ of the internal energy, minus the work $\delta W$
done on the system, by the temperature parameter $\beta$ of the initial state.
The difference between the observed increment of internal energy and the 
work done on the body by changes in an external variable like $V$ can only
be interpreted as arising from the transfer of some of the energy in the
form of heat, absorbed from or given up to the surroundings. We denote a
small quantity of this transferred heat energy by $\delta q = (\delta\bar{E} - \delta W)$ so that
Eq.(278) may be rewritten more suggestively as
\begin{equation}
\delta\sigma = \beta[\delta q],
\end{equation}
emphasising the close relation of uncertainty and heat exchanges. If now
we replace $\beta$ by $1/(k_BT)$ and define thermodynamic entropy $S$ as $k_B\sigma$, we find
that Eq.(279)  becomes identical in form to the usual analytical expression
for entropy change derived from the Second Law of Thermodynamics, i.e.,
\begin{equation}
\delta S = 
{[\delta\bar{E} - \delta W] \over T} = {\delta q \over T}.
\end{equation}
The entropy difference betwen initial and final states $I$ and $F$ can thus
be measured by summing up the above small contributions $\delta S$ as calculated
from observations of temperature and heat transfer taken during a quasi-static
and reversible process connecting the two states. In the limit of
extremely small steps the result may be written as the integral
\begin{equation}
S(F) - S(I) = \int_I^F {dq \over T}.
\end{equation}
Our derivation of thermodynamics from statistical mechanics is complete.
We have succeeded in showing that thermodynamic entropy and statistical
uncertainty are essentially identical quantities; but we have also shown
that entropy can be determined from macroscopic thermal measurement only
up to an additive constant signifying the entropy of a reference state.

The result in Eq.(281) can be generalised so as to apply to {\it any\/} processes
which begin and end with the body in thermal equilibrium. To do this, we
imagine that the body may be coupled with a reservoir large enough to be
considered as undergoing quasi-static and reversible changes as heat is
transferred to or from the body of interest. When the resultant composite
system is thermally isolated the processes occurring in it are adiabatic
and we have proved generally that the total entropy $S_T$ can not decrease.
Hence, writing $S$ and $S_R$, respectively, for the entropies of the body and 
reservoir, and using the additivity property $S_T = S + S_R$, we deduce that
$S(F) + S_R(F) \geq S(I) + S_R(I)$. The small heat increments $\delta q$ of the body at
each stage are equal in size, but opposite in sign, to those taken up by
the reservoir; thus, from Eq.(281), $S_R(F) - S_R(I) = \int_I^F dq/T$, in which $T$
denotes the varying temperature of the reservoir. The final outcome is
\begin{equation}
S(F) - S(I) \geq \int_I^F {dq \over T},
\end{equation}
which is known as the Clausius inequality.

The equality sign in Eq.(282) holds only when the body itself, as well as the 
reservoir supplying the heat, undergoes quasi-static and reversible
changes and, in that case, the body will have a definable temperature at
all stages, which must always be closely equal to the temperature of the
reservoir when the two systems happen to be in thermal contact. For this
very special case, therefore, the $T$ in Eq.(282)  can be interpreted as the
temperature of the body and we recover the result in Eq.(281).

If a body is taken round an arbitrary closed cycle which finishes in the
initial equilibrium state, then the total entropy change in the body is
zero. When the necessary heat transfers are supplied by a reservoir with
a definite $T$ at any stage we have from Eq.(282) that
\begin{equation}
\oint {dq \over T} \leq 0.
\end{equation}
If, further, the body itself undergoes only quasi-static and reversible
changes throughout the cycle then $T$ is interpretable at all intermediate
stages as the varying temperature of the body and equality holds, i.e.,
\begin{equation}
\oint {dq \over T} = 0.
\end{equation}
A special example of this is called a Carnot cycle, in which the working
substance composing the body is taken round a closed cycle involving two
reservoirs at different, fixed temperatures. We shall now revert to using 
the canonical temperature parameter, $\beta$, since we wish to explain how the 
concept of a Carnot cycle provides an operational definition of absolute
temperature $T$, showing directly how $\beta$ and $T$ are related independently of
the properties of particular substances.

In this type of cycle a body starts from a condition of complete thermal
equilibrium and is made to undergo very specific kinds of quasi-static
and reversible change before regaining its original state. Only one way
round the cycle will be described though the sequence can, of course, be
traversed in the opposite direction. The system is first allowed to take
in energy $q_1$, as heat, from a reservoir with fixed temperature $\beta_1$, i.e.,
{\it isothermally\/}, meanwhile expanding and possibly performing work on its
surroundings. Secondly, the body is thermally isolated and permitted to
expand adiabatically, thus doing further external work, until it reaches
a condition with temperature represented by $\beta_2$, the value of which must
be larger than the magnitude of $\beta_1$. This follows because the performance
of adiabatic work by the body will lead to a fall in the internal energy
$\bar{E}$ and we have already proved that $\partial\bar{E}/\partial\beta < 0$.
In the next stage, the body is compressed isothermally while losing heat energy $q_2$
to a reservoir at constant temperature $\beta_2$. This process continues until a point is
reached from which the final leg of adiabatic compression completes the circuit.

In terms of $\beta$, the integral result of Eq.(284) takes the form
\begin{equation}
\oint\beta dq = 0,
\end{equation}
and, since the heat transfer of energies $q_1$ {\it to\/} the body and $q_2$ {\it from\/} the
body occur only at the respective constant reservoir temperatures $\beta_1$ and
$\beta_2$, we deduce easily that
\begin{equation}
\beta_1q_1 = \beta_2q_2.
\end{equation}
No reference has been made here to the nature of the working substances;
so, for any body executing a Carnot cycle, the ratio of heat supplied at
$\beta_1$ to heat subtracted at $\beta_2$ is determined by the reciprocal ratio of the
$\beta$ values and is independent of the internal structure. It was noticed by
Kelvin that this independence, though derived by him in a different way,
could be used to {\it define\/} a universal temperature scale. His treatment led
him to write $q_1/q_2 = T_1/T_2$ where $T_1$ and $T_2$ denote the standard absolute
temperatures of the two reservoirs. Hence we see from Eq.(286) that
\begin{equation}
{q_1 \over q_2} = {T_1 \over T_2} = {\beta_2 \over \beta_1},
\end{equation}
and it is clear that $\beta = 1/(k_BT)$, where $k_B$ is a constant. This operational
relation for the $T$-value ratios is turned into a numerical scale by defining
the temperature of the triple point of water, at which vapour, water and 
ice are in mutual equilibrium, to have the value 273.16 degrees Kelvin. 
On this scale the ice point is very close to $T=273.15$ K. The modern
Celsius scale $t$, which is almost a centigrade scale, has been defined by 
$t=T-T_0$; the steam point being assigned equal to $t=100^{\circ}$C.

The constant, $k_B$, appearing in the relation
\begin{equation}
\beta = {1\over k_BT},
\end{equation}
could now  be found in principle from measurements with a gas thermometer
using $\bar{P}V=N/\beta = Nk_BT$. In practice, there are better ways of determining
the value of Boltzmann's constant $k_B$, which is given accurately by
\begin{equation}
k_B = 1.380662 \times 10^{-23}\ {\rm Joules\ per\ Kelvin}.
\end{equation}
It is useful at this point to summarise the somewhat intricate arguments
connecting the statistical parameter $\beta$ to the universal temperature $T$ of
thermodynamics. We first indicated that $\beta$ could be measured directly, by
use of a gas thermometer with readings extrapolated to ideal conditions.
This step required a result from the statistical theory of an ideal gas,
which we have not yet proved. The relation is $\beta = 1/(k_B\Theta)$, where $\Theta$ denotes
the empirical gas scale temperature. It was then proved that $\beta \propto 1/T$, in
which $T$ is the Kelvin temperature defined thermodynamically. It follows,
therefore, that the perfect gas temperature $\Theta$ can be chosen identical to 
the Kelvin parameter $T$, but the route to this equality has gone through
the intermediary of the statistical theory. In order to confirm that the
whole scheme is consistent, and to close the conceptual cycle, we should
also show that $\Theta$ can be chosen identical to $T$ as a direct consequence of
the analytical formulation of the Second Law in Eq.(280). This final step
will also free the possibility of direct measurement of $\beta$, for all types
of system, from any dependence on a theoretical model of the ideal gas.

To accomplish this aim we consider the infinitesimal version of Eq.(280),
namely $dS = [d\bar{E} - dW]/T$, and use the definition of $dW$ in Eq.(275) to write
\begin{equation}
TdS = d\bar{E} + \bar{P}dV,
\end{equation}
which is a general differential relation between functions of state that
applies to any body whatever. It follows that
\begin{equation}
\left ( {\partial\bar{E} \over \partial S} \right )_V = T \ \ {\rm and} \ \ 
\left ( {\partial\bar{E} \over \partial V} \right )_S = -\bar{P},
\end{equation}
and, from the irrelevance of order in second derivatives, we find that
\begin{equation}
\left ( {\partial^2\bar{E} \over \partial V \partial S} \right )  = 
\left ( {\partial T \over \partial V} \right )_S =
-\left ( {\partial\bar{P} \over \partial S} \right )_V,
\end{equation}
a well-known Maxwell relation. Applying now the mathematical identity
\begin{equation}
\left ( {\partial x \over \partial y} \right )_z \left ( {\partial y \over \partial z} \right )_x
\left ( {\partial z \over \partial x} \right )_y = -1,
\end{equation}
valid when $z=z(x,y)$, the second equality in Eq.(292) takes the form
\begin{equation}
\left ( {\partial T \over \partial S} \right )_V \left ( {\partial S \over \partial V} \right )_T =
\left ( {\partial\bar{P} \over \partial S} \right )_V,
\end{equation}
and, on recalling the further differential identity symbolised by
\begin{equation}
{\left ( {\partial u \over \partial z} \right )_y  \over \left ( {\partial x \over \partial z} \right )_y } =
\left ( {\partial u \over \partial x} \right )_y,
\end{equation}
which holds for functions $u=u(x,y)$ and $z=z(x,y)$ of two variables, we
obtain from Eq.(294) another thermodynamic Maxwell relation, expressed by
\begin{equation}
\left ( {\partial S \over \partial V} \right )_T = \left ( {\partial\bar{P} \over \partial T} \right )_V.
\end{equation}
The possibility of making the gas scale identical to the Kelvin scale is
a consequence of applying this result to the empirically determined gas
laws of Joule and Boyle concerning, respectively, the internal energy of
a quantity of gas and the product of pressure and volume. We state these
laws again here for convenience: as the pressure of a gas tends to zero 
at constant temperature then both the energy $\bar{E}$ and the product $\bar{P}V$ become
independent of $\bar{P}$ and $V$.

From Joule's law we know that at low pressures $(\partial\bar{E}/\partial V)_T = 0$ and we deduce
immediately from Eqs.(290) and (296) that
\begin{eqnarray}
\left ( {\partial\bar{E} \over \partial V} \right )_T &=& T\left ( {\partial S \over \partial V} \right )_T - \bar{P} \nonumber \\
&=& T\left ( {\partial\bar{P} \over \partial T} \right )_V - \bar{P} = 0,
\end{eqnarray}
i.e.,
\begin{equation}
\left ( {\partial\bar{P} \over \partial T} \right )_V = {\bar{P} \over T },
\end{equation}
implying that
\begin{equation}
\bar{P} = Tf(V),
\end{equation}
where $f(V)$ is some so far undetermined function of $V$. But from Boyle's
law we also know that, at low pressure, $\bar{P}V=$ constant for fixed $T$, which
leads at once to $f(V) \propto 1/V$. Thus we find $\bar{P}V \propto T$. Since the gas scale is
defined by $\bar{P}V = R\Theta$, $R$ being constant for a given mass of gas, it follows
that $\Theta \propto T$ and they may be made identical by choosing the same numerical
values for them at a reproducible reference state, e.g., the triple point
of water. We have now shown, by way of two empirical laws and the purely
macroscopic concept of a Carnot cycle, that $\beta(=1/k_BT)$ may be found
experimentally from the readings of a dilute gas thermometer satisfying
\begin{equation}
\bar{P}V = Nk_BT.
\end{equation}
The former step involving a model calculation for a perfect gas has thus
been avoided, though it remains as a useful and entirely correct device
for developing the theory. Its main drawback is that fairly complicated
calculations are required to arrive at the simple formula $\bar{P}V = N/\beta$ given
earlier. The advantage of our alternative derivation is that no specific
assumptions about the microscopic constitution of the gas are invoked.

However, from the point of view of statistical mechanics, there exists a
serious potential difficulty in the idea of a Carnot cycle. This problem 
has to do with how a quasi-static and reversible adiabatic change can be
described in the statistical theory. Such processes are required for two
stages of the Carnot cycle; but the difficulty arises whenever this type
of adiabatic operation on a system needs to be considered. According to
the account given earlier, the entropy of a body changes only when heat
transfers are involved, and the formalism seems to predict unambiguously
that slow adiabatic alterations conserve entropy. As can be seen clearly
from Eq.(290), if the internal energy increment of the system is due only
to work done on it, so that $d\bar{E} = -\bar{P}dV$, then it appears to be inescapable
that the change of entropy $dS$ will be zero. This is in fact correct, but is
nevertheless in apparent conflict with the general statistical proof,
to be given later, that almost all systems will {\it gain entropy\/} as a result
of {\it any\/} kind of adiabatic process, not even excluding the special type we
have called quasi-static and reversible. Failure to resolve this seeming
contradiction would undermine our attempts to derive thermodynamics
from statistical mechanics. We shall discuss the matter again in  subsection 6.3
and prove that no paradox emerges from either discipline.
 
\subsection{Grand canonical method}

\noindent We shall now develop a simple extension of the statistical formalism in
which we consider the number of particles in the system as an estimated
quantity rather than as being definitely known. This modification is not
of the same degree of theoretical importance as the earlier transition
from actual energy to estimated energy, which led to the introduction of
temperature into the probabilistic description of bodies, but is nevertheless
a useful calculational device for many systems and has a direct
significance for bodies in {\it diffusive\/} equilibrium with each other.

One possible way to interpret the new formalism is to admit frankly that
measurements of particle number yield only a statistical estimate, most
usually taken to be the mean number $\bar{N}$. We recognise that the true number
$N$ may depart from $\bar{N}$, but require that the deviation be small compared to
$\bar{N}$ if the experimental results are to be reproducible. Alternatively, we
may assume that a body could in principle, or even in fact, be placed in
diffusive contact with a much larger system (a particle bath) with which
it can exchange particles. If the two bodies were in mutual equilibrium,
with no {\it net\/} transfer of energy or particles, then the system of interest
can often be taken to have a sharply defined mean particle number, $\bar{N}$, as
well as a definite mean energy $\bar{E}$. In either interpretation we treat the
measurable particle number as having a knowable average value $\bar{N}$.

We must now revise the probabilities for occurrence of stationary states
of the system to include a dependence on the actual particle number, $N$,
since specification only of $\bar{N}$ leaves open the possibility of deviations
and we need a rational way of taking this into account. The final result
contains a fresh statistical parameter $\mu$, called the chemical potential,
bearing a somewhat similar relation to $\bar{N}$ as $\beta$ does to $\bar{E}$. Another analogy
to $\beta$ arises from the fact that two bodies with equal $\mu$ retain their mean
particle numbers when in diffusive contact. Hence, just as equality of $\beta$
implies no net transfer of energy between thermally coupled bodies, $\mu$ is 
a sort of diffusive temperature signalling the absence of a net particle
transfer between two bodies having the same value of this parameter.

We assume as usual that the system Hamiltonian $\hat{H}(V, N)$ is known for any $N$
and that we can construct the orthonormal energy eigenfunctions $\psi_j(V, N)$
belonging to the eigenvalues $E_j(V,N)$. The basic problem is to assign a value
for the probability $p_{N,j}$ that the body has exactly $N$ particles and that,
given $N$, the eigenstate $\psi_j(V, N)$ occurs at equilibrium. The probabilities
$\{p_{N,j}\}$ then refer to exhaustive and mutually exclusive possibilities and
must therefore sum to unity. Given in addition that the mean energy and
mean particle number are specified, the least committal assignment of
probabilities is achieved by maximising the uncertainty function
\begin{equation}
\sigma = -\sum_{N,j} p_{N,j}\ln{(p_{N,j})},
\end{equation}
subject to the constraints
\begin{equation}
\sum_{N,j} p_{N,j}=1, \ \ \sum_{N,j} p_{N,j}E_j(V,N)=\bar{E} \ \ {\rm and} \ \
\sum_{N,j} p_{N,j}N = \bar{N}.
\end{equation}
We proceed by varying the probabilities $\{p_{N,j}\}$ and equating the induced
variation of $\sigma$ to zero. This gives
\begin{equation}
-\delta\sigma = \sum_{N,j}[1 + \ln{(p_{N,j})}]\delta p_{N,j} = 0,
\end{equation}
in which the allowed increments $\{\delta p_{N,j}\}$ are constrained by
\begin{equation}
\sum_{N,j} \delta p_{N,j}=0, \ \ \sum_{N,j} \delta p_{N,j}E_j(N,V)=0 \ \ {\rm and} \ \
\sum_{N,j} \delta p_{N,j}N = 0.
\end{equation}
Multiplying these restriction equations, respectively, by $\alpha,\ \beta$ and $\gamma$ and
adding them to Eq.(303) then gives
\begin{equation}
\sum_{N,j}[\alpha + 1 +\beta E_j(V,N) + \gamma N + \ln{(p_{N,j})}]\delta p_{N,j} = 0,
\end{equation}
which, by the now familiar argument, may be satisfied by setting all the
bracketed coefficients of the $\delta p_{N,j}$ to zero. Thus, for all $(N,j)$,
\begin{equation}
p_{N,j} = \exp{(-\alpha-1)}\exp{(-\beta E_j(V,N)-\gamma N)}.
\end{equation}
The factor $\exp{(-\alpha-1)}$ is easily eliminated by use of the first constraint
of Eq.(302). The result for the maximising probabilities then appears in
the clearly normalised form
\begin{equation}
p_{N,j} = {\exp{(-\beta E_j(V,N)-\gamma N)} \over \Xi(\beta, \gamma, V) },
\end{equation}
where
\begin{equation}
\Xi(\beta, \gamma, V) = \sum_{N,j} \exp{(-\beta E_j(V,N)-\gamma N)}.
\end{equation}
The other constraints on the $\{p_{N,j}\}$  lead to implicit equations for $\beta$ and
$\gamma$ in terms of the presumed data $\bar{E}$ and $\bar{N}$. Substitution of Eqs.(307) and (308)
into the second and third of Eqs.(302), followed by obvious manipulations,
soon shows that
\begin{equation}
\bar{E} = -{\partial\ln{[\Xi(\beta, \gamma, V)]} \over \partial\beta }
\end{equation}
and
\begin{equation}
\bar{N} = -{\partial\ln{[\Xi(\beta, \gamma, V)]} \over \partial\gamma }
\end{equation}
We also record, for completeness, that the mean pressure is given by
\begin{equation}
\beta\bar{P} = {\partial\ln{[\Xi(\beta, \gamma, V)]} \over \partial V }.
\end{equation}
This is derived using the result that $P_j = -\partial E_j(V,N)/\partial V$ in state $\psi_j(V,N)$.

The value of the maximised uncertainty is found by inserting Eq.(307) into
Eq.(301) and making use of Eq.(302) once more. This gives
\begin{eqnarray}
\sigma_{\rm max} &=& -\sum_{N,j} p_{N,j} \ln{(p_ {N,j})} \nonumber \\
&=& -\sum_{N,j}p_{N,j} \{ -\beta E_j(V,N) - \gamma N - \ln{[\Xi(\beta, \gamma, V)]} \} \nonumber \\
&=& \beta\bar{E} +\gamma\bar{N} + \ln{[\Xi(\beta, \gamma, V)]}.
\end{eqnarray}
The proof that this is the {\it absolute\/} maximum of the uncertainty function,
under the given constraints, follows the same
lines as given for the canonical uncertainty. We still need, however, to
establish the physical significance of the grand canonical parameters $\beta$
and $\gamma$. This is most easily done by a minor extension of the arguments of
the last two subsections. As anticipated by the notation, the new $\beta$ is just
the same directly accessible temperature parameter that we found before.
The extra statistical parameter $\gamma$ determines, by its values, whether any
particle transfer occurs between diffusively  coupled bodies. We start by
considering how $\sigma_{\rm max}$ changes as the input data $\bar{E}$ and $\bar{N}$ are varied.

On differentiating Eq.(312) with respect to $\bar{E}$, we obtain
\begin{equation}
\left ( {\partial\sigma_{\rm max} \over \partial \bar{E} } \right )_{\bar{N}, V} = \beta 
+\bar{E}\left ( {\partial\beta \over \partial \bar{E} } \right ) 
+\bar{N}\left ( {\partial\gamma \over \partial \bar{E} } \right ) 
+\left ( {\partial\ln{\Xi} \over \partial \beta } \right ) \left ( {\partial\beta \over \partial \bar{E} } \right ) 
+\left ( {\partial\ln{\Xi} \over \partial \gamma } \right ) \left ( {\partial\gamma \over \partial \bar{E} } \right ),
\end{equation}
which, on using Eqs.(309) and (310), becomes
\begin{equation}
\left ( {\partial\sigma_{\rm max} \over \partial \bar{E} } \right )_{\bar{N},V} = \beta.
\end{equation}
By a similar process we also derive easily that
\begin{equation}
\left ( {\partial\sigma_{\rm max} \over \partial \bar{N} } \right )_{\bar{E},V} = \gamma.
\end{equation}
Now apply these results to two bodies, labelled 1 and 2, each separately
in equilibrium, which remain equilibrated if placed in close contact and
permitted to exchange both energy and particles. If the composite system
is assumed to be isolated then the total energy $\bar{E}$ and particle number $\bar{N}$
are constant and we have, from
\begin{equation}
\bar{E} = \bar{E}_1 + \bar{E}_2 \ \ {\rm and} \ \ \bar{N} = \bar{N}_1 + \bar{N}_2,
\end{equation}
that changes in the respective quantities must satisfy
\begin{equation}
\delta\bar{E}_1 + \delta\bar{E}_2 = 0 \ \ {\rm and} \ \ \delta\bar{N}_1 + \delta\bar{N}_2 = 0.
\end{equation}
The total uncertainty at complete equilibrium should be given by
\begin{equation}
\sigma_{\rm max} = \sigma_{\rm max}(1) + \sigma_{\rm max}(2),
\end{equation}
but this can only be a maximised uncertainty if small variations of the
energies and particle numbers, consistent with Eqs.(317), imply no change
in $\sigma$ to first order of small quantities. Hence, using Eqs.(317), we find
\begin{eqnarray}
\left ( {\partial\sigma_{\rm max} \over \partial \bar{E}_1 } \right ) &=&
\left ( {\partial\sigma_{\rm max}(1) \over \partial \bar{E}_1 } \right ) + 
\left ( {\partial\sigma_{\rm max}(2) \over \partial \bar{E}_1 } \right ) \nonumber \\
&=& \left ( {\partial\sigma_{\rm max}(1) \over \partial \bar{E}_1 } \right ) -
\left ( {\partial\sigma_{\rm max}(2) \over \partial \bar{E}_2 } \right ) = 0.
\end{eqnarray}
From this last equality, and the similar one involving $\bar{N}_1$ and $\bar{N}_2$, we see
that Eqs.(314) and (315) lead to the equilibrium relations
\begin{equation}
\beta_1 = \beta_2 \ \ {\rm and} \ \ \gamma_1 = \gamma_2,
\end{equation}
Thus the equality of $\beta$-values implies there is no net transfer of energy
between the two bodies, while $\gamma_1 = \gamma_2$ signals the absence of systematic
particle transfers. If the composite body is confined to a fixed volume,
as well as being otherwise isolated, and the subsystems are permitted to
do work on each other, then a consideration of small changes in $V_1$ and
$V_2$, taken in conjunction with Eq.(311), proves that the state of complete
equilibrium will be attained only when $\bar{P}_1 = \bar{P}_2$ also.

Here, we will not consider any further the physics of particle transfer
between bodies which have different initial values of $\gamma$. We shall regard
the grand canonical method merely as a very useful tool to be applied as
necessary to bodies with definite particle number $N$. In general, we can
only estimate the true number, but it is often quite sensible to replace
this imperfectly known quantity by an average value $\bar{N}$, or the equivalent
statistical parameter $\gamma$. In any particular application of the method, of 
course, we will need to check that the expected deviations from the mean
are small compared with $\bar{N}$. We will also occasionally want to investigate
the effect on the properties of a {\it single\/} system of changes in the number
of particles estimated to be in it.

Subsequent steps in the identification of $\beta$ exactly parallel the earlier
arguments given in connection with the canonical method and it would be
tedious to repeat them. The result is of course that $\beta = 1/(k_BT)$, as before.
The extra parameter is usually written in the form $\gamma = -\mu\beta$ and, with this
replacement, the probability of finding $N$ particles in state $\psi_j(V,N)$ is
\begin{equation}
p_{N,j} = {\exp{[-\beta(E_j(V,N) - \mu N)]} \over \Xi(\beta, \mu, V) },
\end{equation}
where
\begin{equation}
\Xi(\beta, \mu, V) = \sum_{N,j} \exp{[-\beta(E_j(V,N) - \mu N)]} .
\end{equation}
The corresponding maximised uncertainty then takes the form
\begin{equation}
\sigma_{\rm max} = \beta(\bar{E} - \mu\bar{N}) + \ln{[\Xi(\beta, \mu, V)]}.
\end{equation}
\vfil
\newpage

\section{Temperature and statistical physics}
\subsection{Thermodynamic potentials}

\noindent Now that the machinery of statistical mechanics has been assembled, and
its parallels with thermodynamics firmly established, it is useful to
summarise and extend the relationships between the two theories. In this
subsection, therefore, we will write down and discuss the equations which
connect statistically calculable quantities to some of the most useful
functions of thermodynamics. The functions alluded to are often called
potentials since many of the physically significant properties of bodies
can be found from them by differentiation.

As stated before, a pleasing feature of statistical mechanics is that
parameters introduced for mathematical convenience turn out to have
direct experimental meaning in thermal physics. In particular, the
quantities arising in the three different approaches we have developed,
namely, the microcanonical, canonical and grand canonical methods, are
expressed automatically in terms of variables which are natural ones for
the associated thermodynamic functions. A potential function is said to
be expressed in its natural variables when these are such as to enable
the maximum amount of thermodynamic information to  be extracted from it.
Knowledge of any one of the naturally expressed functions we discuss is
then sufficient for construction of the others. In what follows we shall
present equations relating the macroscopic and experimentally accessible
functions of state to quantities derived  by probabilistic methods from
the known microscopic structure and given equilibrium data. We will drop
arbitrary additive constants associated with the experimental starting
configurations since our central concern will be to study differential
properties of potentials. The initial state constants become important 
only when absolute values of the statistically calculated entropy are in
question and not just its variation from state to state. This involves
delicate, and still controversial, discussion of the status of the Third
Law and its implications in low temperature physics, an extensive topic
which is largely beyond our scope.

For our first example of the physical interpretation of the statistical
theory we recall the microcanonical probability method, as applied to a
simple homogeneous body which is adiabatically isolated and observed to
be in thermal equilibrium. The  body is taken to have volume $V$, particle
number $N$ and an estimated energy $\bar{E}$ whose value is assumed to be confined
within a prescribed small energy interval, i.e., such that $E \leq \bar{E} \leq E + \Delta E$.
For fixed $\Delta E$ the number of accessible energy eigenstates is a calculable
function, $\Omega(\bar{E}, V, N)$, of the given data, provided that the Hamiltonian is
known and that the stationary state Schr\"odinger equation can be solved.
We showed in subsection 5.1 that under these conditions
the maximised uncertainty is given by $\sigma_{\rm max} = \ln{[\Omega(\bar{E}, V, N)]}$ and we further
made a case for the proposition that the absolute thermodynamic entropy
is numerically equal to $k_B\sigma_{\rm max}$, where $k_B$ is Boltzmann's constant. Hence
we find that the macroscopic function called entropy appears in the form
\begin{equation}
S(\bar{E}, V, N) = k_B\ln{[\Omega(\bar{E}, V, N)]}.
\end{equation}
Now the fundamental equation governing small changes, whether reversible
or not, in the equilibrium state of a thermodynamic system is written as
\begin{equation}
\delta\bar{E} = T\delta S -\bar{P}\delta V + \mu\delta N,
\end{equation}
where $\delta\bar{E}$ is the increase in the internal energy and, in the reversible
case, $T\delta S$ denotes the heat added, $-\bar{P}\delta V$ is the work done {\it on\/} the body and
$\mu\delta N$ indicates the energy increment brought about by a small increase in 
the number of particles. The last term generalises the equation we gave 
before so that bodies of varying $N$ can be considered. The temperature $T$,
the pressure $\bar{P}$ and the chemical potential $\mu$ are variables which have the
same values throughout a body in complete equilibrium and are called
{\it intensive\/}, as is also the particle number density of the system. In
contrast, the energy $\bar{E}$, volume $V$ and entropy $S$ are additive over
subdivisions of the body and can be assumed proportional to the size of the
body as measured by the value of $N$. Such quantities are called {\it extensive\/}
variables. We observe in passing that we have here made a tacit choice
of the arbitrary zero of energy as that of an empty cavity which would
exactly enclose the given system.

A simple rearrangement of Eq.(325) gives
\begin{equation}
\delta S = {\delta\bar{E}\over T} + {\bar{P}\delta V\over T} - {\mu\delta N\over T},
\end{equation}
and hence we may calculate the intensive parameters from $S$ as follows,
\begin{equation}
{\partial S\over \partial\bar{E} } = {1\over T}, \ \ {\partial S\over \partial V } = {\bar{P}\over T} \ \
{\rm and} \ \ {\partial S\over \partial N } = {-\mu\over T}.
\end{equation}
Thus we see that the theoretical formula for $S$ provided by statistical
mechanics, displayed in Eq.(324), is already expressed in just the right, 
or natural, variables for deriving other quantities of interest.

The intensive variables $T$, $\bar{P}$ and $\mu$ were introduced in our development of
the statistical theory, but it is useful at this point to discuss their
significance again from the viewpoint of pure thermodynamics. Although
derived here in connection with the description of an isolated system at
equilibrium, their full physical import emerges only when we ask whether
two such bodies will remain in equilibrium when they are placed in loose
contact and are free to exchange energy, volume and particles. Envisage,
then, that the two bodies, labelled 1 and 2, are brought together inside
a rigid box with adiabatic and impermeable walls which exactly enclose
them, so that they form an isolated composite supersystem. If they are
not initially in equilibrium with each other, then by the Second Law of
Thermodynamics their individual energies, volumes and particle numbers
will change until a mutual equilibrium is reached, at which the total
entropy will be a maximum and will be the sum of the separate entropies,
i.e.,
\begin{equation}
S(\bar{E}, V, N) = S_1(\bar{E}_1, V_1, N_1) + S_2(\bar{E}_2, V_2, N_2).
\end{equation}
The given condition of isolation also implies the conservation equations
\begin{equation}
\bar{E} = \bar{E}_1 + \bar{E}_2, \ \ V = V_1 + V_2, \ \ {\rm and} \ \ N=N_1 + N_2.
\end{equation}
The thermodynamic requirement of maximised total entropy means that any
imagined (or virtual) small changes in $(\bar{E}_1, V_1, N_1)$ or $(\bar{ E}_2, V_2, N_2)$ will, to
first order, give zero change in $S$. Since, therefore, Eqs.(329) imply that
\begin{equation}
\delta\bar{E}_1 + \delta\bar{E}_2 = 0, \ \ \delta V_1 + \delta V_2 = 0, \ \ 
{\rm and} \ \ \delta N_1 + \delta N_2 = 0,
\end{equation}
we find easily that
\begin{equation}
{\partial S\over \partial\bar{E}_1 } = {\partial S_1\over \partial\bar{E}_1 } - {\partial S_2\over \partial\bar{E}_2 } = 0,
\end{equation}
\begin{equation}
{\partial S\over \partial V_1 } = {\partial S_1\over \partial V_1 } - {\partial S_2\over \partial V_2 } = 0,
\end{equation}
and
\begin{equation}
{\partial S\over \partial N_1 } = {\partial S_1\over \partial N_1 } - {\partial S_2\over \partial N_2 } = 0.
\end{equation}
Reference to Eqs.(327) now shows that the conditions for complete thermal, 
mechanical and diffusive equilibrium are given by
\begin{equation}
T_1 = T_2, \ \ \bar{P}_1 = \bar{P}_2 \ \ {\rm and} \ \ \mu_1 = \mu_2.
\end{equation}
Hence the intensive variables are indicators of various types of mutual
equilibrium between systems. Further discussion, along the lines given
earlier for the temperature, points to the possibility of ordering their
values so as to be able to predict the direction of spontaneous change
when bodies with different values of the intensive parameters are placed
in contact. The final result is of course that if $T_1 > T_2$ then the first
system loses energy, if $\bar{P}_1 > \bar{P}_2$ it loses volume and if $\mu_1 > \mu_2$, it loses
particles.

The entropy maximisation principle can also be applied to derive some
useful conclusions about a {\it single\/} homogeneous isolated body, regarded as
made up of a large number of small subsystems in mutually equilibrated
contact with each other. Each subregion can be thought of as having its
own entropy, energy, volume and particle number which add up to give the
total $S$, $\bar{E}$, $V$ and $N$ of the isolated body. By arguments similar to those
given above, it is easy to see that at complete mutual equilibrium the
temperature,  pressure and chemical potential will be uniform throughout
the body. Furthermore, from the additivity of the extensive quantities,
it is clear that if the body is augmented in size by a factor $\lambda$, keeping
all the intensive parameters constant, then we can deduce at once that
\begin{equation}
S(\lambda\bar{E}, \lambda V, \lambda N) = \lambda S(\bar{E}, V, N),
\end{equation}
i.e., that $S$ is a homogeneous function of order unity in its variables.

Applying Euler's theorem on such functions we have that
\begin{equation}
\bar{E}\left ({\partial S\over \partial\bar{E} } \right ) + V\left ({\partial S\over \partial V } \right )
+ N\left ({\partial S\over \partial N } \right ) = S,
\end{equation}
which, on reference to Eqs.(327), yields easily that
\begin{equation}
\bar{E} = TS - \bar{P}V + \mu N.
\end{equation}
This result is useful in connection with the grand canonical method and
its associated potential function, which will be derived later.

We expect generally that once the entropy of a system has been measured,
or calculated, as a function of $(\bar{E}, V, N)$, then many other properties and
relations can be deduced confidently from the thermodynamical formalism.
However, the microcanonical method is only of somewhat limited interest,
for two reasons. One is that measurements are more usually made of the
temperature rather than the energy. The other is that it is very hard to
calculate accurately the number, $\Omega$, of accessible states, even in simple
models, since it is difficult to handle the constraint of fixed energy.
The canonical method is almost always easier because the definite energy
condition is replaced by knowledge of the simply measurable temperature,
which still corresponds to a sharply defined mean energy $\bar{E}$. Fortunately,
both methods, as well as the grand canonical treatment, when they can
all be carried to a conclusion, give essentially identical answers for
the properties of a macroscopic system. The reason is that nearly all
the quantities calculated as averages turn out to have extremely small 
expected deviations from their means.

We come now to the canonical formalism, in which the given equilibrium
data include the temperature $T$, the volume $V$ and the particle number $N$. 
Again, it is necessary for performing the statistical calculations that
we know also the Hamiltonian of the system and its energy spectrum. From
this information, the canonical method assigns probabilities for finding
the energy eigenstates $\vert\psi_j\rangle$, belonging to energies $E_j$, in the form
\begin{equation}
p_j = {\exp{[-E_j(V,N)/k_BT]} \over Z(T,V,N) },
\end{equation}
where
\begin{equation}
Z(T, V, N) = \sum_j \exp{[-E_j(V, N)/k_BT]}.
\end{equation}
The maximised uncertainty is given by $\sigma_{\rm max} = -\sum_j p_j\ln{p_j}$, and substitution
of Eqs.(338) and (339) into this expression soon yields, using $\bar{E} = \sum_j p_jE_j$,
\begin{equation}
k_B\sigma_{\rm max} = S = {\bar{E}\over T} +k_B\ln{[Z(T, V, N)]}.
\end{equation}
This suggests the definition of a new thermodynamic potential given by
\begin{equation}
F(T, V, N) =  \bar{E} - TS = -k_BT\ln{[Z(T, V, N)]}.
\end{equation}
The new function is called the Helmholtz Free Energy and we see that the canonical
formalism provides a theoretical expression for it in terms of $(T, V, N)$.
But these are precisely the natural variables for extracting the maximum
amount of information on the thermodynamic properties of the system. To
see this, consider small changes in $F = \bar{E} - TS$ to a nearby state, i.e.,
\begin{equation}
\delta F = \delta\bar{E} - T\delta S -S\delta T,
\end{equation}
and compare this result with the fundamental thermodynamic  equation
\begin{equation}
\delta\bar{E} = T\delta S -\bar{P}\delta V +\mu\delta N.
\end{equation}
Adding Eqs.(342) and (343) gives at once that
\begin{equation}
\delta F = -S\delta T -\bar{P}\delta V + \mu\delta N,
\end{equation}
which implies the differential relations
\begin{equation}
{\partial F\over \partial T} = -S, \ \ {\partial F\over \partial V} = -\bar{P}, \ \ 
{\rm and} \ \ {\partial F\over \partial N} = \mu.
\end{equation}
Hence, from a knowledge of $F$ and its natural derivatives, all quantities
of interest can be inferred. For example, since both $F$ and $T$ are assumed
known, and $S$ can be found by differentiation, the value of the energy $\bar{E}$ 
follows from the definition in Eq.(341). The above demonstration also 
shows, incidentally, that the results of the statistical treatment are
completely summarised by presenting the partition function, $Z(T, V, N)$,
and it is this function which is central to later developments.

The second of the Eqs.(345) is the thermodynamical version of the {\it Equation of
State\/} of the system, that is, a relation between $\bar{P}$, $V$, $T$ and $N$. Putting
$F = -k_BT\ln{[Z(T, V, N)]}$ into that equation we get
\begin{equation}
\bar{P} = k_BT{\partial\ln{[Z(T, V, N)]} \over \partial V},
\end{equation}
which shows how the equation of state arises in statistical mechanics.
This derivation is, however, based partially on thermodynamics, so it is
of some interest to demonstrate that the same equation also emerges more
directly from the probability formulation. To do this, we recall that $\bar{P}$,
the average pressure, is calculated from the pressures $P_j$ implied by the
individual energy eigenstates by means of the formula
\begin{equation}
\bar{P} = \sum_j P_jp_j = \sum_j \left ( {-\partial E_j \over \partial V} \right ) p_j.
\end{equation}
Substituting for $p_j$ from Eq.(338) we find
\begin{equation}
\bar{P} = \sum_j \left ( {-\partial E_j \over \partial V} \right ){\exp{[-E_j/k_BT]} \over Z}
\end{equation}
and an easy manipulation then gives
\begin{equation}
\bar{P} = {k_BT\over Z}\left ( {\partial \sum_j \exp{[-E_j/k_BT]} \over \partial V} \right ).
\end{equation}
But $\sum_j \exp{[-E_j/k_BT]} = Z$, so we finally obtain the result
\begin{equation}
\bar{P} = k_BT {\partial\ln{[Z(T, V, N]} \over \partial V},
\end{equation}
in agreement with Eq.(346).

This is further confirmation of the general consistency between thermodynamics
and statistical mechanics, under the assumption that physically 
measurable quantities should be identified with statistical averages. It
is indeed frequently  easier to use the statistical method to explore the
connections between experimental observables than to embark on the often
esoteric manipulations required by a purely thermodynamic approach. Even
for those cases in which mathematical difficulties prevent evaluation of
the sums (the great majority), it is still possible to gain insight.

To conclude this subsection we discuss the grand canonical method and its
associated thermodynamic potential. The assumed data are now considered
to comprise the temperature $T$, equivalent to a definite mean energy for
the system, the volume $V$ and the chemical potential $\mu$, which implies a
sharp estimated value of $\bar{N}$ for the number of particles in the body. The
macroscopic parameters $T$, $V$ and $\mu$ appear explicitly in the eigenstate
probability distribution, which is expressed by
\begin{equation}
p_{N,j} = {\exp{[-(E_j(V,N) - \mu N)/k_BT]} \over \Xi(T, \mu, V) },
\end{equation}
where
\begin{equation}
\Xi(T, \mu, V) = \sum_{N,j} \exp{[-(E_j(V,N) - \mu N)/k_BT]},
\end{equation}
in which the $E_j(V, N)$ are the energy eigenvalues of a system with volume
$V$ and definite particle number $N$. The maximised uncertainty is now given
by $\sigma_{\rm max} = -\sum_{N,j} p_{N,j}\ln{p_{N,j}}$ and on using Eqs.(351) and (352) together with
\begin{equation}
S = k_B\sigma_{\rm max}, \ \ \bar{E} = \sum_{N,j} p_{N,j}E_j(V,N) \ \ {\rm and} \ \
\bar{N} = \sum_{N,j} p_{N,j}N,
\end{equation}
we quickly arrive at the result
\begin{equation}
S= {\bar{E} \over T} - {\mu\bar{N} \over T} + k_B\ln{[\Xi(T, \mu, V)]}.
\end{equation}
If we now define a grand potential function by
\begin{equation}
\Omega_G = \bar{E} - TS - \mu\bar{N},
\end{equation}
then Eq.(354) shows that $\Omega_G$ is related to the grand partition function by
\begin{equation}
\Omega_G(T, \mu, V) = -k_BT\ln{[\Xi(T, \mu, V)]}.
\end{equation}
Yet again, the statistical method has led to a theoretical formula for a
potential function and, as we shall show, expressed it automatically in
its natural variables. It is also interesting that the given data and
the assumed information on the microscopic structure are all subsumed in
the grand partition function, which once more highlights the importance
of the probability normalisation factors.

Addition of the differential version of Eq.(355), namely
\begin{equation}
\delta\Omega_G = \delta\bar{E} - T\delta S - S\delta T - \mu\delta\bar{N} -\bar{N}\delta\mu,
\end{equation}
to the fundamental equation of thermodynamics,
\begin{equation}
\delta\bar{E} = T\delta S -\bar{P}\delta V +\mu\delta N.
\end{equation}
gives
\begin{equation}
\delta\Omega_G = -S\delta T -\bar{P}\delta V - \bar{N}\delta\mu,
\end{equation}
from which it follows that
\begin{equation}
{\partial\Omega_G \over \partial T} = -S, \ \ {\partial\Omega_G \over \partial V} = -\bar{P} \ \
{\rm and} \ \ {\partial\Omega_G \over \partial \mu} = -\bar{N}.
\end{equation}
Hence all quantities of thermodynamic interest are available when $\Omega_G$, or
equivalently $\Xi$, is known as a function of $T$, $\mu$ and $V$. Again, the results
have been derived in part by use of thermodynamics, so it is reassuring that,
for example, the second and third of Eqs.(360), with $\Omega_G$ replaced by
$-k_BT\ln{\Xi}$, can also be constructed directly from the probability functions
given above in Eqs.(351) and (352).

The grand potential $\Omega_G$ is a relatively little known thermodynamic state
function, but it can be expressed in terms of more familiar quantities.
This follows from the definition given in Eq.(355) and the general result
given in Eq.(337). Adding those two equations shows immediately that
\begin{equation}
\Omega_G = -\bar{P}V,
\end{equation}
and so, from Eq.(356), we deduce that
\begin{equation}
\bar{P}V = k_BT\ln{[\Xi(T, \mu, V)]},
\end{equation}
an equation of obvious utility in the theory of gases.

\subsection{Partition functions}

\noindent In the last subsection we saw that all thermodynamical deductions from the 
structure information used in the canonical and grand canonical methods
involved only a knowledge of the respective partition functions $Z(T, V, N)$
and $\Xi(T, V, \mu)$. Since these apparently trivial probability normalisation
factors actually embody all the relevant microscopic details, it is
useful to discuss their general properties. Also of interest is the way
they are related to the accessible state number $\Omega$ (another probability
normalisation factor) and to each other. We consider this aspect first.

We begin by exhibiting once more the characteristic bridging expressions
that connect thermodynamics and statistical theory. The standard methods
known as (i) the Microcanonical, (ii) the Canonical and (iii) the Grand
canonical formalisms lead logically to the propositions that
\begin{itemize}
\item{(i)} $S(\bar{E}, V, N) = k_B\ln{[\Omega(\bar{E}, V, N)]}$,
\item{(ii)} $F(T, V, N) = -k_BT\ln{[Z(T, V, N)]}$,
\item{(iii)} $\Omega_G(T, V, \mu) = -k_BT\ln{[\Xi(T, V, \mu)]}$,
\end{itemize}
with the entropy $S$, the free energy $F$, and grand potential $\Omega_G$, all being
expressed in their natural thermodynamic state variables.

The function $\Omega(\bar{E}, V, N)$ represents, for an isolated system, the number of
accessible energy eigenstates with energies in a small fixed interval $\Delta E$
surrounding the estimated total energy $\bar{E}$. The canonical method nominally
refers to a body in equilibrium with a heat reservoir at temperature $T$,
so that all energy eigenstates are theoretically possible, but for which
a sharply defined average energy exists, also denoted by $\bar{E}$. The quantity
at the centre of interest is now
\begin{equation}
Z(T, V, N) = \sum_j \exp{[-(E_j(V, N)/k_BT)]}.
\end{equation}
This is related  to the concept of accessible state number as follows. We
think of the energy variable as being dissected into intervals, each one
labelled by the energy $\bar{E}_k$ at its centre. If the interval $\Delta E_k$, around $\bar{E}_k$,
is taken small enough, then all the energy eigenvalues in that range can
be regarded as having the value $\bar{E}_k$ and the number of such states written
as $\Omega(\bar{E}_k, V, N)$. In other words, all the eigenstates with energies falling
in the interval $\Delta E_k$ are looked on as being effectively degenerate, with
degeneracy $\Omega(\bar{E}_k, V, N)$. The canonical partition function then appears as
\begin{equation}
Z(T, V, N) = \sum_k \Omega(\bar{E}_k, V, N) \exp{[-(E_k(V, N)/k_BT)]}.
\end{equation}
It seems from this that the calculation of $Z$ is much more difficult than
evaluation of just the single degeneracy factor called for in the microcanonical
method, since we now need to find the $\Omega(\bar{E}_k)$ numbers for many
energies and also perform the indicated sum for fixed particle number $N$.

The mathematical difficulties seem to be even further increased in the 
grand canonical theory. Reference to Eq.(352) of the last subsection and to 
Eq.(363) above shows that the grand partition function can be rewritten in 
terms of the canonical functions $Z(T, V, N)$, i.e.,
\begin{equation}
\Xi(T, V, \mu) = \sum_N \exp{(\mu N/k_BT)} Z(T, V, N).
\end{equation}
Hence we need to calculate $Z(T, V, N)$ for all values of $N$ and then perform
the sum in Eq.(365). So it is difficult to believe at first sight that for
many models in statistical physics the calculations actually become much
easier as we go from $\Omega \to Z \to \Xi$, since each successive stage requires all
that went before and more summations. The calculations would indeed be
considerably harder if they had to be done in that way. The constraints
of fixed $\bar{E}$ and $N$ are analytically inconvenient to maintain, but we see
that in the canonical theories the conditions are progressively relaxed,
being replaced by assignment of the physically significant parameters $T$
and $\mu$. The summations representing $Z$ and $\Xi$  can then often be rearranged
from the groupings displayed in Eqs.(364) and (365), so that the resulting 
unrestricted sums over all $E_j$ and all $N$ become much more tractable and,
in some cases, can be performed exactly.

Another useful remark concerns the factorisation properties of our basic
expressions. Consider first two bodies, labelled 1 and 2, in a state of 
mutual equilibrium mediated through surfaces allowing thermal and
possibly diffusive exchange. Observation establishes the existence of a 
well-defined energy $\bar{E}$ and a definite entropy $S$ for the combined system.
On separation, the two bodies are found to have individually observable
energies and entropies $(\bar{E}_1, S_1)$ and $(\bar{E}_2, S_2)$ which, to good approximation,
satisfy $\bar{E} = \bar{E}_1 + \bar{E}_2$ and $S=S_1+S_2$. The question now arises as to how these
facts are reflected in the formalism. If the couplings responsible for
the attainment of mutual equilibrium are only small perturbations on the 
separate Hamiltonians, then the energy eigenvalues  of the composite body
are the sums of the energies of the two subsystems and the total eigenstates 
are products, i.e.,
\begin{equation}
E_{j,k} = E_j(1) + E_k(2)
\end{equation}
and
\begin{equation}
\vert \chi_{j,k}\rangle = \vert\psi_j\rangle\vert\varphi_k\rangle.
\end{equation}
Using the microcanonical method, we have $\Omega(\bar{E}_1)$ and $\Omega(\bar{E}_2)$ as the numbers
of accessible states of the two bodies in intervals $\Delta E_1$, $\Delta E_2$ near $\bar{E}_1$ and
$\bar{E}_2$. Consequently, since any accessible state of the first body may occur
with any accessible state of the second one, there are $\Omega(\bar{E}) = \Omega(\bar{E}_1) \Omega(\bar{E}_2)$
states accessible to the combined system in the interval $\Delta E= \Delta E_1+\Delta E_2$
near $\bar{E} = \bar{E}_1 + \bar{E}_2$. We deduce that the total entropy $S$ is given by
\begin{eqnarray}
S &=& k_B\ln{[\Omega]} = k_B\ln{[\Omega_1\Omega_2]} \nonumber \\
&=& k_B\ln{[\Omega_1]} + k_B\ln{[\Omega_2]} \nonumber \\
&=& S_1 + S_2
\end{eqnarray}
Thus the additivity of entropy follows from the factorisation properties
of the accessible state numbers. The additivity of energies, of course, 
is an immediate consequence of Eq.(366), connecting the energy eigenvalues
of the combined and individual bodies in the corresponding small energy
intervals near $\bar{E}$, $\bar{E}_1$ and $\bar{E}_2$. All this arises naturally from the assumed
smallness of the interactions responsible for equilibrium, compared with
the bulk energies of the bodies.

The canonical method also implies a correspondence between thermodynamic
additivity and statistical factorisation. With the same assumptions and
notations as above, we now regard an observable energy as a well-defined
average over all states and recall that bodies in mutual equilibrium are
assigned the same temperature $T$. Writing $\beta= 1/k_BT$, the canonical method
leads to a partition function for the composite system in the form
\begin{eqnarray}
Z(T) &=& \sum_{j,k} \exp{[-\beta \{E_j(1) + E_k(2) \} ] } \nonumber \\
&=& \sum_j \exp{[-\beta E_j(1)]} \sum_k \exp{[-\beta E_k(2)]} \nonumber \\
&=& Z_1(T)Z_2(T).
\end{eqnarray}
Since the total energy $\bar{E}_1$ and free energy $F$, are given by
\begin{equation}
\bar{E} = -{\partial \ln{[Z]} \over \partial\beta }
\end{equation}
and
\begin{equation}
F = -k_BT\ln{[Z]}
\end{equation}
we have easily that
\begin{equation}
\bar{E} = \bar{E}_1 + \bar{E}_2
\end{equation}
and
\begin{equation}
F = F_1 + F_2.
\end{equation}
Also, from the definition $F = \bar{E} - TS$, and the fact that the systems have
the same temperature $T$, we can clearly conclude that
\begin{equation}
S = S_1 + S_2.
\end{equation}
It is equally straightforward to demonstrate the factorisability of the
grand partition function under the given conditions and hence to derive
the additivity of the grand potential and other thermodynamic functions.
The multiplicative properties of partition functions and the extensivity 
of the corresponding potentials apply also to a {\it single\/} system when it is
regarded as a  set of spatially disjoint subdivisions in thermal contact.

A further extension of these ideas is possible even for subsystems of a 
body which are not spatially separable. This arises when the degrees of
freedom in the body can be grouped into sets which interact with each
other only very weakly compared to the interactions within the sets. It
is then permissible to consider that each subset of dynamical variables,
labelled by $q$, $q=1\to Q$, has its own Hamiltonian and its own spectrum
of eigen-energies $\epsilon_q$. A typical total energy eigenvalue is written as
\begin{equation}
E = \sum_q \epsilon_q
\end{equation}
and, if all degrees of freedom are in mutual equilibrium at temperature
$T = 1/k_BT$, the partition function $Z=\sum_E \exp{[-\beta E]}$ factorises in the form
\begin{equation}
Z(\beta) = \prod_q Z_q(\beta).
\end{equation}
Hence, under these conditions, each subset contributes additively to the 
entropy, energy and thermal capacity since they all depend on $\ln{[Z(\beta)]}$.

One example of this is a metallic body for which, to a large extent, we
can discuss separately the vibrations of the crystal lattice, the motion
of the free electrons through the system and the atomic and nuclear spin
variables. Another good example is a molecular gas, in which the kinetic
motion of the molecules, and their electronic, vibrational, rotational
and spin degrees of freedom, are only weakly coupled. The residual small
interactions are, of course, necessary for the attainment of a complete
thermodynamic equilibrium, but they can often be ignored in explanations
of the gross characteristics of thermal behaviour in a physical system.

The experimental utility of these simple observations lies mainly in the 
fact that contributions of the various quasi-independent subsystems of a
body to its thermal properties very often become separately important in
different ranges of temperature. Measurement of thermal capacity, over a 
wide interval of $T$, frequently provides valuable information on several
types of microscopic structure. 

\subsection{Expected deviations}

\noindent The main assumption connecting the statistical theory with macroscopic
thermodynamics is that mean value estimates of observable quantities can be
identified with the experimentally reproducible values. It is clearly 
necessary therefore that expected deviations from the estimated numbers
should be small. We start the investigation of this point by indicating
how it can happen that observables like the total energy usually possess
well-defined mean values even under isothermal conditions. That is, we
wish to explain why a macroscopic system, with definite temperature,
retains a reproducibly constant energy when in contact with a heat
reservoir, in spite of the possibility of energy transfer.

The canonical description of a system at isothermal equilibrium assigns
a probability \hfil
$\exp{(-E_j/k_BT )}/Z(T)$ to the occurrence of an energy eigenstate
with eigenvalue $E_j$, so the {\it most probable\/} state is always the
ground state, whatever the temperature. This does not lead to a useful
estimate of the observable energy. The reason is that energy measurement
always has some error associated with it, so we should ask instead what
is the summed probability for finding eigenstates with energies in a 
small interval $\Delta E_k$, surrounding an energy value $\bar{E}_k$. As in the preceding
subsection, we write $\Omega(\bar{E}_k)$ for the number of such states near $\bar{E}_k$, each
with probability $\exp{(-\bar{E}_k/k_BT )}/Z(T)$. The total probability of finding such a state
with energy in the interval $\Delta E_k$ is then given by $\Omega(\bar{E}_k)\exp{(-\bar{E}_k/k_BT )}/Z(T)$.
For a macroscopic system the number $\Omega(\bar{E}_k)$ is a very rapidly increasing
function of $\bar{E}_k$. On multiplying it by the decreasing exponential factor,
we find that the grouped probability expression will have its maximum
value well away from the ground state energy, except at the very lowest
temperatures. This maximum is so extremely sharply peaked as a function 
of energy that the resulting estimate of the energy most likely to be 
observed can just as well be replaced by the mean value expression
\begin{equation}
\bar{E} = \sum_j E_j {\exp{(-\bar{E}_j/k_BT )} \over Z(T) } \approx
\sum_k \bar{E}_k\Omega(\bar{E}_k){\exp{(-\bar{E}_k/k_BT )} \over Z(T) }.
\end{equation}
The expected deviation then measures the reliability of this estimate.

The simple calculation required has already been given in our treatment
of energy transfer between bodies at different temperatures, but will be
repeated here for convenience. We revert to the notation $\beta = 1/(k_BT)$ for the
temperature parameter and write
\begin{equation}
\bar{E}(\beta) = \sum_j E_j {\exp{(-\bar{E}_j/k_BT )} \over Z(T) } =
-{\partial\ln{[Z(\beta)]} \over \partial\beta}.
\end{equation}
We also employ the convention of taking the zero of energy as that of an
empty cavity which would exactly enclose the system so that, at constant
values of the intensive parameters, $\bar{E}$ is proportional to $N$, the particle
number. The mean square deviation of the energy from the average is now
\begin{eqnarray}
(\Delta E)^2 &=& \sum_j {(E_j-\bar{E})^2 \exp{(-\beta E_j)} \over Z} \nonumber \\
&=& \sum_j {E_j^2 \exp{(-\beta E_j)} \over Z} - \bar{E}^2.
\end{eqnarray}
Differentiating Eq.(378) with respect to $\beta$ we find
\begin{equation}
-{\partial \bar{E} \over \partial \beta} = \sum_j {E_j^2 \exp{(-\beta E_j)} \over Z} + 
{1 \over Z}\left ( {\partial Z \over \partial\beta} \right )
\sum_j {E_j \exp{(-\beta E_j )} \over Z },
\end{equation}
which shows, using Eq.(378) again, that
\begin{equation}
-{\partial \bar{E} \over \partial \beta} = {\sum_j E_j^2 \exp{(-\beta E_j)} \over Z} - \bar{E}^2
\end{equation}
From Eqs.(379) and (381) we deduce that
\begin{equation}
(\Delta E)^2 = -{\partial \bar{E} \over \partial \beta} = {\partial^2\ln{[Z]} \over \partial\beta^2}
\end{equation}
and we see that yet another physically significant quantity is derivable
from a knowledge of the partition function. A measure of the reliability
of our energy estimate in Eq.(377) is now provided by the square of the 
{\it relative\/} energy deviation, i.e.,
\begin{equation}
{(\Delta E)^2 \over \bar{E}^2 } = -\left ( {1 \over \bar{E}^2} \right ) {\partial \bar{E} \over \partial \beta}
= \left ( {k_BT^2 \over \bar{E}^2 } \right ) {\partial \bar{E} \over \partial T},
\end{equation}
which has been expressed in terms of $\partial\bar{E}/\partial T = C_V$, the thermal capacity of
the body at constant volume.

Since the $T$ is independent of the size of the system and $\bar{E}$ is extensive,
we can conclude from its definition that the thermal capacity will be
proportional to $N$. The relative energy deviation must therefore satisfy
\begin{equation}
{\Delta E \over \bar{E} } \propto {1 \over \sqrt{N} }.
\end{equation}
For macroscopic systems, which have $N = 10^{22}$ or greater, we can safely
conclude that deviations from the estimated average will be negligible.
A similar argument shows that the expected energy deviation given by the
grand canonical method for a macroscopic body is also small compared to
the mean energy. The result for $(\Delta E)^2$ has the same form as above, i.e.,
\begin{equation}
(\Delta E)^2 = -{\partial \bar{E} \over \partial \beta} = {\partial^2\ln{[\Xi]} \over \partial\beta^2}
\end{equation}
where $\Xi$ is the grand partition function. It gives again a result of the 
form of Eq.(384), though with $N$ replaced by the mean particle number $\bar{N}$.

When using the grand canonical method we should also check that expected
deviations away from the calculated mean particle number $\bar{N}$ are small, so
that our results correspond to reproducible observables. We have that
\begin{equation}
\beta\bar{N} = \beta\sum_{N,j} {N\exp{[-\beta(E_j-\mu N)]} \over \Xi} = {\partial \ln{[\Xi]} \over \partial\mu},
\end{equation}
while the mean square deviation is easily shown to be given by
\begin{equation}
\beta^2(\Delta N)^2 = \beta{\partial\bar{N}\over\partial\mu} = {\partial^2\ln{[\Xi]} \over \partial\mu^2},
\end{equation}
where the steps in the derivation parallel those given for the energy in
the canonical method and the grand partition function is defined by
\begin{equation}
\Xi = \sum_{N,j} \exp{[-\beta\{E_j(N) - \mu N\} ]}.
\end{equation}
Finally, since $\mu$ and $\beta$ are both intensive quantities and $\bar{N}$ is clearly
extensive, we can quickly deduce that
\begin{equation}
{\Delta N \over \bar{N}} \propto {1 \over \sqrt{\bar{N}} },
\end{equation}
which is completely negligible for a macroscopic system.

Similar considerations apply to estimates of other extensive quantities
and generally support the reliability of statistical calculations. It is
also of some interest that expected deviations from the calculated mean
values of intensive properties like the pressure can be disregarded. As
a by-product of the discussions in the next subsection we shall show that
the mean square deviation of the pressure away from its average value $\bar{P}$
is given by
\begin{equation}
\sum_j (P_j - \bar{P} )^2p_j = (\Delta P)^2 = {(\kappa_a - \kappa_{\beta})\over \beta V},
\end{equation}
$\kappa_a$ and $\kappa_{\beta}$ being, respectively, the adiabatic and isothermal bulk moduli
of the body. A bulk modulus is defined by $\kappa = -V(\partial\bar{P}/\partial V)$, the inverse of
the corresponding compressibility, and is clearly intensive. Since $\beta$ is
also intensive, while the volume $V$ is extensive, we see that $(\Delta P)^2$  must
be proportional to $1/N$. The estimated relative deviation of the pressure
thus takes on the now familiar standard form
\begin{equation}
{\Delta P \over \bar{P}} \propto {1 \over \sqrt{N} },
\end{equation}
We remark in passing that the result in Eq.(390) shows that the adiabatic 
modulus is always greater than the isothermal modulus, since $(\Delta P)^2$ is
by definition strictly positive for a system having two or more eigenstates
with non-zero probability.

The arguments do sometimes fail for systems near to a phase change. One
example is a fluid kept at the constant temperature and pressure which
mark a point of transition between liquid and vapour. The heat capacity
has a singular behaviour under these conditions and there will be large
and uncontrollable fluctuations in the energy and volume. Another breakdown
of the statistical method occurs in the treatment of an ideal Bose
gas below a critical temperature. Macroscopic numbers of particles can
condense into their ground states and the grand canonical formalism has
to be modified in order to describe the situation adequately. But, apart
from such special conditions, the validity of the statistical estimation
of reproducible observables is backed up by a vast mass of experience.

\subsection{Slow adiabatic changes}

\noindent We turn now to another and more fundamental application of the formalism
of error estimation. The problem to be discussed concerns a crucial link
between thermodynamics and statistical mechanics. It is worth a careful
look since, if not resolved, it throws into doubt the whole statistical
foundation of thermal physics.

The question at issue is whether quasi-static and reversible adiabatic
processes are also isentropic. Processes which are both adiabatic and 
isentropic are easily definable in pure thermodynamics and are routinely
assumed to be possible. They appear, for example, as components of the
Carnot cycle used to set up an absolute temperature scale independent of
particular substances and are thus vital to the phenomenological theory.
But it is not at all clear that the assumption of their existence for an
arbitrary system is consistent with quantum mechanics and with our basic
statistical interpretation of entropy as maximised uncertainty.

The potential difficulty arises from our microscopic explanation of the
Second Law of Thermodynamics. We showed in section 5, in all generality,
that an arbitrary adiabatic process could never decrease the maximised
uncertainty, i.e., the entropy of a body; but this still leaves open the
possibility of adiabatic changes in which the entropy is unaltered. Here
we will investigate the matter again and show that, even under the most 
favourable conditions of quasi-static change, an adiabatic process will
almost always {\it increase\/} the entropy.

For definiteness, we shall consider a thermally isolated system of fixed
particle number $N$ whose thermodynamic state at equilibrium is adequately
described in terms of the variables $S$, $\bar{E}$, $\bar{P}$, $T$ and $V$. It will be assumed
that changes from an initial equilibrium state are induced by externally 
controlled variations of the volume $V$ which cause work to be done on, or
by, the system.

In thermodynamics the differences in the variables describing nearby
equilibrium states satisfy the fundamental equation
\begin{equation}
\delta\bar{E} = T\delta S - \bar{P}\delta V
\end{equation}
to first order in quantities $\delta x$ which represent small but finite changes
in the parameters $x$. This follows also from the statistical formalism if
we identify $S$ with the maximised uncertainty, multiplied by Boltzmann's
constant. To connect this with dynamics it is assumed in addition that a
system can be brought from one equilibrium state to another by means of
a quasi-static and reversible process, during which the system remains
effectively in equilibrium at all times. This is a slightly vague notion
which becomes sharp only in the limit of infinitesimally slow changes.
under these conditions, $-\bar{P}\delta V$ can be interpreted as the work done on the
system in each stage of the process, while $T\delta S$ is taken to represent the
heat added, their sum giving the increment $\delta\bar{E}$ in the internal energy.
(It should be noted that these {\it interpretations\/} are not allowable for any
other kind of process, though Eq.(392) is generally valid for initial and
final states of {\it equilibrium\/}. The difference is that $T\delta S$ and $\bar{P}\delta V$ can then
no longer be regarded as heat and work terms.)

However, for a quasi-static, reversible process performed on a thermally
isolated body, the condition of zero heat transfer is always assumed to
imply that $\delta S = 0$ in all intermediate steps. Thus, from Eq.(392), we deduce
that the system should obey the differential equation
\begin{equation}
\left ( {\partial\bar{E} \over \partial V} \right )_S = -\bar{P},
\end{equation}
which can be integrated to give the final energy in terms of the initial
energy and the values of $\bar{P}$ along the adiabatic path. It is also possible
to obtain, from pure thermodynamics, a differential equation describing
the change of temperature along the path, in terms of other observables.
The result is
\begin{equation}
\left ( {\partial T \over \partial V} \right )_S = -\left ( {T\over C_V} \right )
\left ( {\partial \bar{P} \over \partial T} \right )_V,
\end{equation}
where $C_V$ is the thermal capacity of the body at constant volume, but for
our purposes it is not necessary to give the derivation of this.

The entropy of the final state is, of course, the same as in the initial
state since $\delta S$ has been set equal to zero in the intervening steps. Thus
in thermodynamics the given description of quasi-static, reversible and
adiabatic processes seems to imply the conservation of entropy. All this
is perfectly consistent with the Second Law, which claims only that an
adiabatic change can not decrease the entropy. But the isentropic nature
of this {\it special\/} process is far from obvious when we consider it from the 
point of view of quantum mechanics, yet require to maintain our previous
microscopic interpretation of the entropy.

The quantum treatment of such processes relies on Ehrenfest's Principle,
also called, somewhat confusingly, the Adiabatic Theorem\cite{[Ehrenfest]}. This says that
if a pure work process, as described by a time-dependent modification of
the Hamiltonian, is performed slowly enough, then any initial eigenstate
of energy will remain at later times an eigenstate of the slowly varying
energy operator. We will not pause here to give the rather lengthy proof
of this result, but merely note that it is plausible. The point of this
theorem for statistical mechanics is that during such quasi-static and
adiabatic variations our initial state of knowledge about the system is
preserved. At the start, the equilibrium state is described by assigning
canonical probabilities to the possible energy eigenstates. At the end,
each possible eigenstate will have evolved smoothly and uniquely from a 
definite initial eigenstate, there will have been no quantum transitions
to neighbouring states and we can assign the new canonical probabilities
from a knowledge of the final mean energy. The important observation now
is that, according to the quantum theory, the final mean energy can be
estimated by using the initial probabilities associated with the varying
but smoothly connected eigenstates, since nothing in the process we have
described suggests any revision of those probabilities. The initial and
final uncertainties about which state occurs will thus be equal and, at
first sight, this seems to confirm that the entropy will be unchanged.

The difficulty is that entropy is taken as the {\it maximised\/} uncertainty of
a final equilibrium state and is based solely on the observable data for
that state. Thus, in spite of being able to obtain the final mean energy
from the initial probabilities and the final eigen-energies, there is no
reason to believe that the {\it canonical\/} probability of any final state will
be equal to the probability of the uniquely corresponding initial state.
But if they are not all equal then the final maximised uncertainty will
inevitably be larger than the initial uncertainty  and this, according to
our microscopic interpretation, implies greater entropy.

In order to see this more clearly, consider an initial equilibrium state
of volume $V$ and temperature parameter $\beta = 1/(k_BT)$, with energy eigenstates
and eigenvalues $\vert\psi_j(V)\rangle$ and $E_j(V)$, canonical probabilities
\begin{equation}
p_j(V,\beta) = {\exp{[-\beta E_j(V)]} \over Z(V, \beta)},
\end{equation}
and mean energy
\begin{equation}
\bar{E}(V, \beta) = \sum_j p_j(V,\beta)E_j(V).
\end{equation}
At the end of a slow adiabatic process the volume will have a different
value, say $V_F$, and the smoothly evolved eigenstates and eigenvalues will
be $\vert\psi_j(V_F)\rangle$ and $E_j(V_F)$; but, from the quantum theorem, the probabilities
of the states can be carried over unchanged. The final average energy of
the body can thus be reliably represented by
\begin{equation}
\bar{E}_F(V_F) = \sum_j p_j(V,\beta) E_j(V_F).
\end{equation}
Application of the usual statistical method, assuming the above value of
$\bar{E}_F$, will result in the assignment of equilibrium canonical probabilities
for the final states in the form
\begin{equation}
p_j(V_F,\beta_F) = {\exp{[-\beta_F E_j(V_F)]} \over Z(V_F, \beta_F)},
\end{equation}
in which the final temperature $T_f = 1/(k_B\beta_F)$ is related to the energy by
\begin{equation}
\bar{E}_F(V_F) = \sum_j p_j(V_F,\beta_F) E_j(V_F) = -{\partial \ln{[Z(V_F,\beta_F)]} \over
\partial \beta_F }.
\end{equation}
Now we saw in subsection 5.2 that, given the same data values, a
proposed probability distribution $\{q_j\}$ has the maximum uncertainty only
if every $q_j$ equals the corresponding canonical probability $p_j$. Since the
Eqs.(397) and (399) say that the initial state probabilities and the final
canonical distribution are both consistent with the same final value of
the energy, $\bar{E}_F(V_F)$, we deduce that the entropy will be unchanged after a
slow adiabatic process only if, for {\it every\/} $j$,
\begin{equation}
p_j(V,\beta) = p_j(V_F,\beta_F).
\end{equation}
Using Eqs.(395) and (398), this implies that for any two pairs of corresponding
states, labelled by $j$ and $k$, the energy eigenvalues should obey
\begin{equation}
\beta_F[E_j(V_F) - E_k(V_F)] = \beta[E_j(V) - E_k(V)],
\end{equation}
i.e., that the energy spectrum at the end becomes uniformly expanded or
compressed in the ratio $\beta/\beta_F$ relative to the initial spectrum. It should
be abundantly clear that this will not happen in general, since for an 
arbitrary system the initial and final spectra are completely determined
by the microscopic structure. An interesting exception is the ideal gas,
in which the energy eigenvalues are all proportional to $1/V^{2/3}$, so that
the equations can be satisfied by choosing $\beta_F = \beta(V_F/V)^{2/3}$. For almost all
other systems, there is no single value of $\beta_F$  which can ensure the truth
of Eq.(401). We conclude that Eq.(400) can not hold for all $j$, and hence
that the entropy will have increased.

Even more disturbingly, we see that there is something inconsistent in
the notion of adiabatic reversibility. If the process causes the entropy
to increase as we go from $V$ to $V_F$, then reversibility should imply that
the entropy will decrease to its initial value as we return to volume $V$.
But we have shown that this is certainly not possible in {\it any\/} adiabatic
change, since such a result would contravene the Second Law.

The only conceivable way out of these difficulties is to attempt to show
that the entropy increase in a slow but finite work process is generally
negligible. A consideration of the extreme sharpness of the statistical
estimates of thermodynamic variables provides a clue in this direction.
After an adiabatic change, the overwhelming concentration of probability
attaches to eigenstates with energies near to the final mean energy $\bar{E}_F$.
It is therefore quite plausible that the final canonical distribution in
that region could be a good approximation to the canonical probabilities
for corresponding states at the beginning of the process, which, as seen
earlier, should lead to a reliable prediction of the final energy.

To support this idea we need to derive explicit estimates for changes in the
observables when the volume $V$ of a thermally isolated body is varied
by a small but finite amount $\delta V$. We shall use the notation $(Dx)_a$ for the
adiabatic increment of a quantity $x$, expanded up to order $(\delta V)^2$ in the
slowly varied control parameter $V$. The term adiabatic here refers to the
quantum description and implies that evaluation of $(Dx)_a$ is always to be
based on energy changes obtained by holding the eigenstate probabilities
constant at their initial values.

Starting as before with an internal equilibrium state of mean energy $\bar{E}$,
volume $V$ and canonical temperature parameter $\beta$, the adiabatic change in
energy, induced by a small modification of volume, is represented  by
\begin{equation}
(D\bar{E})_a = \left ( {\partial\bar{E} \over \partial V} \right )_a (\delta V) +
{1\over 2}\left ( {\partial^2\bar{E} \over \partial V^2} \right )(\delta V)^2.
\end{equation}
According to our assumptions, this expression can be interpreted in two
different ways which, for consistency, must agree.

By the adiabatic quantum theorem, the possible energy states and eigenvalues 
will move smoothly to the new ones and corresponding states will
retain their initial canonical probabilities $p_j(\beta, V) = \exp{[-\beta E_j(V) - \ln{Z}]}$.
Hence, with fixed $\beta$ and $V$ in $p_j(\beta,V)$, the change  in $\bar{E} = \sum_j E_j(V)p_j$
can be estimated from the increments in the $E_j(V)$ alone, i.e.,
\begin{equation}
(D\bar{E})_a = \left [ \sum_j  \left ({\partial E_j \over \partial V} \right ) p_j  \right ](\delta V) 
+ {1 \over 2} \left [ \sum_j \left ( {\partial^2E_j\over \partial V^2} \right ) p_j \right ] (\delta V)^2.
\end{equation}
Since the pressure in state $\vert\psi_j(V)\rangle$  is $P_j = -\partial E_j/\partial V$, the
coefficient of $\delta V$ is just $-\bar{P}$, where $\bar{P} = \sum_j P_jp_j$ is the initial
mean pressure. Comparison with Eq.(393) then shows that, {\it at the start of the process\/},
we have
\begin{equation}
\left ( {\partial \bar{E} \over \partial V} \right )_a = -\bar{P} = 
\left ( {\partial \bar{E} \over \partial V} \right )_S,
\end{equation}
and we see that the quantum description of quasi-static adiabatic change
leads to exactly the same {\it initial\/} value of the derivative as the thermodynamic
account, which is based on the assumption that the entropy stays 
constant. This state of affairs does not hold for higher derivatives and
we shall show that discrepancies between quantum statistical theory and
ordinary thermodynamics appear already in the next order.

Using the canonical probability formula $p_j = \exp{[-\beta E_j(V)-\ln{Z(\beta,V)}]}$, the
summation in the second term of Eq.(403) can be rewritten as follows
\begin{eqnarray}
\sum_j \left ( {\partial^2 E_j\over \partial V^2} \right ) p_j &=& 
\left ({\partial [\sum_j (\partial E_j/\partial V)p_j]\over \partial V} \right )_{\beta} -
\sum_j \left ( {\partial E_j \over \partial V} \right )
\left ( { \partial p_j\over \partial V} \right )_{\beta} \nonumber \\
&=& -\left ( { \partial \bar{P}\over \partial V} \right )_{\beta} +
\beta\sum_j P_j(P_j-\bar{P})p_j \nonumber \\
&=& -\left ( { \partial \bar{P}\over \partial V} \right )_{\beta} +
\beta\sum_j (P_j-\bar{P})^2p_j,
\end{eqnarray}
where we have obtained the second equality from the usual definitions of
$P_j$ and $\bar{P}$ and have also recalled the results $\beta\bar{P} = (\partial\ln{Z}/\partial V)_{\beta}$.
The last form comes from the obvious substitution of $(P_j-\bar{P})$ for the factor $P_j$ in the
second term, which is clearly allowable since $\sum_j\bar{P}(P-\bar{P})p_j=0$. Hence,
on writing $(\Delta P)^2$ for the mean square pressure deviation, we find that
\begin{equation}
(D\bar{E})_a = -\bar{P}\delta V + {1\over 2}\left [ \beta(\Delta P)^2 - 
\left ({\partial\bar{P} \over \partial V} \right )_{\beta} \right ] (\delta V)^2.
\end{equation}
The alternative way of viewing the expression for $(D\bar{E})_a$ is to employ our
basic statistical assumption that the intervening equilibrium states can
be described by canonical distributions with varying $\beta$ values. Thus, for
consistency to first order, we should be able to write
\begin{equation}
\left ( {\partial\bar{E} \over \partial V} \right )_a = 
\left ( {\partial\bar{E} \over \partial V} \right )_{\beta} + 
\left ( {\partial\bar{E} \over \partial \beta} \right )_V
\left ( {\partial\beta \over \partial V} \right )_a = -\bar{P}.
\end{equation}
This gives a useful formula for the initial adiabatic derivative of the
canonical temperature. From the definition $\bar{E}(\beta,V)=\sum_jE_j(V)p_j(\beta,V)$ we get
\begin{eqnarray}
\left ( {\partial\bar{E} \over \partial V} \right )_{\beta} &=&
\sum_j \left ( {\partial E_j \over \partial V} \right )p_j +
\sum_j E_j \left ( { \partial p_j\over \partial V} \right )_{\beta} \nonumber \\
&=& -\bar{P} + \beta\sum_j(E_j-\bar{E})(P_j-\bar{P})p_j \nonumber \\
&=& -\bar{P} + \beta\langle(\Delta E)(\Delta P)\rangle,
\end{eqnarray}
where $\langle(\Delta E)(\Delta P)\rangle$ denotes the expected value of the products of pressure
and energy deviations. We also know from Eq.(397) of the last subsection that
\begin{equation}
\left ( {\partial\bar{E} \over \partial \beta} \right )_V = -\sum_j(E_j-\bar{E})^2p_j = 
-(\Delta E)^2.
\end{equation}
Substituting Eqs.(408) and (409) into Eq.(407) now yields
\begin{equation}
\left ( {\partial\beta \over \partial V} \right )_a = \beta{\langle(\Delta E)(\Delta P)\rangle \over
(\Delta E)^2}.
\end{equation}
Under the statistical interpretation of measured quantities, this agrees with the thermodynamic result
for quasi-static and adiabatic temperature change quoted in Eq.(394). Such
agreement is to be expected since the quantum theory and the isentropic
assumption of ordinary thermodynamics both lead to the initial value of
the volume derivative of $\bar{E}$ being equal to $-\bar{P}$, as recorded in Eq.(404). It
should be noted, however, that Eq.(410) holds {\it only\/} at the initial state.

The second order volume derivatives of $\bar{E}$ are not in general equal in the 
two theories. In the quantum treatment we have from Eq.(406) that
\begin{equation}
\left ( {\partial^2\bar{E} \over \partial V^2} \right )_a = 
-\left ( {\partial\bar{P} \over \partial V} \right )_a = \beta(\Delta P)^2 - 
\left ( {\partial\bar{P} \over \partial V} \right )_{\beta},
\end{equation}
while the corresponding derivative taken at constant entropy is given by
\begin{equation}
\left ( {\partial^2\bar{E} \over \partial V^2} \right )_S = 
-\left ( {\partial\bar{P} \over \partial V} \right )_S = 
-\left ( {\partial\bar{P} \over \partial \beta} \right )_V
\left ( {\partial\beta \over \partial V} \right )_S -
\left ( {\partial\bar{P} \over \partial V} \right )_{\beta}.
\end{equation}
A simple manipulation, using the definition $\bar{P} = \sum_jP_jp_j$, quickly shows that
$(\partial\bar{P}/\partial\beta)_V=-\langle(\Delta E)(\Delta P)\rangle$.
Also, since $(\partial\beta/\partial V)_S = (\partial\beta/\partial V)_a$,
as mentioned above, we can use Eq.(410) to rewrite Eq.(412) in the form
\begin{equation}
-\left ( {\partial\bar{P} \over \partial V} \right )_S = 
\beta {\langle(\Delta E)(\Delta P)\rangle^2 \over (\Delta E)^2} 
-\left ( {\partial\bar{P} \over \partial V} \right )_{\beta} 
\end{equation}
Comparison of Eqs.(411) and (413) now leads to a {\it necessary\/} condition for
complete agreement between the quantum and thermodynamic descriptions of
quasi-static and adiabatic processes, namely, that we should have
\begin{equation}
(\Delta P)^2 - {\langle(\Delta E)(\Delta P)\rangle^2 \over (\Delta E)^2}  = 0.
\end{equation}
The expression in Eq.(414) is related to the correlation coefficient for
energy and pressure and it can vanish exactly only if $P_j = aE_j + b$ for all
$j$, where $a$ and $b$ are independent of $j$. This is almost certainly not true
for real bodies, but the undoubted success of thermodynamics does seem 
to imply that the condition holds approximately, with both terms small.

We remark at this point that Eq.(411) provides the justification for our
estimate of the $N$-dependence of the relative pressure deviation given at
the end of the last subsection. Since any bulk modulus of a body is defined
by $\kappa = -V(\partial P/\partial V)$, we have immediately that the difference of adiabatic and
isothermal bulk moduli is related to the expectation value of the square
pressure deviation by
\begin{equation}
{(\kappa_a - \kappa_{\beta})\over\beta V} = (\Delta P)^2,
\end{equation}
which shows clearly that $(\Delta P)$ is proportional to $1/\sqrt{N}$. It is also worthy
of remark that $(\Delta P)^2$ can not be calculated by differentiation of $\ln{Z}$, so
that we have at last a quantity that is not accessible from knowledge of
the partition function alone. All observables of ordinary thermodynamics
{\it are\/} directly related to $\ln{Z}$ and we see that the quantum version of the
adiabatic  bulk modulus must differ from the thermodynamic version which is
evaluated at constant entropy. Indeed, this analogue of Eq.(415) coming
from thermodynamics can be read off from Eq.(413) and takes the form
\begin{equation}
{(\kappa_a - \kappa_{\beta})\over\beta V} = {\langle(\Delta E)(\Delta P)\rangle^2 \over (\Delta E)^2}.
\end{equation}
This is derivable from $\ln{Z}$, but does not lead to an estimate of $(\Delta P)/\bar{P}$.

We have now developed all the equations necessary for the evaluation of
the change in the entropy to second order in the volume increment $\delta V$. We
start from the uncertainty expression
\begin{equation}
\sigma(V,\beta) = \beta\bar{E}(V,\beta) + \ln{[Z(V,\beta)]},
\end{equation}
and require to calculate the differential coefficients in the formula
\begin{equation}
(D\sigma)_a = \left ({\partial\sigma\over\partial V}\right)_a(\delta V) + 
{1\over 2}\left ({\partial^2\sigma\over\partial V^2}\right )_a(\delta V)^2.
\end{equation}
The adiabatic subscript implies that the derivatives are to be computed
from the changes in energy $\bar{E}$ and effective canonical temperature $\beta$ which
are given by the quantum description of the process. It is also implied 
that these coefficients denote rates of change evaluated at the initial
equilibrium values of the parameters $V$ and $\beta$.

The first adiabatic derivative of $\sigma$ is given by
\begin{equation}
\left ({\partial\sigma\over\partial V}\right)_a = 
\beta\left ({\partial\bar{E}\over\partial V}\right)_a +
\bar{E}\left ({\partial\beta\over\partial V}\right)_a +
\left ({\partial\ln{Z}\over\partial V}\right)_{\beta} + 
\left ({\partial\ln{Z}\over\partial \beta}\right)_V
\left ({\partial\beta\over\partial V}\right)_a
\end{equation}
Since $(\partial\ln{Z}/\partial\beta)_V=-\bar{E}$ and, by hypothesis, the adiabatic variation of $\bar{E}$
can be represented by using a canonical distribution with suitably chosen $\beta$,
the second and fourth terms in Eq.(419) will cancel at all stages of the
process. Similarly, we can use the relation $(\partial\ln{Z}/\partial V)_{\beta} = \beta\bar{P}$, where
$\bar{P}$ is the canonically estimated pressure at any stage, to obtain the result
\begin{equation}
\left ({\partial\sigma\over\partial V}\right)_a = 
\beta \left [\left ({\partial\bar{E}\over\partial V}\right ) + \bar{P} \right ].
\end{equation} 
We shall consider the implications of this more generally below, but for
the moment we note only that the {\it initial\/} state adiabatic derivative of $\bar{E}$
has been shown in Eq.(404) to be given by $(\partial\bar{E}/\partial V)_a = -\bar{P}$,
in which $\bar{P}$ is now the initial pressure. Hence the starting value of 
$(\partial\sigma/\partial V)_a$ is zero.

Differentiating again, the second adiabatic derivative takes the form
\begin{equation}
\left ({\partial^2\sigma\over\partial V^2}\right)_a = 
\left ({\partial\beta\over\partial V}\right)_a \left [ \left ({\partial\bar{E}\over\partial V}\right)_a + \bar{P} \right ]
+\beta \left [ \left ({\partial^2\bar{E}\over\partial V^2}\right)_a + 
\left ({\partial\bar{P}\over\partial V}\right)_{\beta} + \left ({\partial\bar{P}\over\partial \beta}\right)_V 
\left ({\partial\beta\over\partial V}\right)_a \right ].
\end{equation}
The expression in square brackets in the first term of Eq.(421) again has
the initial value of zero and we can substitute in the second term the 
initial value results recorded in Eqs.(410) and (411), together with the 
expression for $(\partial\bar{P}/\partial\beta)_V$ given immediately after Eq.(412). Inserting these
derivatives into Eq.(418), we find, to order $(\delta V)^2$ in volume change, that
the increment of uncertainty over its original value is given by
\begin{equation}
(D\sigma)_a = \left ({\beta^2\over 2} \right )
\left [ (\Delta P)^2 - {\langle(\Delta E)(\Delta P)\rangle^2 \over (\Delta E)^2} \right ]
(\delta V)^2.
\end{equation}
This is our basic result for slow adiabatic processes and we deduce once
again that a necessary condition for them to be also isentropic is that
\begin{equation}
(\Delta P)^2 - {\langle(\Delta E)(\Delta P)\rangle^2 \over (\Delta E)^2} = 0
\end{equation}

For real systems, however, it can easily be shown that the expression on
the left hand side of Eq.(423) has a value greater than zero. Denoting it
by $D$ and using the transcriptions of the expectation values in terms of
the initial canonical probabilities, we soon find that it takes the form
\begin{equation}
D = \left [{1\over (\Delta E)^4} \right ] \sum_j [(P_j-\bar{P})(\Delta E)^2 -
(E_j - \bar{E})\langle(\Delta E)(\Delta P)\rangle ]^2p_j.
\end{equation}
This is a sum of squares weighted by the eigenstate probabilities and is
thus strictly positive unless every square vanishes, that is, for all $j$,
\begin{equation}
(P_j-\bar{P})(\Delta E)^2 - (E_j - \bar{E})\langle(\Delta E)(\Delta P)\rangle = 0.
\end{equation}
We have here an exact version of the linear relation between pressures
and energies, mentioned after  Eq.(414), which is required to hold if the
statistical and thermodynamic descriptions are to agree. It is extremely
unlikely that Eqs.(425) are valid for any real system. On the other hand,
it is quite plausible that among all eigenstates which carry appreciable
probability the overwhelming majority will have values of $P_j$ and $E_j$ very
close to the respective mean values $\bar{P}$ and $\bar{E}$. We have indeed argued from
the beginning that this {\it must\/} be true if statistical mechanics is to have
any hope of providing sharply defined estimates of macro-variables. This
applies, of course, only to reproducible properties of a system, but the
observed pressures and energies certainly fall into that category, so we
expect that the criterion in Eq.(423) will be satisfied to quite adequate
approximation. We conclude that for all practical purposes the increase
of entropy in a slow adiabatic process can be neglected.

The above argument is almost, but not entirely, convincing as a solution
to this special problem of reconciling statistical theory with accepted
results of thermodynamics, since there could conceivably be surprises in
higher orders of the type of calculation just given. It is therefore of
some interest to show that linear relations between the pressures $P_j$ and
energies $E_j$ are in fact sufficient, as well as necessary, conditions for a
quasi-static and adiabatic change to be isentropic. To see this, we go
back to Eq.(420), which holds throughout the process, and recall that the
quantum mechanical estimate of the mean energy is always represented by
\begin{equation}
\bar{E}(V) = \sum_j E_j(V) p_j(V_I,\beta_I),
\end{equation}
at all $V$, with the probabilities being held at their initial equilibrium
values $p_j(I)$. As the volume is varied, the value of $\bar{E}(V)$ is used at each
stage to set up a new canonical distribution $\{p_j(V,\beta)\}$ by maximising the
uncertainty. The varying pressure can then be estimated from the formula
$\bar{P}(V,\beta) = \sum_jP_jp_j(V,\beta)$. Substituting Eq.(426) into Eq.(420) 
we now find that
\begin{equation}
\left ( {\partial\sigma\over\partial V} \right )_a = \beta\sum_j 
[\bar{P}(V,\beta) - P_j(V)]p_j(V_I,\beta_I),
\end{equation}
where we have used the result $P_j(V) = -\partial E_j(V)/\partial V$ and
brought $\bar{P}$ inside the bracket by invoking the normalisation of the 
probabilities.

It is easy to see that the derivative will not vanish in general, except 
at the initial values of $V$ and $\beta$, since the probabilities used to define
$\bar{P}$ will have changed from their starting values. But it is also obvious
that a sufficient condition for it to vanish everywhere is that each $P_j$
is related to the corresponding eigenstate energy $E_j$ by $P_j(V) = aE_j(V)+b$,
where $a$ and $b$ can depend on $V$ but not on $j$. This follows because the sum
in Eq.(427) is then proportional to $\sum_j[\bar{E}(V) - E_j(V)]p_j(I)$, which is zero
by the definition of an adiabatic process displayed in Eq.(426).

When the linear relation which is both necessary and sufficient for zero
entropy change is not satisfied, the adiabatic derivative of $\sigma$ is either
zero or positive (for increasing $V$). But segments of the adiabatic path
must exist along which $(\partial\sigma/\partial V)$ is strictly positive, since the entropy
will increase. Still, the underlying rationale of statistical mechanics
does suggest very strongly that the slope of $\sigma$ is unlikely to grow large
enough to be significant. The reason has already been given above and is
just that most states likely to occur will have individual pressures $P_j$
which are very close in value to the mean pressure $\bar{P}(V,\beta)$ evaluated from
the canonical distribution for the given $V$ and $\beta$. It if were not so, the
formalism could not describe reproducible measurements. When it is so,
the derivative will be small whatever the initial state probabilities.
Assuming, then, that the adiabatic path does not cross a phase boundary,
where large deviations are possible, we can ignore the entropy increase.
The argument applies also for a decrease of $V$ from its initial value, in
which case the proved increase of entropy implies that the derivative of
$\sigma$ will be negative. The apparent discrepancy between thermodynamics and
statistical mechanics has thus been effectively resolved.

It is, however, of some theoretical interest that the discrepancy occurs
at all and a few final comments are in order. Just as the quantum theory
estimate of the adiabatically changing energy is $\bar{E}(V) = \sum_jE_j(V)p_j(I)$, the
most reasonable estimate of the actual pressure exerted by the system at
any stage is given by $P(V) = \sum_jP_j(V)p_j(I)$, where $P_j(V) = -\partial E_j(V)/\partial V$. This
will in general be different from the mean pressure $\bar{P}(V,\beta)$ derived using
the canonical equilibrium probabilities at that volume, which have been
assigned from the value of $\bar{E}(V)$. We can now write Eq.(427) in the form
\begin{equation}
\left ( {\partial\sigma\over\partial V} \right )_a = \beta[\bar{P}(V,\beta) - P(V)].
\end{equation}
We know that for increasing $V$ this is never negative and will in fact be
positive on the average, so that $P(V)$ is generally less than $\bar{P}(V,\beta)$. The
implication is that the actual external work done by the body during the
expansion will be a little smaller than the work estimated canonically.
If the body is compressed, then $P(V)$ will on average be greater than the
pressure $\bar{P}(V,\beta)$, since the entropy must still increase, so the work done
on the system will be slightly larger than the canonical estimate. In either
case, the final energy of the system will be greater than that predicted
from the isentropic assumption of thermodynamics.

We have thus shown that the terms $T\delta S$ and $P\delta V$ in Eq.(392) can not strictly
be regarded as representing increments of heat and work in this special
quasi-static process, though such an interpretation is adequate for most
purposes. We have also seen that the process can not strictly be thought
of as reversible. If the internal mechanics of the system were entirely
under control then it would be possible to reverse any changes. But only
a few macroscopic variables are reproducible enough for use in physics
and even these are subject to measurement errors as well as fluctuations
from their estimated mean values. The irreversibility arises because we
are forced to supplement the mechanics with probability arguments, which
are always such that uncertainty will increase as we move away from some
initial equilibrium state. The uneasy blending of dynamics and inference
has been a constant theme in statistical mechanics since its inception.
 
\subsection{Classical formalism}
 
\noindent All the expressions involving the canonical partition function $Z(V,\beta)$ or
the grand canonical partition function $\Xi(V, \beta, \mu)$ were derived using the quantum
theory and these functions are discrete sums over the quantum numbers of
energy eigenstates. What happens when we go over to Classical Mechanics? There are then no
quantum states and all dynamical quantities are continuous variables. In
particular, the system energy can take on any value and we may therefore
expect that our previous sums will be replaced by integrals. \hfil
\\[1\baselineskip]

\leftline{{\underline {BUT:}} How are we to define System States in Classical Mechanics?}

\noindent One possibility is to specify all the coordinates necessary to describe
the system together with all the corresponding momenta which define its
state of motion. The set of current values of these quantities provide a
label $j = \{x_i,p_i\}, i=1,\ldots,3N$, where $N$ is the number of particles. This
denotes a point in what is called the phase space of the system and its
variation in time gives a complete picture of the evolution of the body.

Thus the quantum energy state can be replaced by a phase function, i.e.,
\begin{equation}
E_{\rm quantum}(j) \Rightarrow E_{\rm classical}(\{x_i,p_i\}).
\end{equation}
Now in classical mechanics the energy expressed in terms of coordinates
and momenta is called the Hamiltonian of the system. Hence by analogy
with the quantum treatment we can try to write the canonical formula as
\begin{equation}
{\rm Probability} \propto \exp{[-H(\{x_i,p_i\})/k_BT]}
\end{equation}
\medskip
\leftline{{\underline {BUT:}}  How are we to calculate the Number of States?}
\medskip
\noindent This problem must be solved before we can normalise the probabilities.
The method, going  back to Boltzmann, is to break up Phase-Space into small
cells of Phase-Volume
\begin{equation}
d^{3N}xd^{3N}p \Rightarrow dx_1\ldots dx_{3N}dp_1\ldots dp_{3N},
\end{equation}
and assume that the number of states with values of $\{x_i,p_i\}$ within this
little hypercube is proportional to the given volume element. Hence for
the normalisation constant of probability we are led to write
\begin{equation}
Z = \sum_j \exp{[-\beta E_j]} \Rightarrow \int d^{3N}x d^{3N}p \ \exp{[-\beta H(\{x_i,p_i\})]}.
\end{equation}
To make the classical version of $Z$ dimensionless we need to divide this
integral by a suitable unit of phase-volume. Nothing suggests itself in
classical theory, but the quantum theory of the kinetic motion of atoms
does provide the natural unit $(2 \pi \hbar )^{3N}$. Inserting this we get
\begin{equation}
Z_{\rm classical} = \int {d^{3N}xd^{3N}p \over (2\pi\hbar)^{3N}} \ 
\exp{[-\beta H(\{x_i,p_i\})]}.
\end{equation}
Most physical quantities that we require to calculate may be obtained
from $\ln{[Z_{\rm classical}]}$, so the additive constant $\ln{[(2\pi\hbar)^{3N}]}$ makes no difference.

But the constant is important when calculating the entropy and the other
thermodynamic functions which depend directly on $\ln{[Z]}$. So this remnant
of quantum mechanics is still relevant for some purposes and the first
versions of statistical mechanics only gave an answer for the entropy to
a then unknown constant. Despite this, classical statistical theory
was and is surprisingly successful in explaining many facts of thermal
physics and usually only goes catastrophically wrong at very low values
of the temperature and also where individual quantum effects in atoms or
molecules are still important at room temperature. Historically, it was
certain discrepancies between the classical calculations and observation 
of specific heats of gases and solids that provided strong clues to the
presence of quantum phenomena.

Finally, in calculating the properties of gases, it turns out that there
should be a further trace of quantum mechanics left in the formalism of
classical statistics. The additive property of entropy is only obtained
if $\ln{[Z_{\rm classical}]}$ contains a factor $(1/N!)$.
\vfil
\newpage

\section{Conclusions}

\noindent We have separated the principal concerns of statistical mechanics into two distinct
problems - one concerning the treatment of probability and the other concentrating on
the explicit mechanical aspects of the calculations. By taking a Bayesian view of the 
probability question we have shown that the development of the subject can be expressed in a highly 
logical fashion, much closer to the dictates of ``common sense''. To illustrate this point
we have presented a didactic account of the probability calculus of Cox which is likely to be
unfamiliar to many physicists but is ideal for this development and as such worthy of wider 
dissemination and a place in future text books.

Our main aim has been to make an unbiased assignment of probabilities by maximising Shannon's 
information entropy 
\begin{equation}
\sigma(\{p_j\}) = -\sum_j p_j\ln{(p_j)}, \nonumber
\end{equation} 
subject to the constraints implied by the data $K$ and normalisation. We have then shown that
this function $\sigma_{\rm max}(\{p(j\vert K)\})$ possesses all the properties of the
thermodynamic entropy $S$. It is an extensive function of state, defined for 
equilibrium conditions, that does not decrease in adiabatic processes and is related
experimentally to heat and temperature. 

Once the identification of $\sigma_{\rm max}$ with the phenomenological
entropy $S$ was made it was possible to develop statistical mechanics from it and
to re-derive many well known results and links to thermodynamics. We investigated
the traditional microcanonical, canonical and grand canonical methods with the constraints
on normalisation, mean energy and mean particle number accommodated by means of physically
significant Lagrange multipliers. We not only found a maximum of the uncertainty function
but demonstrated explicitly that it was indeed the abolute maximum of our entropy function.

In the course of further development we came upon one unexpected and truly startling
result concerning the question of whether quasi-static and reversible adiabatic
processes are also necessarily isentropic. The answer in traditional thermodynamics to 
this question is assumed to be definitively ``yes''. However, in our analysis it 
presents more subtle facets and is not a trivial point to answer. We have found that
such quasi-static, reversible processes, requiring the entropy to remain 
constant, are only possible if the final and initial energy eigenspectra scale in the ratio of the 
final to initial temperatures. Although this can be achieved in the simplest idealised systems 
such as the perfect gas, it is not generally possible for more complicated systems. We showed that
it effectively requires the vanishing of the correlation coefficient between energy and pressure 
deviations, which can only happen if $P_j = aE_j + b$ for all states $j$.

\end{document}